\pdfoutput=1
\documentclass[aps,prd,preprint,superscriptaddress,tightenlines,nofootinbib]{revtex4}
\usepackage{graphicx}
\usepackage{epsfig}
\usepackage{bm}
\usepackage{latexsym,amssymb,amsmath,amsfonts,amssymb,txfonts,pxfonts,wasysym,float}
\usepackage{textcomp}
\usepackage{xspace}
\usepackage{color}

\newcommand{\postscript}[2]{\setlength{\epsfxsize}{#2\hsize}
   \centerline{\epsfbox{#1}}}
\newcommand{\PRE}[1]{{#1}}   

\newcommand{\comment}[1]{}

\def\Offline{\mbox{$\overline{\textrm%
{Off}}$\hspace{.05em}\protect\raisebox{.4ex}%
{$\protect\underline{\textrm{line}}$}}\xspace}

\usepackage[usenames,dvipsnames]{xcolor}
\definecolor{orange}{cmyk}{0,0.5,1,0}
\definecolor{rossoCP3}{cmyk}{0,.88,.77,.40}
\definecolor{graa}{rgb}{0.8,0.8,0.8}
\definecolor{blaa}{rgb}{0.2,0.2,0.6}

\begin{document}

\title{\color{rossoCP3}{White paper on EUSO-SPB2
}}

\author{James H. Adams Jr.}
\affiliation{Center for Space Plasma and Aeronomic Research,\\
University of Alabama in Huntsville, Huntsville, AL 35899, USA
\PRE{\vspace*{.05in}}
}

\author{Luis A.~Anchordoqui}

\affiliation{Department of Physics \& Astronomy,\\
Lehman College, City University of  New York, NY 10468, USA
\PRE{\vspace*{.05in}}
}

\affiliation{Department of Astrophysics, American Museum of Natural History, NY
 10024, USA
\PRE{\vspace*{.05in}}
}

\author{Jeffrey \nolinebreak  A. \nolinebreak  Apple}
\affiliation{Marshall Space Flight Center, Huntsville, AL 35812, USA
\PRE{\vspace*{.05in}}
}

\author{Mario \nolinebreak E. \nolinebreak Bertaina}
\affiliation{Dipartamento di Fisica, Universit\'a di Torino, Torino, Italy
\PRE{\vspace*{.05in}}
}

\author{Mark \nolinebreak J. \nolinebreak Christl}
\affiliation{Marshall Space Flight Center, Huntsville, AL 35812, USA
\PRE{\vspace*{.05in}}
}

\author{Francesco Fenu}
\affiliation{Dipartamento di Fisica, Universit\'a di Torino, Torino, Italy
\PRE{\vspace*{.05in}}
}

\author{Evgeny \nolinebreak Kuznetsov}
\affiliation{Center for Space Plasma and Aeronomic Research,\\
University of Alabama in Huntsville, Huntsville, AL 35899, USA
\PRE{\vspace*{.05in}}
}

\author{Andrii \nolinebreak Neronov}
\affiliation{Astronomy Department,\\
University of Geneva, Ch. d’Ecogia 16, 1290, Versoix, Switzerland
\PRE{\vspace*{.05in}}
}

\author{Angela V. Olinto} 
\affiliation{Department of Astronomy \& Astrophysics,
Enrico Fermi  Institute and  Kavli Institute for Cosmological Physics, University of Chicago, Chicago, IL 60637, USA
\PRE{\vspace*{.05in}}
}

\author{Etienne Parizot}
\affiliation{Universit\'e ́ Paris Diderot, Laboratoire Astro Particule
  et Cosmologie, CNRS, Paris, France
\PRE{\vspace*{.05in}}
}
\affiliation{Observatoire de Paris, Sorbonne Paris Cit\'e, 119 75205 Paris, France
\PRE{\vspace*{.05in}}
}

\author{Thomas \nolinebreak C. \nolinebreak  Paul}
\affiliation{Department of Physics \& Astronomy,\\
Lehman College, City University of  New York, NY 10468, USA
\PRE{\vspace*{.05in}}
}

\author{Guillaume Pr\'ev\^ot}
\affiliation{Universit\'e ́ Paris Diderot, Laboratoire Astro Particule
  et Cosmologie, CNRS, Paris, France
\PRE{\vspace*{.05in}}
}

\author{Patrick \nolinebreak Reardon}
\affiliation{Center for Space Plasma and Aeronomic Research,\\
University of Alabama in Huntsville, Huntsville, AL 35899, USA
\PRE{\vspace*{.05in}}
}

\author{Ievgen Vovk}

\affiliation{Max Planck Insitut fur Physik, Foehringer Ring 6, 80805, Munich, Germany 
\PRE{\vspace*{.05in}}
}

\author{Lawrence Wiencke} \affiliation{Department of Physics, Colorado
  School of Mines, Golden, CO 80401, USA
\PRE{\vspace*{.05in}}
}

\author{Roy M. Young}
\affiliation{Marshall Space Flight Center, Huntsville, AL 35812, USA
\PRE{\vspace*{.05in}}
}

\begin{abstract}\vskip 2mm
  \noindent EUSO-SPB2 is a second generation Extreme Universe Space
  Observatory (EUSO) on a Super-Pressure Balloon (SPB). This document
  describes the physics capabilities, the proposed technical design
  of the instruments, and the simulation and analysis software. 
\end{abstract}
\maketitle

\tableofcontents

\newpage

\section{SCIENCE OBJECTIVES}
\subsection{General idea}

We propose to monitor the night sky of the Southern hemisphere with a
second generation of the Extreme Universe Space Observatory (EUSO) instrument, to
be flown aboard a Super-Pressure Balloon (SBP). This mission,
EUSO-SPB2, has several exploratory and scientific objectives.

EUSO-SPB2 will be the first instrument to measure air-showers by
viewing their Cherenkov light from high in the atmosphere. We expect
to observe a rather large sample of cosmic rays in the energy range
$10^7 \alt E/{\rm GeV} \alt 10^8$, with the aim of discriminating 
among primary protons, heavy nuclei, and photons via their
characteristic Cherenkov profiles~\cite{Neronov:2016iax}. The instrument will also be
able to characterize the background for upward going showers initiated
by the decay of tau leptons which are expected to be produced by
Earth-skimming tau neutrinos~\cite{Neronov:2016zou}.

In addition to detection of Cherenkov light, we plan to use
fluorescence light from air showers to measure, for the first time, the evolution of
nearly horizontal extensive air showers, which develop at high
altitude in a nearly constant density atmosphere. Such measurements
will provide a unique channel to tune hadronic interaction models at
ultrahigh energies, and may elucidate the reason why ultrahigh-energy cosmic
ray (UHECR) showers observed by ground-based detectors contain more
muons than expected from existing hadronic interaction models~\cite{Aab:2016hkv}.

Importantly, EUSO-SPB2 will serve as a pathfinder for the more
ambitious space-based measurements by the Probe Of Extreme
Multi-Messenger Astrophysics (POEMMA), selected by NASA for an
in-depth probe mission concept study in preparation for the next
decadal survey.  POEMMA will combine the well-developed Orbiting
Wide-field Light-collectors (OWL) concept~\cite{Stecker:2004wt} with
the recently proposed CHerenkov from Astrophysical Neutrinos Telescope
(CHANT) concept~\cite{Neronov:2016zou} to form a multi-messenger probe
of the most extreme environments in the universe.

EUSO-SPB2 will build upon the experience of flying EUSO-SPB in the
Spring of 2017. A number of upgrades will render EUSO-SPB2 more
powerful, including a Schmidt design reflecting telescope and a faster
ultraviolet (UV) camera to increase exposure to UHECR
observations. The new instrument will detect the fluorescence signal
from UHECR generated air-showers of highly inclined events.  EUSO-SPB2
will be built to view the true horizon of the Earth.  Horizontal
observations will lead to much larger acceptances for inclined UHECRs, with a
distance-dependent energy threshold. We are also prepared to consider
additional nadir observations of the fluorescence based on EUSO-SPB
results. The combination of nadir (EUSO-SPB) and tilted (EUSO-SPB2)
observations will explore the power of space observatories to observe
UHECR of extreme energies.  A long enough flight of EUSO-SPB2
observations will match and complement ground
observations~\cite{Anchordoqui:2013eqa}.

In addition to improving the exposure to UHECRs, EUSO-SPB2 will study
the possibility of detecting tau neutrinos via direct Cherenkov
light~\cite{Neronov:2016zou}.  A coincidence veto will be
developed for EUSO-SPB2 so it can characterize the background for
Cherenkov signals from the decay of tau leptons, produced in charged
current interactions of Earth skimming neutrinos.
EUSO-SPB2 will inform the best strategy for future space missions such
as POEMMA.

The detectors aboard EUSO-SPB2 will  measure the Cherenkov signals
from nearly horizontal air-showers initiated by high-energy cosmic
rays in the upper atmosphere. The instrument will use the technique of
imaging atmospheric Cherenkov telescopes (IACT), which is widely used
in contemporary gamma-ray astronomy. We expect a large statistical
sample which will allow for study of spectral features and composition
in an interesting energy regime. Since good distinguishing power
between baryon and photon induced showers has been shown to be
feasible for an IACT~\cite{Neronov:2016iax} we can theoretically reach
a competitive photon sensitivity with a similar airborne instrument.

EUSO-SPB2 addresses the fourth science goal of the 2011 NASA Strategic
Plan~\cite{nasa1}, to ``Discover how the universe works, explore how
it began and evolved'' and one of the ``Physics of the Cosmos''
questions in NASA's 2010 Science Plan~\cite{nasa2}: ``How do matter,
energy, space, and time behave under the extraordinarily diverse
conditions of the cosmos?''  EUSO-SPB2 directly addresses the sixth
question in the Connecting Quarks with the Cosmos report~\cite{cqc} in
this report's list of ``Eleven Science Questions for the New
Century,'' which is ``How do Cosmic Accelerators Work and What are
They Accelerating?''  EUSO-SPB2 science is in line with the NASA
Astrophysics Roadmap of 2013 missions for the next 3
decades~\cite{roadmap}. Upcoming measurements of EUSO-SPB2 are
essential to achieve the ambitious recommendations of the U.S. HEP
Snowmass planning process: ``The Bright Side of the Cosmic Frontier:
Cosmic Probes of Fundamental Physics''~\cite{Beatty:2013lza}.

\subsection{Observational status of high- and ultrahigh-energy cosmic rays}
\label{sec-CR}

The origin(s) of cosmic rays remains a challenging enigma of particle astrophysics. The energy
spectrum is known to span about eleven decades of energy, $1 \alt E/{\rm GeV}
\alt 10^{11}$. The spectral shape can be described by a broken power law with  three major breaks: the
steepening of the spectrum dubbed the ``knee'' at
$E \approx 10^{6.6}~{\rm GeV}$~\cite{Antoni:2005wq}, a pronounced hardening
of the spectrum at $E \approx 10^{9.6}~{\rm GeV}$, the so-called ``ankle''
feature~\cite{Bird:1993yi, Abbasi:2007sv,Abraham:2010mj}, and 
the high frequency cutoff at $E \approx10^{10.6}~{\rm GeV}$~\cite{Abbasi:2007sv,Abraham:2008ru}.
Three additional more subtle features have been reported over the
years in between the
knee and the ankle: a hardening of the spectrum at 
$10^{7.3}~{\rm GeV}$~\cite{Apel:2012tda,Aartsen:2013wda,
  Knurenko:2013dia, Prosin:2014dxa} followed by two softenings at
$10^{7.9}~{\rm GeV}$~\cite{Apel:2012tda, Aartsen:2013wda} and 
$10^{8.5}~{\rm GeV}$~\cite{AbuZayyad:2000ay, Bergman:2007kn,
  Knurenko:2013dia, Prosin:2014dxa}. The latter softening is usually 
referred to as the ``second knee.''


The variations of the spectral index in the energy spectrum reflect
various aspects of cosmic ray production, source distribution, and
propagation. The first and the second knee reflect characteristic
energy scales of magnetic confinement and/or acceleration capability
of the sources, both of which grow linearly in the charge $Z$ of the
nucleus.  The first knee has been studied by a number of experiments
and its characteristics are well represented by magnetic rigidity
effects for different nuclear composition~\cite{Hoerandel:2002yg}. The
physical significance of the second-knee, however, is less well
established and the energy at which the Galactic extragalactic
transition takes place is an open question.

From existing IACT data, we can estimate the EUSO-SBP2 sensitivity to
high-energy cosmic ray events. If we assume a trigger threshold of 100 photoelectrons with 100 days of
data collection, we find that EUSO-SPB2 can reach a sensitivity
for an energy-squared weighted flux of $\sim 3 \times 10^{-10}~{\rm GeV
  \, cm^{-2} \, s^{-1} \, sr^{-1}}$, in the energy range $10^7 <
E/{\rm GeV} < 10^8$. The average cosmic ray flux
in this decade of energy is $\sim 3 \times 10^{-7}~{\rm GeV \, cm^{-2}
  \, s^{-1} \, sr^{-1}}$~\cite{Olive:2016xmw}. 
Measurement of the electron and muon component of the shower
will be possible via statistical analysis of the full
data sample. Such information will be invaluable for understanding the
nuclear composition providing an
opportunity to clarify the origin of the second-knee. Observation in
this energy regime will also contribute to understanding of other
subtle features reported in the spectrum.

The simplest interpretation of the ankle is that above $10^{9.6}~{\rm
  GeV}$ a new population emerges which dominates the more steeply
falling Galactic population of heavy nuclei. The extragalactic
component can be dominated either by protons~\cite{Bird:1993yi} or
heavies~\cite{Allard:2005ha,Allard:2005cx}, with the highest energy particles being 
subject to photopion production and photodisintegration,
respectively. This is the mechanism behind the well-known {\it
  Greisen-Zatsepin-Kuz'min} (GZK)
cutoff~\cite{Greisen:1966jv,Zatsepin:1966jv}.  It has also been
advocated that the ankle feature could be well reproduced by a
proton-dominated power-law spectrum, where the ankle is formed as a
{\it dip} in the spectrum from the energy loss of protons via
Bethe-Heitler pair production~\cite{Hillas:1967,Berezinsky:2002nc}. In
this case extragalactic protons would already have started to
dominate the spectrum somewhat beyond $10^{8.7}~{\rm GeV}$.  Optical
observations of air showers with fluorescence telescopes or
non-imaging Cherenkov detectors consistently find a predominantly
light composition at around $10^9~{\rm GeV}$~\cite{Kampert:2012mx} and
that the contribution of protons to the overall cosmic ray flux is
$\agt$~50\% in this energy
range~\cite{Aab:2014kda,Aab:2014aea,Aab:2016htd,Aab:2016zth}.  Due to the
absence of a large anisotropy in the arrival direction of cosmic rays
below the ankle~\cite{Auger:2012an,ThePierreAuger:2014nja}, we can
conclude that these protons must be of extragalactic origin.  At
energies above $10^{9.4}~{\rm GeV}$, the high-statistics data from the
Pierre Auger Observatory suggests a gradual increase of the fraction
of heavy nuclei in the cosmic ray flux~\cite{Aab:2014aea,
  Aab:2014kda,Aab:2016htd,Aab:2016zth}. Within uncertainties, the data
from the Telescope Array (TA) are consistent with these
findings~\cite{Abbasi:2014sfa, Abbasi:2015xga}.  In addition, TA has
observed a statistically significant excess in cosmic rays with
energies above $57~{\rm EeV}$ in a region of the sky spanning
about $20^\circ$, centered on equatorial coordinates (R.A. =
$146.7^\circ$, Dec. = $43.2^\circ$)~\cite{Abbasi:2014lda}. This is
colloquially referred to as the TA hot spot. The absence of a
concentration of nearby sources in this region of the sky corroborates
other experimental evidence for heavy nuclei, whereby a few local
sources within the GZK sphere can produce the hot spot through
deflection in the extragalactic and Galactic magnetic
fields.\footnote{Beyond the GZK energy threshold, observable sources
  must lie within about 100~Mpc, the so called GZK horizon, or GZK sphere.} 

The Galactic to extragalactic transition is likely to extend over a wide range of
energies. For protons, the transition is thought to occur in the KASCADE-Grande
 light ankle at $E \approx 10^8~{\rm GeV}$~\cite{Apel:2013ura}. The relative abundance of Galactic nuclei decreases gradually, with nuclei of larger $Z$ decreasing in abundance more slowly than those of lower $Z$. There are a number of models proposed for this interesting energy range where a transition from Galactic to extragalactic may occur~\cite{Aloisio:2013hya,Unger:2015laa,Globus:2015xga}. An increase in observations of the varying components in this energy range will help determine the correct model.

\begin{figure}[tbp]
\begin{minipage}[t]{0.49\textwidth}
    \postscript{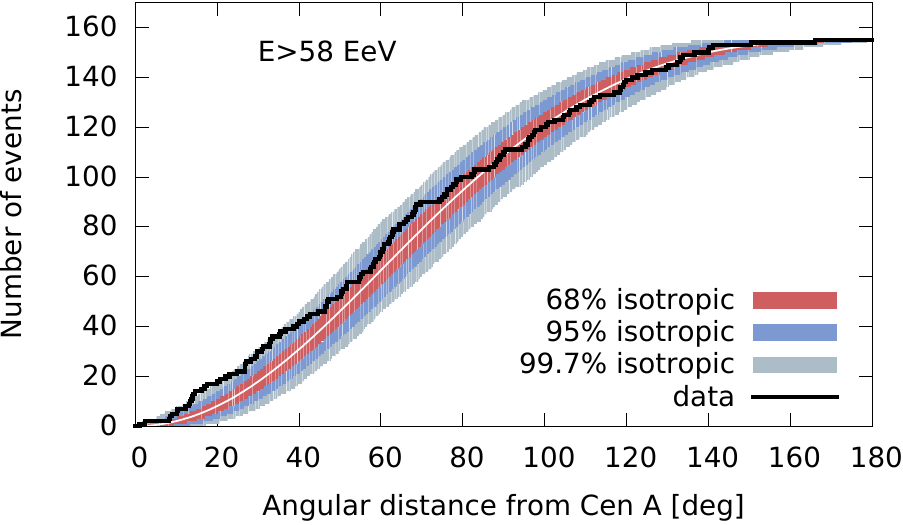}{0.8}
\end{minipage}
\begin{minipage}[t]{0.49\textwidth}
    \postscript{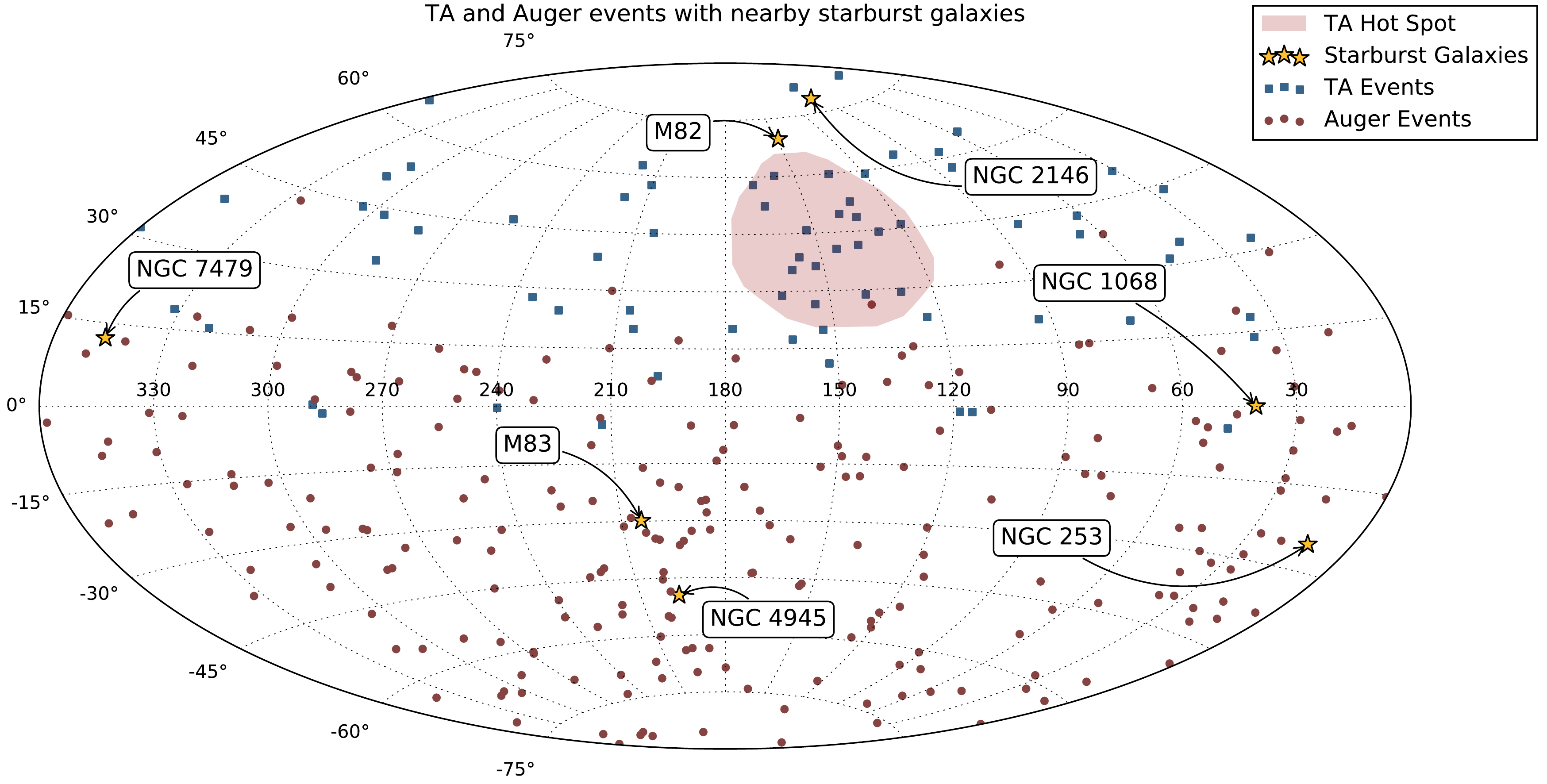}{0.99}
\end{minipage}
\caption{{\it Left.} Correlation of Auger events with Cen
  A. The black line indicates the cumulative number of events as a
  function of the angular range,  for the threshold $E_{\rm th} =
  58~{\rm EeV}$~\cite{PierreAuger:2014yba}. {\it Right.}~Comparison of UHECR event locations with nearby starburst galaxies in equatorial coordinates, with R.A. increasing from right to left. The
circles indicate the arrival directions of 231 events with $E >
52~{\rm EeV}$ and zenith angle $\theta < 80^\circ$ detected by the
Pierre Auger Observatory from 2004 January 1 up to 2014 March
31~\cite{PierreAuger:2014yba}. The squares indicate the arrival
directions of 72 events with $E > 57~{\rm EeV}$ and $\theta <
55^\circ$ recorded from 2008 May 11 to 2013 May 4 with TA~\cite{Abbasi:2014lda}. The
stars indicate the location of nearby (distance $< 50~{\rm Mpc}$)
starburst galaxies in the \textit{Fermi}-LAT
catalog~\cite{Ackermann:2012vca}, with flux emission (or upper limit)
in the gamma ray band $0.1 < E_\gamma/{\rm GeV} < 100$ bigger than $5 \times
10^{-9}~{\rm cm^{-2} \,  s^{-1}}$. The shaded region delimits the TA hot-spot.
\label{fig:1}}
\end{figure}

A plethora of source candidates have been poposed, among the most
popular being active galactic nuclei (AGNs), starburst galaxies, and
gamma-ray bursts (GRBs)~\cite{Torres:2004hk,Kotera:2011cp}. AGNs are
actively-accreting super-massive black holes and are sometimes
associated with jets terminating in lobes, which can be detected in
radio. The so-called ``radiogalaxies'' are a sub-class of AGNs, which
contain localized regions of intense synchrotron emission known as hot
spots.  These regions are presumably produced when the bulk kinetic
energy of the jets ejected by a central AGN is reconverted into
UHECRs~\cite{Biermann:1987ep,Rachen:1992pg}. UHECR acceleration is
also possible in polar cap regions of the black hole
magnetosphere~\cite{Neronov:2007mh,Moncada:2017hvq}. Centaurus A (Cen A) is the
closest radiogalaxy to Earth and has long been suspected to be a
potential UHECR accelerator~\cite{Cavallo:1978,Romero:1995tn}. The
Pierre Auger Collaboration has searched for anisotropies in the
direction of Cen A scanning the energy threshold between
$10^{10.6}~{\rm GeV}$ and $10^{10.9}~{\rm GeV}$ and counting events in
angular radii ranging from $1^\circ$ to
$30^\circ$~\cite{PierreAuger:2014yba}.  The strongest departure from
isotropy (post-trial probability $\sim 1.4\%$) has been  observed
for $E > 58~{\rm EeV}$ in a window of $15^\circ$, see Fig.\ref{fig:1};
14 events (out of a total of 155) have been observed in such an
angular window while 4.5 are expected on average from isotropic
distributions. Starbursts are galaxies undergoing a large-scale star
formation episode. Their characteristic signatures are strong infrared
emission (originating in the high levels of interstellar extinction), a
very strong HII-region-type emission-line spectrum (due to a large
number of O and B-type stars), and a considerable radio emission
produced by recent supernova remnants. UHECRs could be efficiently
accelerated at the terminal shock of a galactic-scale superwind, which
is driven by the collective effect of supernovae and massive star
winds~\cite{Anchordoqui:1999cu}.  Because of the high prevalence of
supernovae, starbursts should posses a large density of newly-born
pulsars.  Due to their important rotational and magnetic energy
reservoirs these young neutron stars have been explored as a potential
engine for UHECR
acceleration~\cite{Blasi:2000xm,Fang:2012rx,Fang:2013cba}. A recent
study~\cite{Kotera:2015pya} demonstrates that for the most reasonable
range of neutron star surface temperatures ($T < 10^7~{\rm K}$), a
large fraction of heavy nuclei survive photo-disintegration losses in
the hostile environment sustained by the thermal radiation field from
the star.  The spectrum of accelerated UHECRs is determined by the
evolution of the rotational frequency: As the star spins down, the
energy of the cosmic ray particles ejected decreases. As a
consequence, the total fluence of UHECRs accelerated in the neutron
star magnetosphere is very hard, typically $\propto
E^{-1}$~\cite{Blasi:2000xm}.  The arrival directions of the highetst
energy cosmic rays recorded by the Yakutsk, Fly's Eye, and AGASA
experiments can be traced back to the two nearest starbursts: M82 and
NGC 253~\cite{Anchordoqui:2002dj}. The possible association of the TA hot
spot with  M82 has not gone
unnoticed~\cite{Anchordoqui:2014yva,He:2014mqa,Pfeffer:2015idq}, and the possible
association of Auger events with  NGC 253 and NGC 4945
did not escape attention either~\cite{Nemmen:2010bp}. However, as can be seen in Fig.~\ref{fig:1}, existing data neither favor nor exclude the possibility of starbursts as UHECR emitters. GRBs are
short-lived, luminous explosions at cosmological distances, thought to
originate from relativistic plasma launched at the deaths of massive
stars. The widely accepted interpretation of GRB phenomenology is that
the observable effects are due to the dissipation of the kinetic
energy of a relativistically expanding fireball~\cite{Piran:1999kx}. The physical
conditions in the dissipation region imply that cosmic rays can be
accelerated to energies $\agt 10^{11}~{\rm
  GeV}$~\cite{Waxman:1995vg,Vietri:1995hs}. UHECR acceleration at GRB
internal shocks may also yield a hard source
spectrum~\cite{Globus:2014fka} and consequently  accommodate cosmic
ray observations~\cite{Globus:2015xga,Globus:2017ehu}. 

Even if cosmic rays include a significant component of heavy nuclei we
still expect to observe an anisotropy associated with the heavy
component at the highest energies, due to the anisotropic distribution
of matter within the GZK sphere~\cite{Anchordoqui:2002dj}. In the
event that a correlation with astrophysical sources is present in the
data it is important to predefine a search prescription in order to
assign a meaningful {\it a priori} statistical significance to a
potential observation. We follow the approach taken by the Auger
Collaboration, in which we assume an ``interesting'' anisotropy result
requires a pre-specified chance probability~\cite{Clay:2003pv}.  Given
the exploratory nature of the SPB missions we will consider a 1\%
significance to constitute such an interesting result. For
our  prescription, we adopt a low energy cutoff of $10^{10.7}~{\rm
  GeV}$, which is in the range where TA observes the hot spot.  We
chose our candidate objects subject to the following considerations:
Even though we do not know the exact declination observable during the
flight the majority of candidate objects will be in the southern
sky. On the basis of Auger and TA data we assume EUSO-SPB2 will not
observe small-scale clustering of events, rather we chose to search
for excesses in $20^\circ$ regions of the sky centered at the source
targets. We further assume that UHECRs are accelerated in nearby sources. In
particular, we partition the probability budget equally between
starbursts and radiogalaxies. More specifically, we consider the
nearby radiogalaxy Cen A with a 0.5\% budget and starbursts NGC 253,
NGC 4945, M83, NGC 1068 (assuming the latter is within the balloon
exposure) the remaining 0.5\%.  If NGC 1068 is not in the field of
view we retain the partition of the probability budget equally
between starbursts and radiogalaxies. Including GRBs in the
prescription would introduce a considerable complexity in the analysis
due to their transit nature. Note that this prescription has been
designed in time for the launch of the first EUSO-SPB in late March
2017, implying that we can also use data gathered in this flight in
our search.

\subsection{Properties of nearly horizontal air showers developing at high
  altitude}

When the incident cosmic radiation interacts with atomic nuclei of air
molecules, it produces fluxes of secondary, tertiary, and subsequent
generations of particles.  All these particles together create a
cascade, called air shower. As the cascade develops longitudinally the
particles become less and less energetic since the energy of the
incoming cosmic ray is  redistributed among more and more
participants. The transverse momenta acquired by the secondaries cause
the particles to spread laterally as they propagate through the
atmospheric target. Most of the air shower particles excite nitrogen
molecules in the atmosphere, which fluoresce in the UV. Fast UV
cameras aboard EUSO-ESP2 will record the fluorescence light produced
by the particle cascades.

If the primary cosmic ray is a baryon, hundreds to thousands of secondary particles are usually produced at
the interaction vertex, many of which have energies above the
highest accelerator energies~\cite{Anchordoqui:1998nq}. These secondary products are hadrons,
mostly pions with a small admixture of kaons and nucleons.
When the $\pi^0$'s (with a lifetime of $\simeq 8.4 \times10^{-17}~{\rm
  s}$) do decay promptly to two photons, they feed the electromagnetic
component of the shower.  Charged mesons because of a longer lifetime, not only decay but also interact strongly with
atmospheric nuclei. The competition between the two processes depends
essentially on the balance between interaction mean free path
(dependent on the cross-section and the density of the medium
transversed) and the mean decay length. Both vary substantially with
energy and become equal at a critical energy $\xi_c$. For a vertical
transversal of the atmosphere, such a critical 
energy is found to be $\xi_c^{_{\pi^\pm}} \sim 115~{\rm GeV}$ for
charged pions and $\xi_c^{_{K^\pm}} \sim 850~{\rm GeV}$,
$\xi_c^{_{K^{0,L}}} \sim 210~{\rm GeV}$, $\xi_c^{_{K^{0,S}}} \sim 30~{\rm TeV}$ for 
kaons~\cite{Gondolo:1995fq}. Hence, below the critical energies the
decay probability becomes larger than the interaction probability. Charged
pions and kaons give rise to muons and muon-neutrinos  in the
shower. Neutrinos escape detection carrying roughly 2\% of the primary energy,
while the highly relativistic muons propagate to the ground.

The number of particles as a function of the amount of atmosphere
penetrated by the cascade ($X$ in ${\rm g/cm^2}$ ) is known as the
longitudinal profile. A well-defined peak in the longitudinal
development, $X_{\rm max}$, occurs where the number of $e^\pm$ in the
electromagnetic shower is at its maximum. $X_{\rm max}$ increases with
primary energy, as more cascade generations are required to degrade
the secondary particle energies. Evaluating the mean and the
dispersion of the $X_{\rm max}$ distribution is a
fundamental part of many of the composition analyses done when
studying air showers. The generic shower properties can be
qualitatively well understood using the superposition principle, which
states that a shower initiated by a nucleus with $A$ nucleons and
energy $E$ behaves to a good approximation as the superposition of $A$
proton showers with initial energy $E/A$~\cite{Anchordoqui:2004xb}. This phenomenological
assumption relies on the fact that the effect of nuclear binding is
 negligible compared to the extremely high energies of the incoming
 cosmic rays. Thus, for a given total
energy $E$, showers initiated by a heavy nucleus have smaller $X_{\rm
  max}$ than proton induced showers. Shower-muon-richness is also
sensitive to the nuclear composition~\cite{Anchordoqui:2004xb}.

EUSO-SPB2 will be able to measure for the first time showers which
develop nearly horizontally at high altitude, where the density of the
atmosphere is low and nearly constant. For such showers, the
competition between interaction and decay significantly modifies the
average and spread of the $X_{\rm max}$ distribution compared to the
distributions characterizing more vertical showers. This results from
the hadrons spending more time in the tenuous atmosphere where they
are more likely to decay than interact as compared to a shower
developing through an atmosphere of progressively increasing
density. At present, hadronic interaction models extrapolated from LHC
data can be vetted at higher energies only via air shower observables
(like the mean and spread of $X_{\rm max}$ and muon richness) for showers that
impact the Earth~\cite{Ulrich:2010rg}. EUSO-SPB2 observations of high-altitude horizontal
showers will provide a complementary handle on hadronic interaction
models due to the unique environment in which the particle cascades
take place.

\subsection{The hunt for astrophysical tau-neutrinos using upgoing air showers}

The announcement by the IceCube Collaboration of the observation of 53
astrophysical neutrino candidates in the energy range $10^{4.5} \alt
E_\nu/{\rm GeV} \alt 10^{6.3}$ has been greeted with a great deal of
justified
excitement~\cite{Aartsen:2013bka,Aartsen:2013jdh,Aartsen:2014muf,Aartsen:2014gkd,Aartsen:2015zva}. With
these events, a purely atmospheric explanation for the neutrino flux is rejected at more
than $5.7\sigma$~\cite{Aartsen:2015zva}. IceCube's discovery represents the ``first light'' in the nascent
field of neutrino astronomy.

A nearly guaranteed neutrino flux originates in the decay of charged
pions, which are expected to be produced in $pp$ or $p \gamma$ collisions
near the cosmic ray acceleration sites, either in Galactic or
extragalactic sources~\cite{Gaisser:1994yf,Learned:2000sw,Halzen:2002pg,Anchordoqui:2009nf,Anchordoqui:2013dnh}.  Since  cosmic
rays and neutrinos may originate at the same sites,  searches for
correlations in the arrival directions of IceCube events and UHECRs
have been carried out~\cite{Aartsen:2015dml}. In particular, a
possible association between the TA hot spot and IceCube neutrinos
has been studied~\cite{Fang:2014uja}.  In $pp$ and $p \gamma$
collisions only muon and electron neutrinos are produced. If we assume
the hypothesis of maximal mixing~\cite{GonzalezGarcia:2007ib}, the flavor ratios arriving at Earth
should be \mbox{$\nu_e:\nu_\mu: \nu_\tau \sim 1:1:1$.} Within errors these
ratios are consistent with IceCube's
observations~\cite{Aartsen:2015ivb,Palomares-Ruiz:2015mka}. However,
the identification of tau neutrinos has remained elusive: no
(``double-bang''~\cite{Learned:1994wg}) events
were found in three years of IceCube data, in agreement with the
expectation of 0.5 signal events~\cite{Xu:2017yxo}.

EUSO-SPB2 will search for up-going air showers produced by tau leptons
originating from neutrino interactions below the Earth's
surface~\cite{Fargion:2000iz,Feng:2001ue,Bertou:2001vm}.  If no
neutrino candidate is observed, EUSO-SPB2 measurements will help to
establish the background for future experiments of this type, like the
POEMMA space mission.

There are three types of events which could occasionally be
misinterpreted as neutrino induced up-going air showers: {\it
  (i)}~random coincidences of excesses of the night Earth background
may occasionally exceed the detection threshold of the telescope
camera; {\it (ii)}~signals from high-energy cosmic ray induced
showers; {\it (iii)} shower-like signals produced by the direct
interactions of cosmic rays with the focal plane of the
detector.\footnote{The night Earth backgrund is a combination of the
  atmospheric airglow and scattered emission from the starlight.} A
detailed discussion of background events has been presented
elsewhere~\cite{Neronov:2016zou}.

In summary, the primary objective of using EUSO-SPB2 for neutrino
studies is to evaluate in detail the background with which future more
ambitious space-based missions will have to contend. An optimist might
also hope for detection of a few tau neutrino candidates.

\subsection{Cosmic ray physics using Cherenkov light }

\begin{figure}[tbp]
\begin{minipage}[t]{0.49\textwidth}
    \postscript{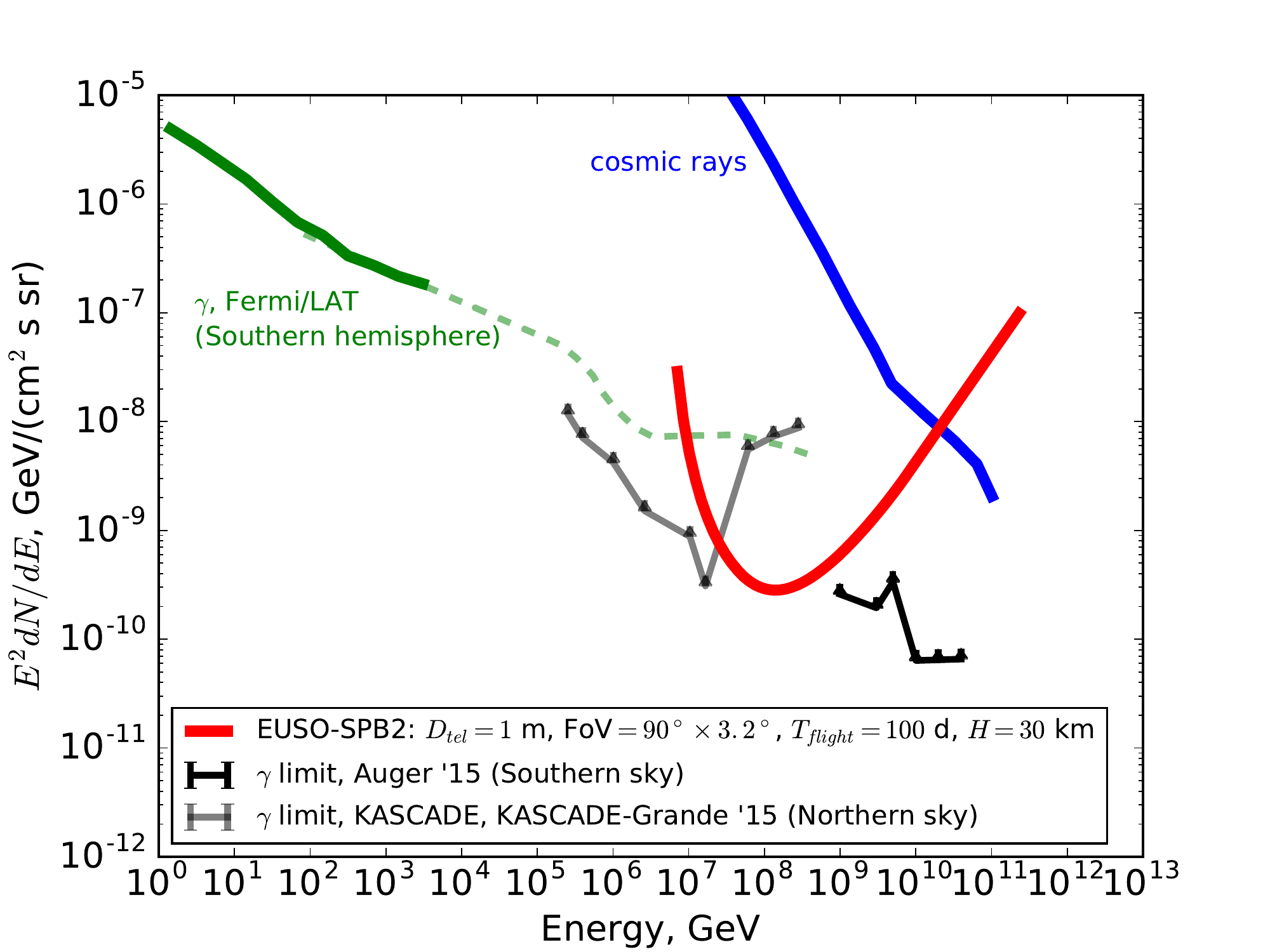}{0.99}
\end{minipage}
\begin{minipage}[t]{0.49\textwidth}
    \postscript{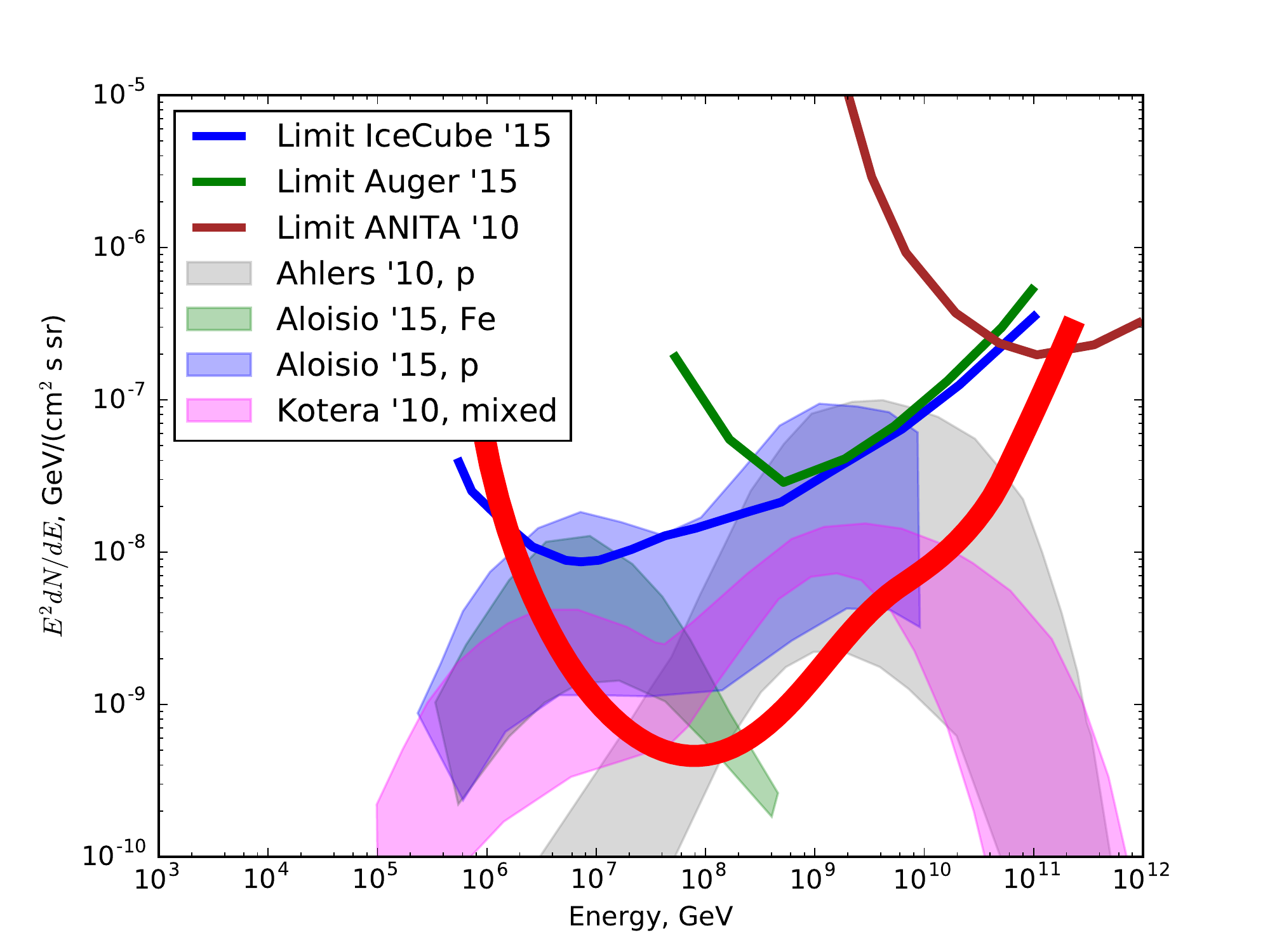}{0.99}
\end{minipage}
\caption{{\it Left.} Sensitivity of EUSO-SPB2 to gamma rays. The red
  line shows the best achievable sensitivity to the photon flux based
  on the assumption photons can be uniquely identified applying the
  techniques described in~\cite{Neronov:2016iax}. The green solid line
  shows the square energy weighted flux of gamma-rays observed by
  \textit{Fermi}-LAT~\cite{Ackermann:2014usa,TheFermi-LAT:2015ykq} and
  the dashed line shows a power-law extrapolation of the square energy
  weighted spectrum, modified by absorption on cosmic microwave
  background over the distance scale 8~kpc. The blue curve indicates
  the cosmic ray intensity~\cite{Olive:2016xmw}. The grey and black
  lines show current photon limits from
  KASCADE-Grande~\cite{Kang:2015gpa} and
  Auger~\cite{Aab:2015bza,Aab:2016agp}, respectively. {\it
    Right.}~Comparison of the sensitivity of space borne POEMMA to the
  expected levels of cosmogenic neutrino flux, modelled under
  different assumptions about UHECR flux composition and radiation
  backgrounds in the intergalactic medium~\cite{Kotera:2010yn,Ahlers:2010fw,Aloisio:2015ega}. Limits from IceCube~\cite{Aartsen:2015zva}, Auger~\cite{Aab:2015kma} and ANITA~\cite{Gorham:2010kv} are shown for
  comparison.\label{fig:2}}
\end{figure}

It is also possible to exploit Cherenkov light detection to estimate
the muon richness of air showers. In an air shower, the electrons, positrons
and muons all generate Cherenkov light. In the case of highly inclined
showers, however, muons propagate over much larger distances than the
electrons and photons, resulting in Cherenkov light topologies
exhibiting ``halos'' or ``tails'' which are distinct from the
spatiotemporal distribution of Cherenkov light emerging from the
quickly developing electromagnetic shower component, as discussed
in~\cite{Neronov:2016iax}.  This can enable searches for photon
primaries in an energy regime complementary to other experiments,
potentially yielding the best bounds for $10^8 \alt E_\gamma/{\rm GeV}
\alt 10^9$, as illustrated in Fig~\ref{fig:2}. Achieving such an
exceptional sensitivity will, of course, require refinement of
previously explored techniques to reject the cosmic ray
background~\cite{Neronov:2016iax}.  Sufficient rejection power is not
unprecedented in other cosmic ray experiments; indeed the Auger
Collaboration has reported photon bounds at the level of 0.1\% of the
total cosmic ray flux~\cite{Aab:2015bza,Aab:2016agp}.

The experiment will have sensitivity to the muon component of the
air showers, for primary cosmic rays in the energy range $10^7 \alt
E/{\rm GeV} \alt 10^{10}$~\cite{Neronov:2016iax}. This will allow us to
test hadronic interaction models beyond colliders energies. The Pierre
Auger Collaboration has reported an excess in the number of muons of a
few tens of percent over expectations computed using extrapolation of
hadronic interaction models tuned to accommodate LHC
data~\cite{Aab:2016hkv}. This has been interpreted as a possible
signal of new physics at sub-fermi
distances~\cite{AlvarezMuniz:2012dd,Farrar:2013sfa,Anchordoqui:2016oxy,Tomar:2017mgc}. The
hypothesis of a new physics process is consistent with the
non-observation of the muon excess at $E \sim 10^8~{\rm GeV}$ in data
collected with the Moscow State University Extensive Air Shower
(EAS-MSU) array~\cite{Fomin:2016kul}. However, in this same energy
range, $10^8 \alt E/{\rm GeV} \alt 10^9$,  a possible excess of muons
has been observed with HiRes-MIA~\cite{AbuZayyad:1999xa} and
KASCADE-Grande~\cite{Haungs}. Therefore, corroborating evidence for the muon excess at lower
energies is essential for our understanding of fundamental physics. As discussed in~\cite{Neronov:2016iax}
EUSO-SPB2 will have the potential to provide such evidence.

A point worth noting at this juncture is that because EUSO-SPB2 will not observe Cherenkov light in stereo, it will
not be feasible to measure all the physical characteristics of the
shower simultaneously. With a mono system, we will be sensitive to the
number of photons originating from electrons versus muons. However,
reconstruction of the primary particle energy for an individual event
depends not only on signal size, but also on impact parameter and the
depth of the first interaction. In addition, the relation between the
number of electrons and muons and the strength of the Cherenkov signal
in the two components depends on the impact parameter. Therefore, our
analysis of the muon content of showers as a function of energy cannot
be done on an event-by-event basis, but will require statistical
analysis of our event ensemble. We envision using a maximum likelihood
approach. We will construct a model for the behavior of the ensemble
of measurements that includes free parameters which describe the
energy spectrum and the average muon content. The likelihood that this
model matches the ensemble of data will be maximized by adjusting
these parameters. The result will be a determination of the parameters
and the errors of these determinations.

\subsection{NASA's POEMMA}

POEMMA will measure orders-of-magnitude more UHECR events
than attained by ground-based observatories at $E \agt 10^{10.8}~{\rm
  GeV}$. POEMMA has been designed to reach unprecedented geometrical
apertures $>10^6~{\rm km}^2 \, {\rm sr} \, {\rm yr}$, which, after
duty cycle corrections, correspond to annual exposures of more than
$10^5~{\rm km} \, {\rm sr} \, {\rm yr}$ at the highest
energies. POEMMA will also have high angular resolution (about 
$1^\circ$) and $X_{\rm max}$ determination ($\sim 20~{\rm
  g/cm^2}$). The unprecedented POEMMA exposure will also provide full
coverage of the Celestial Sphere. This will enable far more sensitive
sky maps leading to the discovery of the brightest sources of UHECRs
in the sky.

POEMMA will search for astrophysical and cosmogenic neutrinos with two
techniques.\footnote{A diffuse flux of neutrinos originates in the
  energy losses of UHECRs {\it en route} to Earth.  UHECR interactions
  with the cosmic microwave and infrared backgrounds generate pions
  and neutrons, which decay to produce
  neutrinos~\cite{Beresinsky:1969qj,Stecker:1978ah}. The accumulation
  of these neutrinos over cosmological time is known as the cosmogenic
  neutrino flux.} With the same system designed to observe UHECRs
based on the OWL design, POEMMA can detect deeply penetrating
horizontal showers initiated by all flavors of EeV neutrinos in the
atmosphere. In addition, a Cherenkov telescope based on the CHANT
concept combined with Cherenkov measurements by the fluorescence
telescope can observe the signal produced from tau neutrinos beginning
$10^7~{\rm GeV}$ (where astrophysical IceCube neutrinos are expected)
to $10^{10}~{\rm GeV}$ (where cosmogenic neutrinos can be discovered,
as illustrated in Fig.~\ref{fig:2}).

POEMMA will also be sensitive to ultrahigh-energy photons. The
ultrahigh-energy photon flux is highly model dependent for
astrophysical sources, being highly sensitive to the location of the
closest sources~\cite{Decerprit:2011qe}. Ultrahigh-energy photons are
the dominant component of models based on relic decays from the early
universe, including super-heavy dark
matter~\cite{Kuzmin:1998uv,Dubovsky:1998pu,Berezinsky:1998ed,Birkel:1998nx,Blasi:2001hr,Sarkar:2001se,Aloisio:2015lva}. A
clear detection of these photons would be momentous discovery.

In summary, POEMMA will provide a new window on the Universe and on its most
energetic environments and events.

\section{TECHNICAL DESIGN OF THE INSTRUMENTS}

\subsection{Observation and measurement strategies}

Advancements in the study of extensive air showers over the past
decades are largely based on larger apertures and exposures to attain
more detailed information on the nature and sources of the UHECR flux
and its role in the universe. We can now observe these rare particles
from balloon altitudes with apertures approaching that envisioned for
 space-based missions ({\it cf.} 10\% of the original EUSO design
envisioned to fly onbard the Internations Space Station (ISS)) and with
exposures lasting $\sim 100$~days with a favorable duty cycle, making
NASA's SPB balloon program an excellent opportunity for further
advances. 

The new SPB platform in Wanaka ($45^\circ~{\rm S}, 169^\circ~{\rm
  E}$), New Zealand, is designed to fly at near constant pressure
altitude through day-night cycles by maintaining a positive internal
pressure, thus enabling Ultra-Long Duration mid-latitude Balloon
(ULDB) flights. The edge of the atmosphere provides
a distinct vantage point for looking at the developing showers and
enhances observations such as the muon generated signals relative to
that of the electromagnetic cascade, making possible measurements that
can add significantly to the study of the UHECR primary composition.

\begin{figure}[tbp]
    \postscript{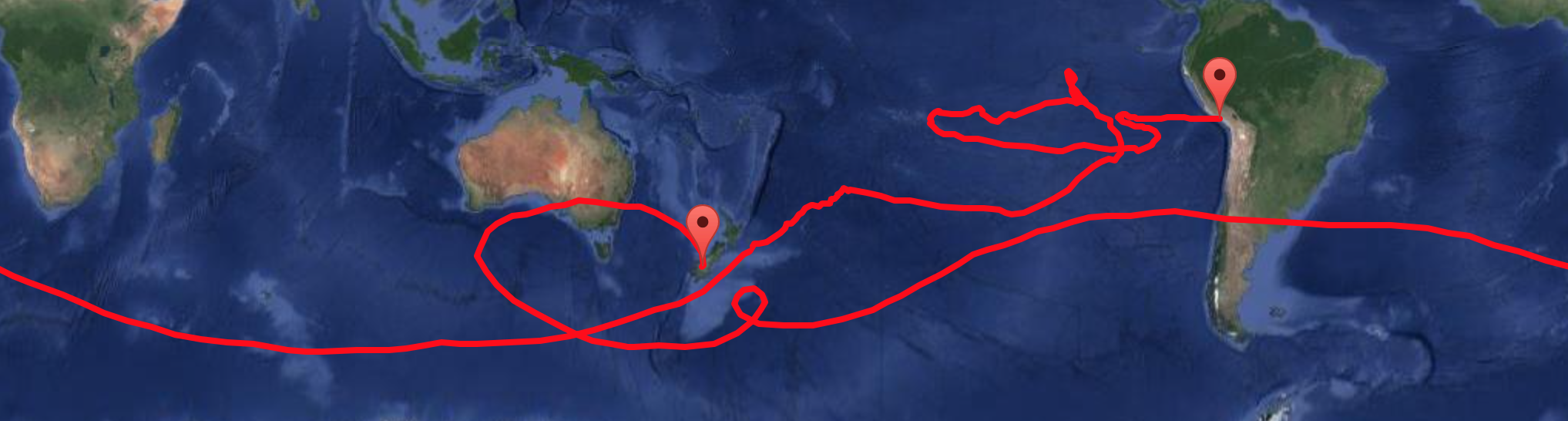}{0.9}
    \postscript{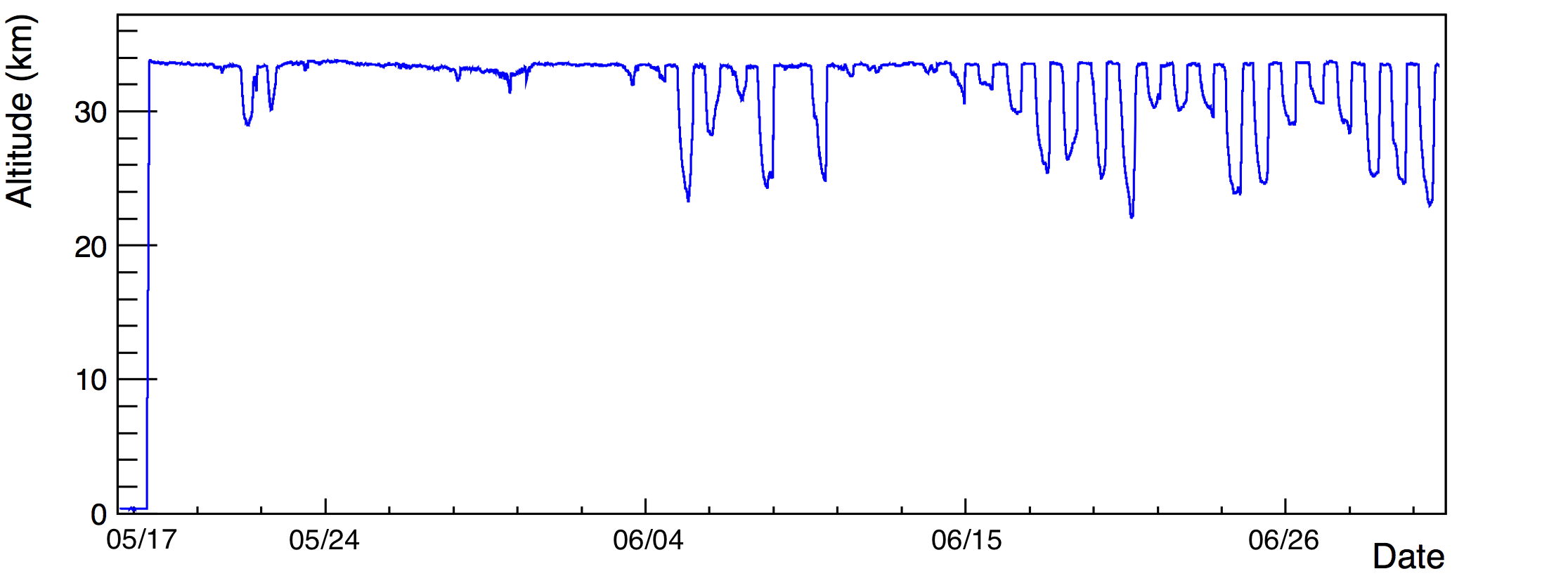}{0.9}
\caption{Flight path and altitude of the first SPB science flight, lifting the
  Compton Spectrometer and Imager (COSI)~\cite{Kierans:2017bmv}.  {\it Top.}~Flight path
  from launch in Wanaka, New Zealand, to termination in Arequipa, Peru. COSI was
  afloat for 46 days and spent much of its time over the Southern
  Pacific Ocean. {\it Bottom.} Altitude profile over the duration of
  the flight. After the night of June 5th, large altitude drops during
  the cold nights were seen. \label{fig:3}}
\end{figure}

Two balloon flights are proposed: a 1 night CONtinental USa (CONUS)
test flight and an ULDB science flight. The test flight in 2020 will
be used to verify performance of the instruments and analyze the data
collected on background signal levels. The test flight payload will
have the same functionality of the ULDB instruments, but the focal
surface will be only partially populated to minimize risks that could
impact the schedule for a ULDB science flight in 2022. The data
acquired during the flight will provide direct knowledge on the
anticipated backgrounds during the ULDB mission and allow optimizing
the instrument parameters for the mission. It also prepares the team
for the challenges of the SPB campaign in New Zealand. The ULDB is
well suited to address the science objectives. The mid-latitude flight
provides a balance between day/night for powering the instrument and
carrying out the night-time observations over a 100 day mission, at an
average altitude of 30~km (see Fig.~\ref{fig:3}).

\begin{figure}[tbp]
 \postscript{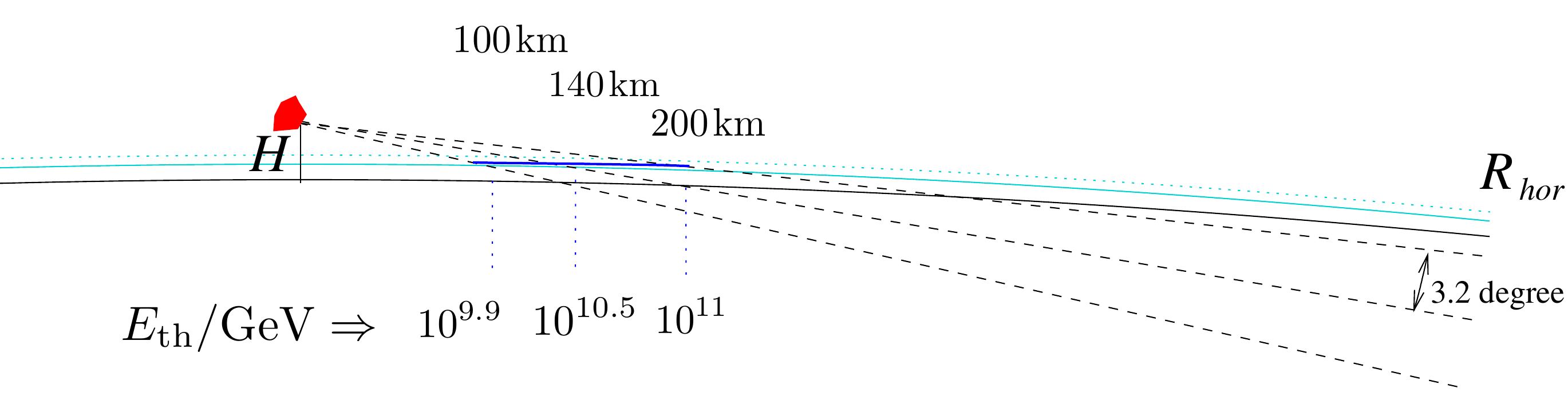}{0.99}
\caption{Schematic illustration of the shower energy and distance
  detection thresholds. \label{fig:4}}
\end{figure}

Three independent telescopes designed for specific measurements 
provide the data needed to meet the science objectives. A pair of
large telescopes will observe the Cherenkov signal from horizontal events
and the upward directed Cherenkov signal from air showers produced in the
atmosphere. Both telescopes have a field of
view (FoV)
$3.5^\circ \times 45^\circ$ and they
image an annulus centered on the balloon location and a $>100~{\rm km}$ radius
of Earth's surface. Data is acquired at different elevation angles
from the horizon to $\sim 10^\circ$ below, see Fig.~\ref{fig:4}. This
amounts to an observational area
\begin{equation}
A_{\rm EUSO-SPB2} \sim \pi \, [(200~{\rm km})^2 - (140~{\rm km})^2]/4 = 1.60
\times 10^4~{\rm km}^2 \,,  
\label{Aspb}
\end{equation}
which is about 10\% of the area it would have been observed by EUSO onboard
the ISS:
\begin{equation}
A_{\rm EUSO} \sim  \pi \, [400~{\rm km} \, \tan (\pi/6)]^2 = 1.67 \times 10^5~{\rm km}^2 \,,
\label{Aeuso}
\end{equation}
where the factor of 1/4 in (\ref{Aspb}) accounts for the fact that
EUSO-SPB2 will observe $\sim 90^\circ$ of a full circle and where in
(\ref{Aeuso}) we have taken the nominal EUSO FoV of $60^\circ$ and a
circular orbit for ISS at an altitude of 400~km~\cite{Adams:2012hr}.
A third telescope with a smaller FoV $3.2^\circ \times 28.8^\circ$ will measure the
intensity of the slower fluorescence signals produced by extensive air
showers, imaging the event trajectory as the cascade develops down through the
atmosphere. The fluorescence sensor will be configured  using the components recovered from EUSO-SPB.

No imaging is required for the upward Cherenkov flashes that are
focused to a small spot on the focal plane at the position of the
event within the FoV. The signal amplitude corresponds to the number
of photons collected by the telescope, which depends on impact
parameter and energy. A fast coincidence between multiple detectors is
needed to suppress false triggers from the charged particle
background. These upward Cherenkov flashes arrive at the balloon
platform with pulse durations of $\sim 10~{\rm ns}$ and essentially parallel. The
telescope's FoV insures the primary particle trajectory passed through
Earth.

\begin{figure}[tbp]
\begin{minipage}[t]{0.49\textwidth}
    \postscript{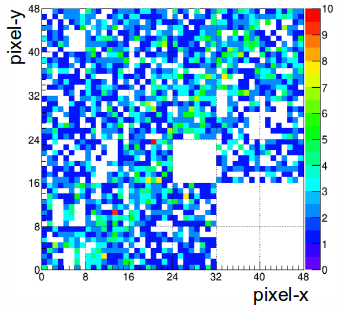}{0.99}
\end{minipage}
\begin{minipage}[t]{0.49\textwidth}
    \postscript{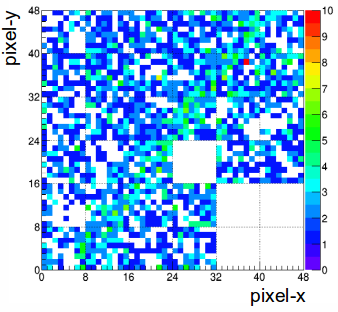}{0.99}
\end{minipage}
\caption{Comparison of a shower measured by  the EUSO-TA pathfinder on
  May 13, 2015 (left) and a shower simulated by {\sc CONEX}~\cite{Bergmann:2006yz} (using the
  energy, zenith angle, and impact parameters measured by BRM-TA
  Collaboration for that event) and processed by the \Offline software (right). The color-scale
  indicates the number of  counts integrated over all the gate time
  units ($2.5~\mu {\rm s}$  time bins) during which the shower was
  crossing the EUSO-TA FoV.  \label{fig:5}}
\end{figure}

The Cherenkov and fluorescence signal have greatly differing signal
characteristics and therefore require separate focal sensors. For
fluorescence measurements, the angle of the telescope defines the
nearest distance within the FoV that in turn determines the observable
energy threshold for showers; see Fig.~\ref{fig:4}. The atmospheric
cascade can be imaged as they propagate down through the atmosphere
producing a video clip of the air shower evolution at $2.5~\mu {\rm
  s}$ frame rate; see Fig.~\ref{fig:5}. Similar measurements will be
made by the EUSO-SPB flight this year looking at the nadir. The
EUSO-SPB2 mission would increase the event rate by a factor of $5$ as
a result of the balloon-payload design and observation strategy.  The
time evolution of the fluorescence signals extends over 10's to 100's
microseconds. Spatial and temporal data are used to determine the
direction and energy of the primary particle.

\begin{figure}[tbp]
\begin{minipage}[t]{0.49\textwidth}
    \postscript{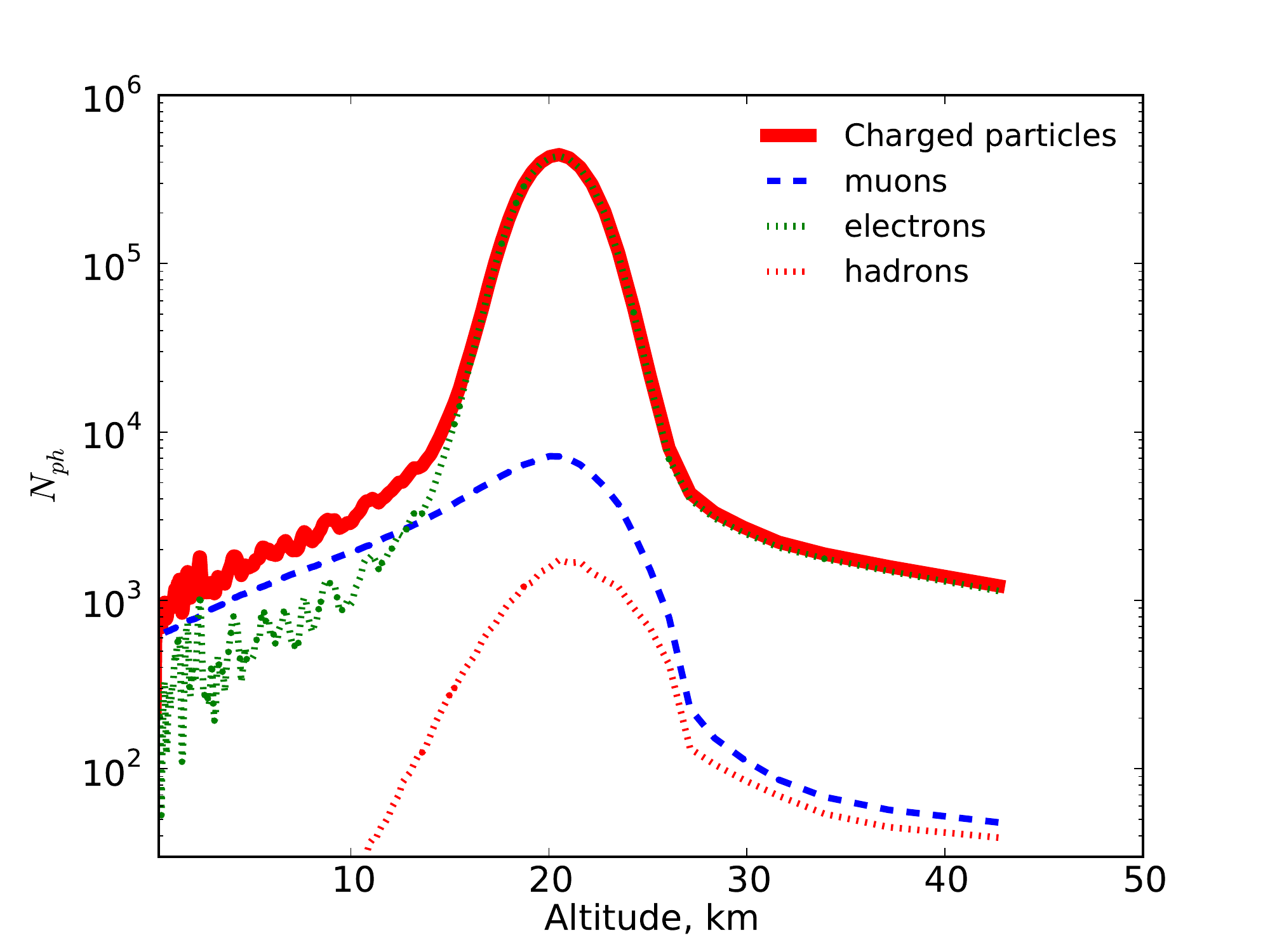}{0.99}
\end{minipage}
\begin{minipage}[t]{0.49\textwidth}
    \postscript{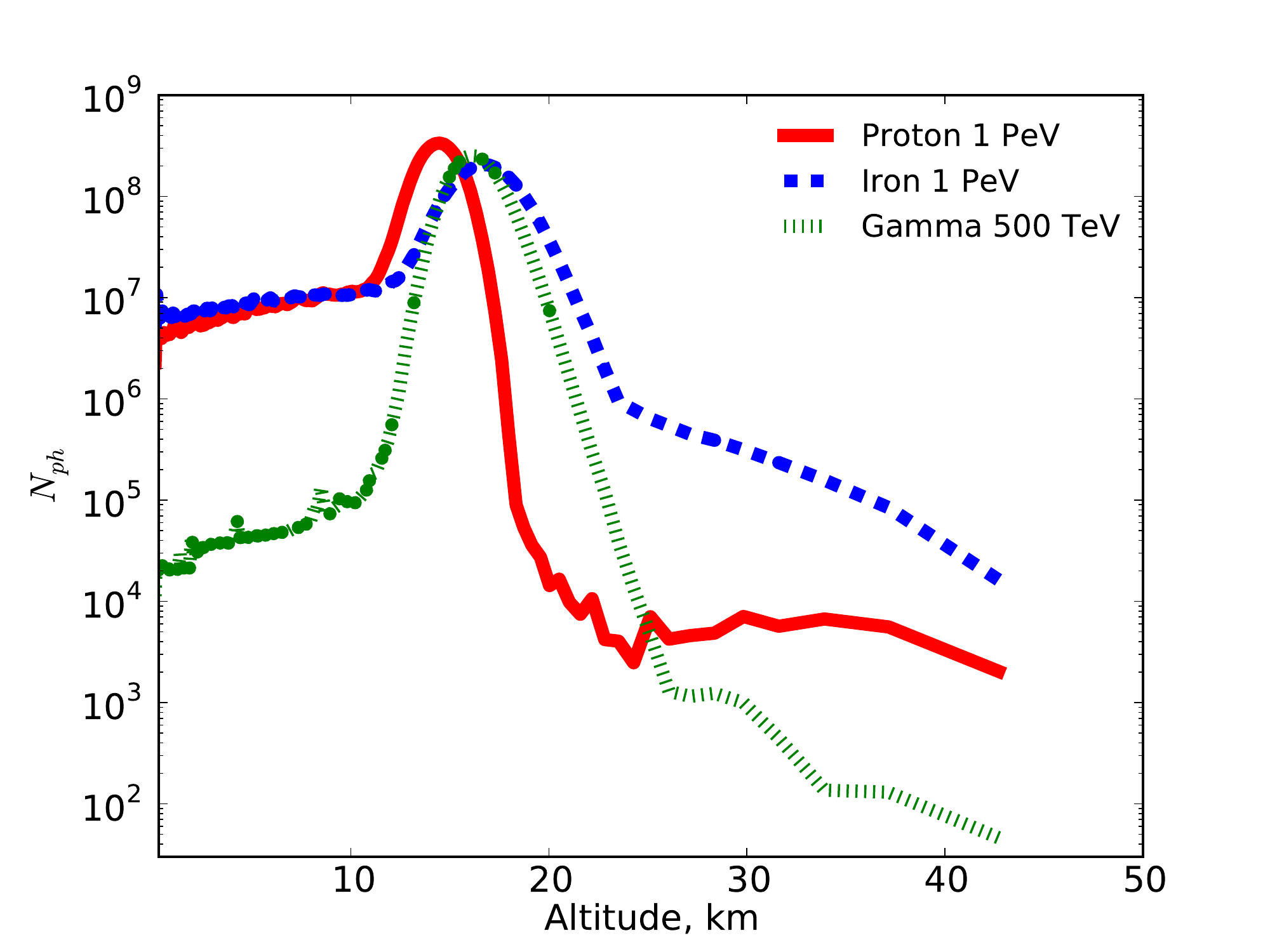}{0.99}
\end{minipage}
\caption{{\it Left.}  Longitudinal profiles of charged particle
  distribution in an air shower initiated by a 1~PeV energy proton
  incident at a zenith angle $\theta = 87^\circ$. {\it Right.}~Longitudinal profiles of
    Cherenkov light emission of an air shower initiated by a proton
    (red solid line), an iron nucleus (blue dashed line), and a gamma-ray
    (green dotted line), with $\theta =87^\circ$~\cite{Neronov:2016iax}. \label{fig:6}}
\end{figure} 

\begin{figure}[tbp]
 \postscript{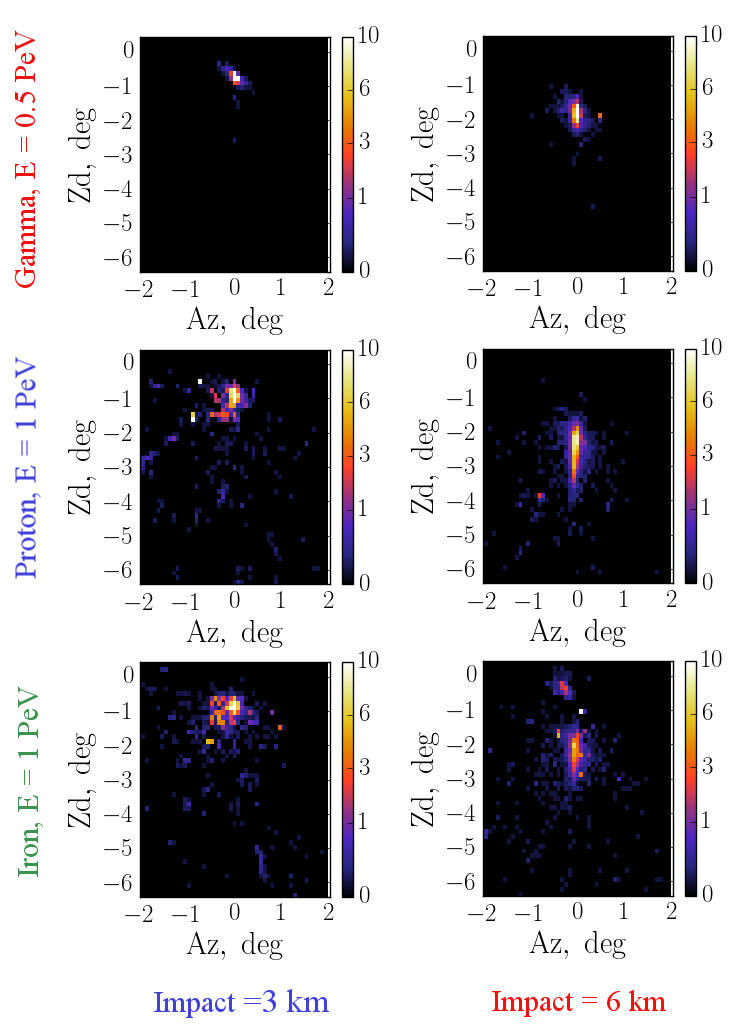}{0.85}
 \caption{Images and time profiles of 0.5~PeV gamma-ray, 1~PeV proton
   and 1~PeV iron induced air showers with $\theta
   =87^\circ$. The left
   column shows the images for showers with impact parameter
   $d =3$~km, and the right panels are for
   $d =6$~km~~\cite{Neronov:2016iax}. \label{fig:7}}
\end{figure}

The observable Cherenkov signals generated from nearly horizontal showers differ
greatly from lower inclination events. The charge particle
distributions generated from nearly horizontal showers observed at
large distances ($\sim 300~{\rm km}$) are shown in Fig.~\ref{fig:6}  and the
longitudinal Cherenkov emissions for  gamma-ray,  proton, and iron
primaries are also compared in Fig.~\ref{fig:6}. In the region of shower maximum
electron emissions are dominant producing fluorescence and Cherenkov
photons, but beyond this region the long lived muon component is seen
to exceed the Cherenkov emission rates of electrons. We exploit this
feature of the muon Cherenkov signal at large distances from the
shower, together with the small Cherenkov angle ($< 1^\circ$ to $1.5^\circ$) to
discriminate the primary particle identity from the image formed on
the focal plane. Simulations of the image formed by the Cherenkov
signal from showers generated by protons, photons, and iron primaries
are shown in Fig.~\ref{fig:7}. In these images, the muon component (or lack
of) is responsible for the details in each image. The gross properties
of the photons reaching the telescope are within the Cherenkov cone
angle $\sim 1^\circ$ and form an image on the focal plane at the azimuth angle of
the event, which varies little event to event due to the short vertical
projections of a few degrees. However, the detail structure within the
image of the event, as recorded on a focal plane, depends on the impact
parameter (distance of closest approach) that can produce a ``halo'' or
``tail'' feature within the image. The arrival time of the photons
creating this image depends on the distance away they were
generated. Since the muon speed exceeds that of the Cherenkov photons,
photons produced early in the event arrive the latest and
visa-versa. 

\subsection{Integrated gondola}

\subsubsection{Overview}

\begin{figure}[tbp]
\begin{minipage}[t]{0.49\textwidth}
    \postscript{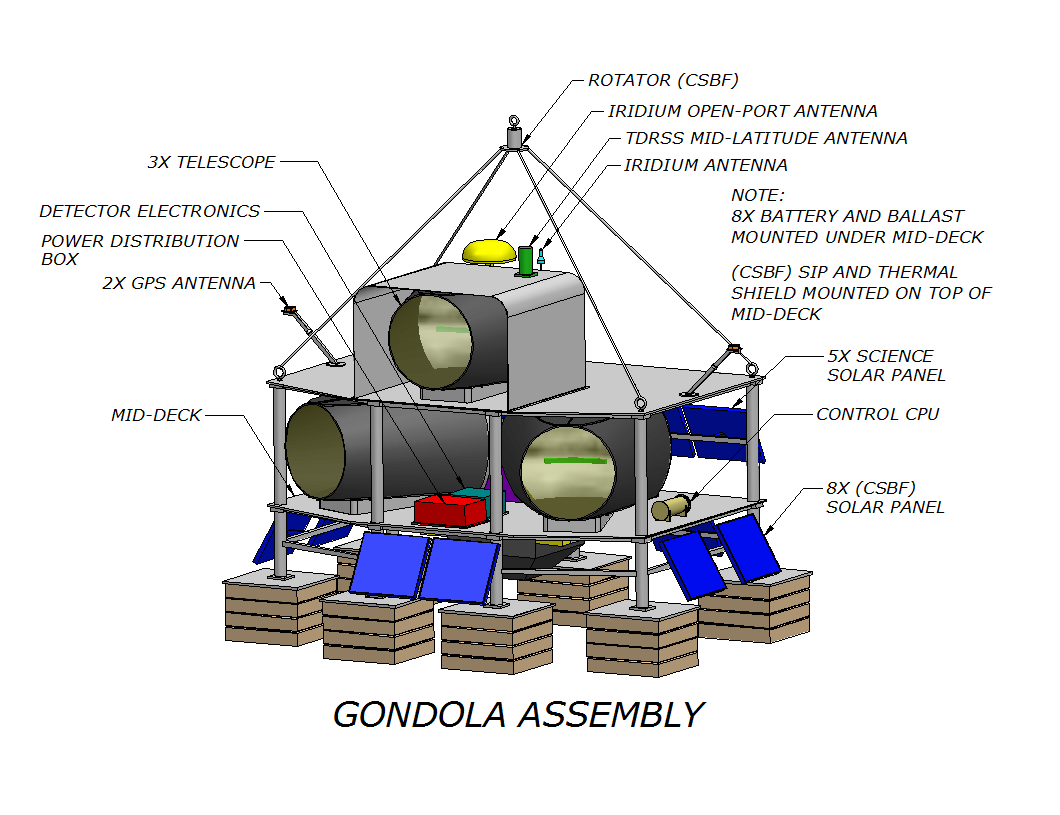}{0.99}
\end{minipage}
\begin{minipage}[t]{0.49\textwidth}
    \postscript{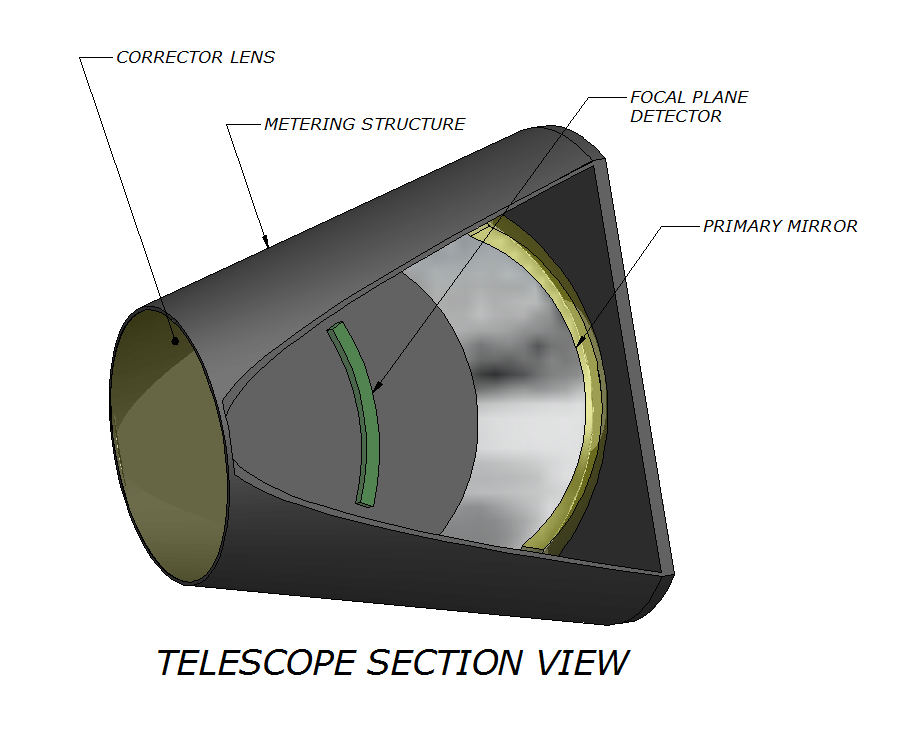}{0.99}
\end{minipage}
\caption{Concepts of the integrated balloon gondola
  (left) and one telescope assembly (right). \label{fig:8}}
\end{figure}

The gondola provides mounting for the telescopes, instrument
subsystems, and balloon equipment for the mission. It is suspended
from a pointing rotator used to maintain illumination of the solar
array panel during daytime and to point the instruments favorably
during nightime operations. The solar array panel serves as a sunshade for the
instrument deck. The telescope focal surfaces are further protected
from accidental exposure to direct sunlight by tilting the telescope
to the stowed position which acts as a full-aperture stop. The
electronics boxes are consolidated in the interior of the deck plate
and in close proximity for ease of cable routing. A schematic view of
the gondola is shown in Fig.~\ref{fig:8}. The mechanical design will be
refined to meet load limits, volume constraints and safety
requirements levied by NASA, the launch vehicle and range
operations.

\subsubsection{Pointing control}

\begin{figure}[tbp]
 \postscript{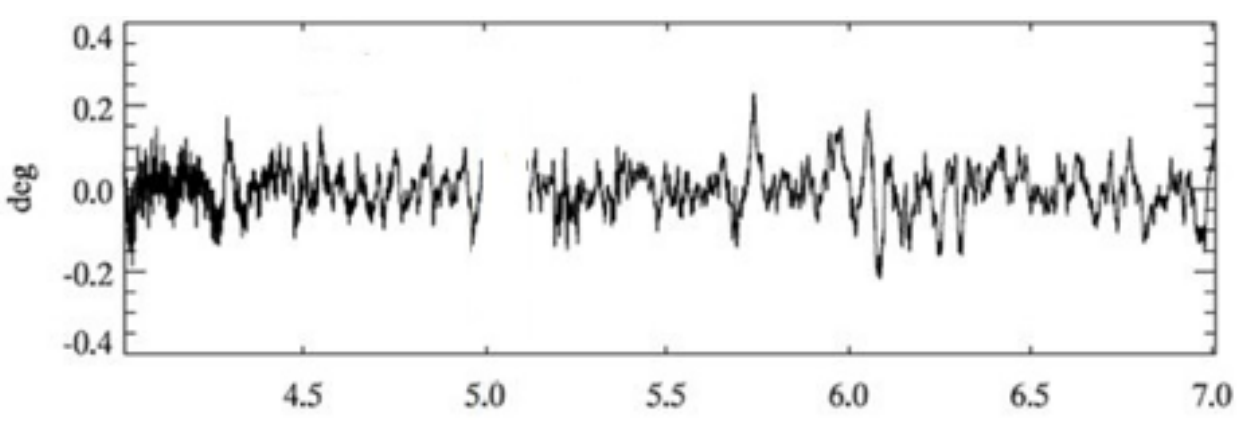}{0.99}
\caption{Measured pitch angle of a passive gondola at float altitude over a 3 hour period. \label{fig:9}}
\end{figure}

The gondola requires modest pointing control to maintain stability for
the required observations. Analysis of data from past flights shows
that the stability in the elevation angle will be within $\pm
0.1^\circ$  over the course of a night, see Fig.~\ref{fig:9}. Low frequency
variations will not impact light collection during observations as 
the amplitude is $<5\%$ of the full FoV. Any offset of the gondola due
to mass distribution or cable tension will be compensated by tilting
the telescope and monitoring them through inclinometers. A single
ballast hopper will be used and centered below the gondola to avoid
inducing torques when ballast releases are made to control the balloon
flight path.

Horizontal (azimuth) control of the FoV is used to optimize exposures
by steering toward regions with reduced clouds and scenes with lower
levels of moonlight. The preferred method for night time pointing
control makes use of the Columbia Scientific Balloon Facility (CSBF)
rotator, but using Global Positioning System (GPS) as feedback instead
of the sun-sensor. This can be commanded from the ground but will
include the onboard computer in the loop to insure reliable
communication with the rotator control equipment. The onboard computer
will also use the GPS system for data on pointing knowledge that will
be downloaded as part of the telemetry stream for analysis. The
resolution in pointing knowledge exceeds stability requirements.

Adjustments to the telescope FoV will follow the science observation
plan developed for the mission. The plan calls for extended exposures
lasting several hours and potentially all night minimizing variations
seen in Fig.~\ref{fig:9}. Default pointing directions will be loaded on the
computer prior to launch and these values are updated throughout the
mission. The operations team will monitor satellite based images to
determine preferred viewing directions throughout the mission. Typical
observations will seldom have multiple adjustments during a
night. 

\subsubsection{Power system}

The instrument will be powered from batteries at night that are
recharged each day using solar panels. A single solar panel will be
used as it will be pointed at the sun by the rotator. From our
EUSO-SPB experience, we have found that the optimum solar panel tilt
for  flights from Wanaka, NZ is $15^\circ$ from vertical.

The solar power system will be designed based on experience gained
with EUSO-SPB. We propose to use custom-made $27'' \times 31''$ solar panels
with a 6 by 5 configuration of SunPower cells including bypass
diodes. The solar cells will be encapsulated within EVA based
laminates and subsequently mounted to custom aluminum core, FRP faced
honeycomb sandwich substrates to form each panel. Each panel will
produce 100 watts at float. We will employ six panels wired in two
series strings of three panels each to provide the voltage needed to
charge a 24 volt battery pack.

For the battery pack, we propose to use lithium-Ion batteries from
Valence U1-12XP. These 40 Ah batteries weigh 13~kg and have been
flight-proven by CSBF. 20 of these batteries will provide
400~amp-hours, enough capacity to provide 600~watts of nighttime
power for up to 16 hours of darkness. The battery pack will consist of
10 strings, each with two batteries in series.

The batteries will be charged during the day using Morning Star
S-MPPT-30 charging controller. This system will operate the solar
panel at its peak power point to harvest the maximum power from the panel
while charging the battery pack.

\subsubsection{Telemetry}

Primary command and telemetry will be through the Iridium Pilot or
Certus system which can support non-continuous data rates of 100~kbps
(Pilot) to 1.4~Mbps (Certus).  On-board storage will be used to buffer
data between downlink opportunities. An independent secondary command
and telemetry path will use the Iridium Short-Burst Data system for
continuous limited bandwidth (255~bytes/minute) command and state
monitoring.  During the CONUS engineering flight a continuous 740~kbps
line-of-sight transmitter will be flown which will allow extensive
science and engineering data to be collected for analysis of system
performance.

\subsection{Optics design}

Previous EUSO optics pursued wide fields of view (WFoV) by employing
purely refractive designs. For this new balloon observatory, the WFoV
is only in one direction; namely, along the horizon.  The vertical
field of view (FoV) is only $3.2^\circ$.  A well-known design form for
achieving good image quality over a wide field is the Schmidt
telescope: a catadioptric design, consisting of both reflective and
refractive optical elements.  The Schmidt design utilizes a spherical
primary mirror with the stop at its center of curvature which
eliminates coma and astigmatism, and a refractive plate placed at the
stop which corrects the spherical aberration.  The impact of field
curvature is eliminated by curving the array of detectors.

\begin{figure}[tbp]
 \postscript{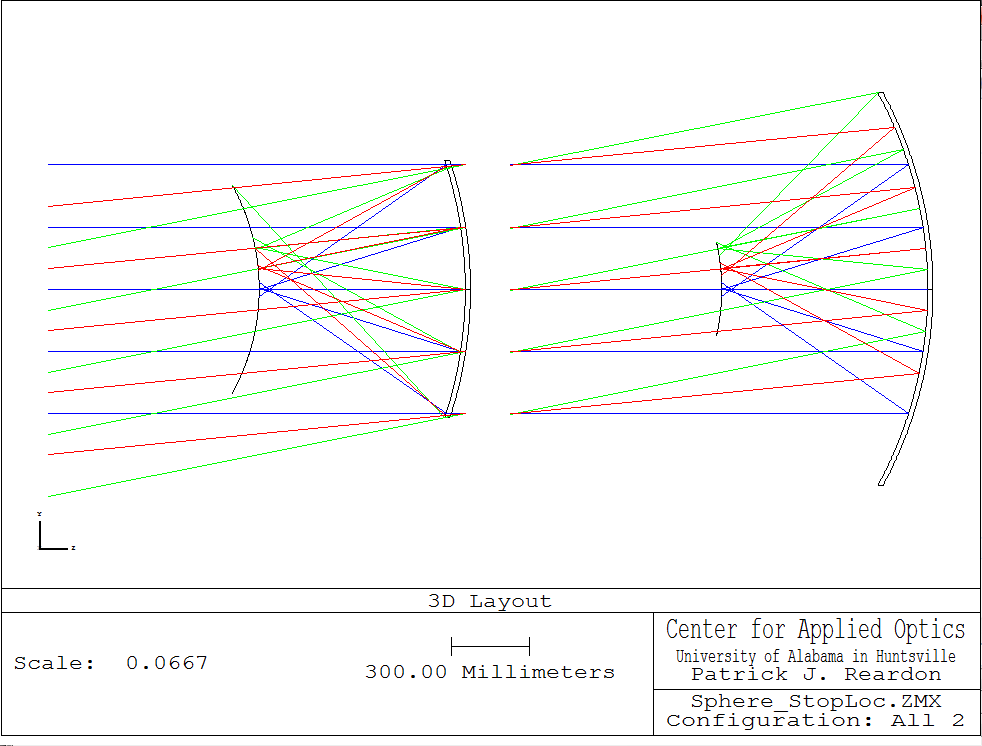}{0.99}
 \caption{Imaging of a distant object with a spherical mirror with the
   stop at the mirror (left) and the stop at the center of curvature
   of the same mirror (right).  Placing the stop at the center of
   curvature eliminates coma and astigmatism purely by symmetry, but
   the cost is a larger diameter mirror.
\label{fig:o1}}
\end{figure}

In Fig.~\ref{fig:o1}, we show a spherical mirror which focuses
collimated light emanating from the left.  The red (green) rays show
how light from an object point on-axis (off-axis) will intersect the
mirror and define the location and size of the stop, and therefore the
entrance pupil for the system. Off-axis rays, top and bottom,
intersect the mirror at far different angles of incidence, unlike the
rays for the on-axis beam.  The off-axis light will suffer from coma
and astigmatism in addition to the spherical aberration found in both
beams.  However, as shown in  Fig.~\ref{fig:o1}, by placing the stop at the center of curvature of the
spherical mirror,  the off-axis rays intersect the mirror in
the exact same manner as the on-axis beam. Thus, stop position
eliminates the coma and astigmatism, but the image still suffers from
spherical aberration.  Spherical aberration is a rotationally
symmetric deviation from perfect focusing wavefront that follows a
$4^{\rm th}$, $6^{\rm th}$, and higher even-powered radial polynomial
form, where the higher orders appear as the aperture diameter
increases.

To eliminate the spherical aberration, a refractive plate is placed at
the stop which provides the opposite wavefront deviation generated by
the spherical mirror.  Since the deviation, at the lowest
order, follows a $\rho^4$ form, one surface of the plate will have
this form.  However, for broad spectrum applications, the addition of
this refracting component creates a wavelength dependent aberration.
Though the mirror creates the same geometrical error for all
wavelengths, the refractive corrector will only perfectly correct the
error for one wavelength.  To reduce the image degradation from
spherochromatism, a very small amount of optical power, $\rho^2$, is
typically added to the corrector surface which balances some of the
residual chromatic power and spherochromatism.

\begin{table}
 \caption{Specifications of the design. \label{tabla}}
\begin{tabular}{|c|c|c|c|}
\hline
\hline
Element & Material & Shape & Dimensions \\
\hline
~~~Corrector~~~&~~~UV Transmitting PMMA~~~&~~~Sphere $+ A R^2 + B R^4$~~~&
Thickness = 20~mm \\
&  &  &  Diameter = 1260~mm \\
\hline
Stop & & Flat with hole &~~~441~mm from corrector~~~\\
\hline
Primary & CFRP & Spherical, concave & $1.8~{\rm m} \times 1.1~{\rm m}$\\
& & $R = -1604~{\rm mm}$ & 1602~mm from corrector \\ 
\hline
Image & & Spherical, convex & 828~mm from corrector \\
surface & & $R = 830.6~{\rm mm}$ & \\ 
\hline
\hline
\end{tabular}
\begin{tabular}{|c|c|c|c|}
  ~~~~~~ Surface ~~~~~~  & $R ({\rm mm})$ & $A$ & $B$ \\
\hline
1 & ~~~~~781.648~~~~~ & ~~~~~$-4.17888 \times 10^{-4}$~~~~~ & ~~~~~$-2.15028 \times
10^{-10}$~~~~~ \\
2 & 738.415 & $-4.92112 \times 10^{-4}$ & $-1.75838 \times
10^{-10}$\\
\hline
\hline
\end{tabular}
\end{table}
 
\begin{figure}[tbp]
\begin{minipage}[t]{0.49\textwidth}
    \postscript{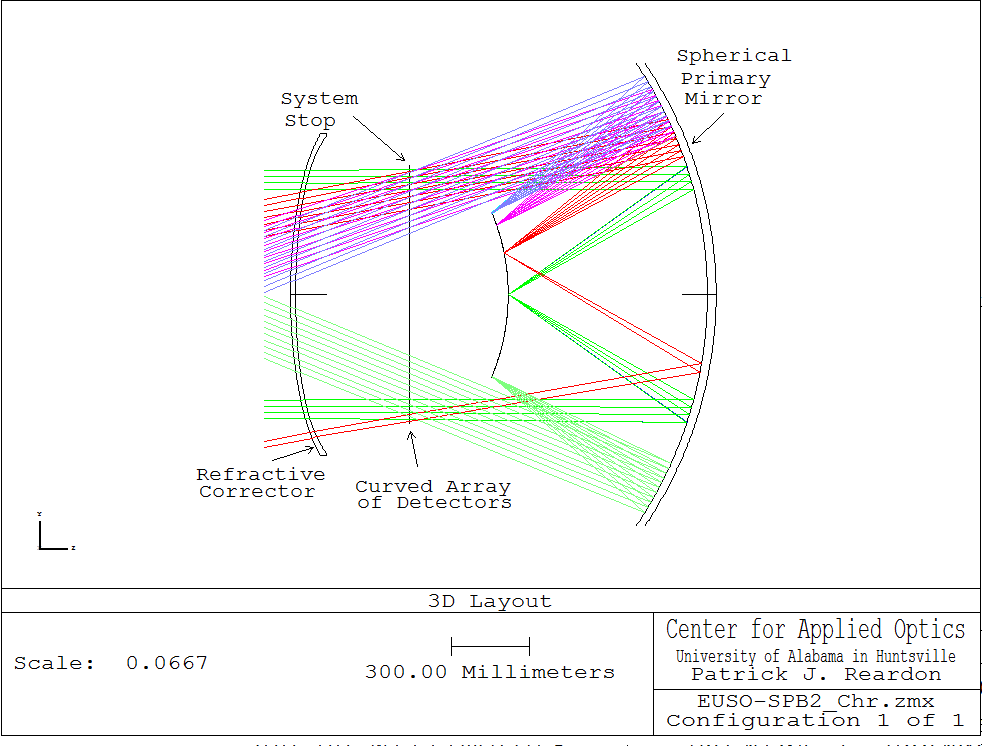}{0.98}
\end{minipage}
\begin{minipage}[t]{0.49\textwidth}
    \postscript{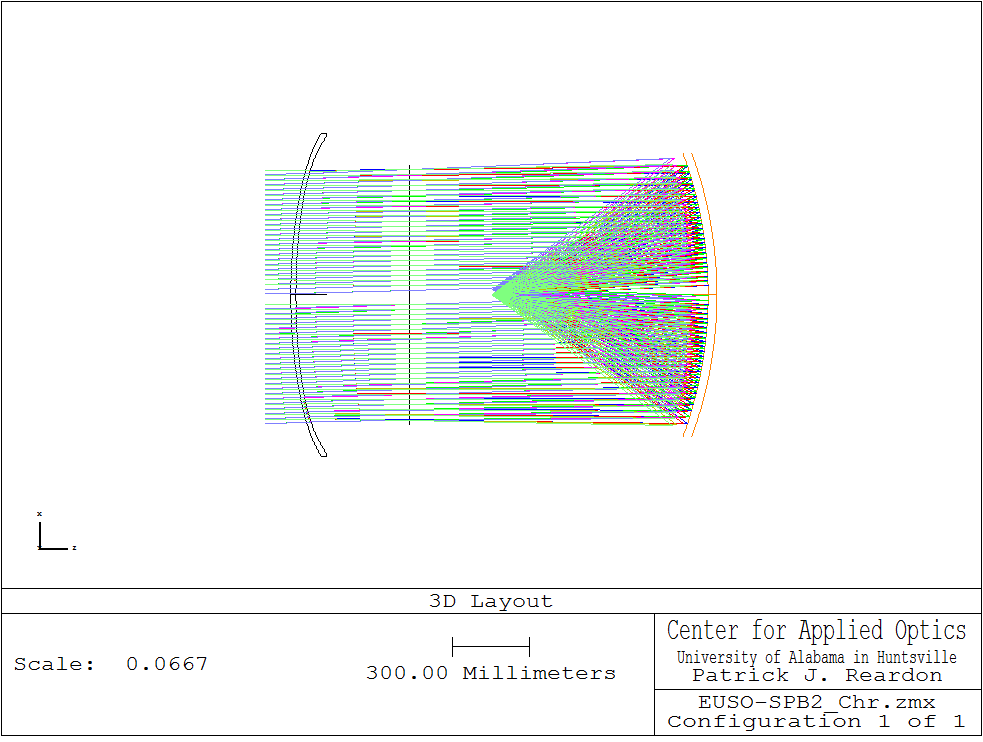}{0.99}
\end{minipage}
\caption{{\it Left.} Wide-field cross section showing, from left to
  right, corrector, physical stop, curved image and spherical primary
  mirror.  Obscured rays are blocked by detector 2.  {\it
    Right.}~Narrow-field cross section showing, left to right,
  corrector, physical stop, curved image and spherical primary mirror.
  Obscured rays are blocked by detectors. \label{fig:pat}}
\end{figure}

\begin{figure}[tbp]
 \postscript{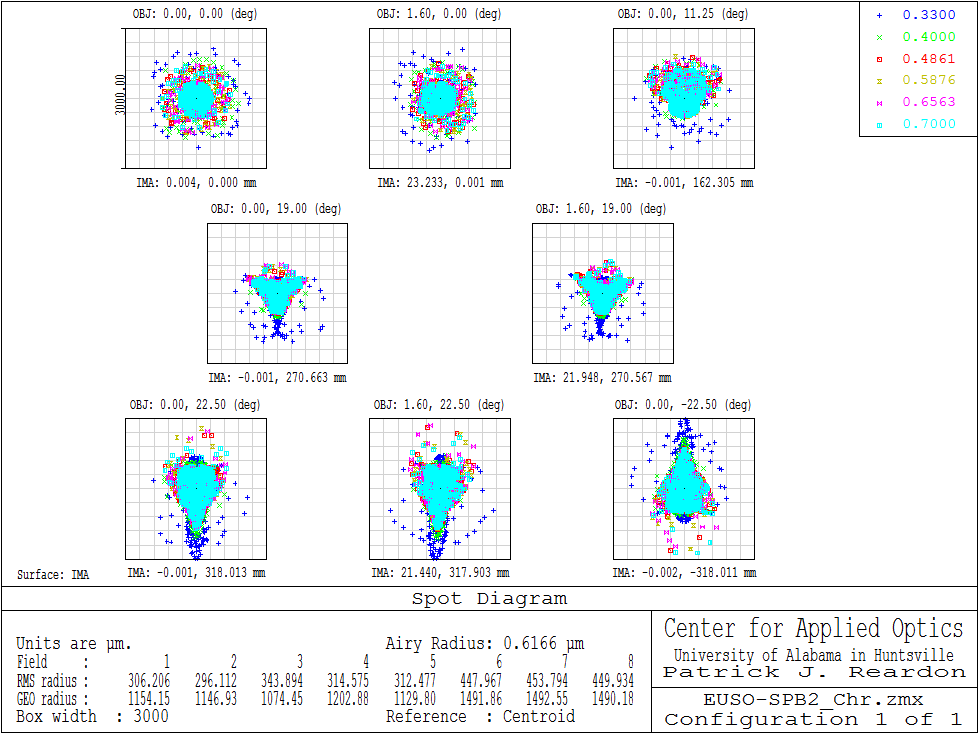}{0.9}
 \caption{Spot diagrams across one half the horizontal field, all
   fitting in the $3~{\rm mm}^2$  detectors.   \label{fig:o3}}
\end{figure}

The optic specifications are derived from the science requirements and
sensor resolution.  For both the Cherenkov and fluorescence imaging
systems, the sensors have position resolution of 3~mm, and requiring
angular resolution of $0.2^\circ$ degrees.  The required aperture area
has been computed to be $0.61~{\rm m}^2$, which includes obscuration
by the sensors.  The full FoV is $45^\circ \times 3.2^\circ$ for the
two Cherenkov telescopes and $28.8^\circ \times 3.2^\circ$ for the
fluorescence telescope system. These specifications lead to an
equivalent focal length (EFL) for the fluorescence imager of 0.83~m,
and for the larger Cherenkov imager, an entrance pupil diameter of
0.967~m, meaning that the optic is operating at F/0.86.

\begin{figure}[tbp]
 \postscript{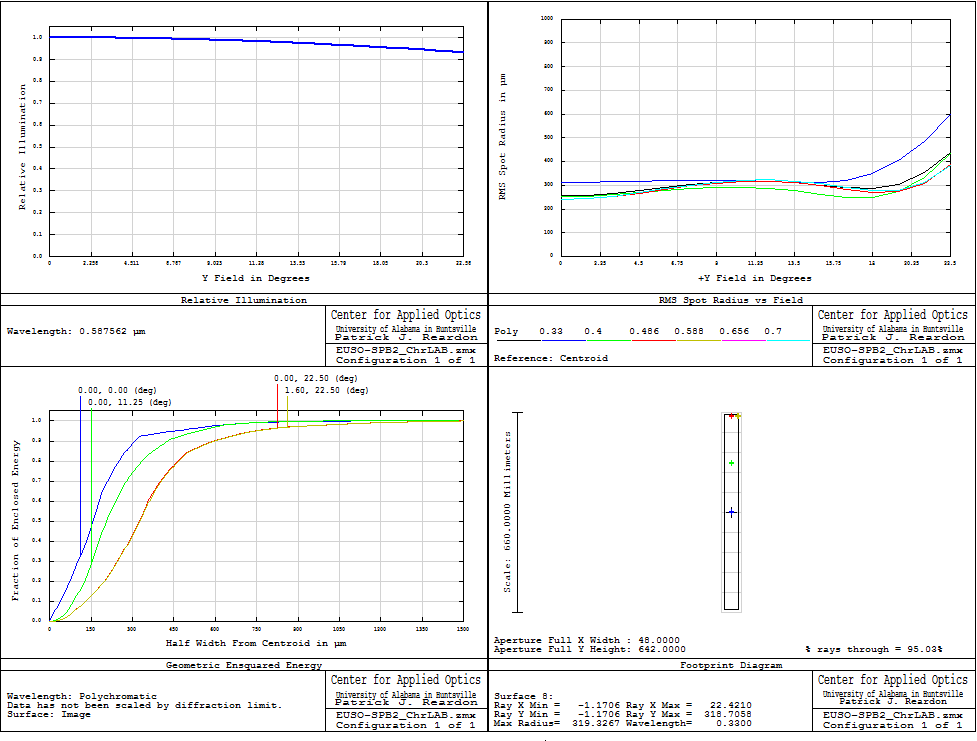}{0.9}
 \caption{{\it Up-left}.  Relative throughput only shows minor cosine falloff
over full $54^\circ$ FoV.  {\it Up-right.}~RMS spot radius as a
function of FoV.  {\it Down-left.}~The ensquared energy shows that
nearly all the energy  fits inside squares of $1.8~{\rm mm}$ width. {\it Down-right.}~Location of analyzed spots on the detector
array. \label{fig:o4}}
\end{figure}

A baseline design was completed using Zemax~\cite{Zemax}, with the
specifications listed in Table~\ref{tabla} and shown in
Fig.~\ref{fig:pat}.  The material selected for the corrector is an UV
transmitting poly methyl methacrylate (PMMA) well known to the EUSO
team and successfully used in several prototype systems.  To achieve
sufficient image quality, the design optimization varied: {\it
  (i)}~the radius of the primary mirror; {\it (ii)}~the spacing
between the primary mirror and the corrector; {\it (iii)}~the location
of the physical stop, {\it (iv)}~the shape of both corrector surfaces
including the spherical radius and even powered radial deformations
out to the 6th order term; and {\it (v)}~defocus. This design study concluded
successfully easily meeting the clear aperture and the 3~mm spot size
for both applications. The expected performance is shown in
Figs.~\ref{fig:o3} and \ref{fig:o4}.  Given the successful baseline
performance, the innovative approach for acquiring interlaced
horizontal and vertical detection required a modified optical
solution.  The desire is to create two horizontally separated spots,
separated by 25~mm (1 multi-anode photo-multiplier (MAPMT) width).
This ``bi-focal'' solution is to split the spherical mirror along a
horizontal line and rotate the top and bottom halves about the
vertical axis.  The amount of rotation of the two spherical sections
is $\pm 0.4^\circ$, yielding a relative angular difference of $\pm
0.8^\circ$.

A preliminary tolerance analysis has been performed that indicates
this system is extremely insensitive to misalignments.  For each
parameter, doubling the spot size requires: {\it (i)}~lateral
decenters of the corrector by more than 11~mm vertically or 7~mm
horizontally; {\it (ii)}~tilts of the corrector exceeding $6^\circ$;
{\it (iii)}~mirror slope errors must exceed 1~arc minute (a 0.010~mm
amplitude error with a period of 63~mm); {\it (iv})~corrector surface
slope errors exceeding 4~arc minutes of slope; {\it (v})~radius error on the
spherical mirror must exceed 20~mm; and finally {\it (vi)}~refractive index
errors, either isotropic, or spatially varying, must exceed any
reasonably expected deviations to yield a measurable performance
error.

Thermal soaks and gradients can impact the form of the optical
elements, the refractive index of the corrector and the positions of
the elements due to impacts on the structure.  Misalignments tend to
have minimal impact on the image quality, but large misalignments in
flight could shift the apparent position of the detector array which
could either result in a pointing uncertainty or an aberration if the
motion results in significant defocus. These optic considerations will
be included in the thermal analysis of the gondola.

\subsection{Focal plane detectors}

\subsubsection{Muon Cherenkov sensor}

The baseline focal surface detector uses a segmented linear
architecture to indentify the primary particle from its shower
characteristics. This architecture fits within the
scope of the project. A technology development effort aimed at improving on
the focal plane architecture and demonstrating it 
is also included to meet future measurements needs, including potentially
additional flights of the EUSO-SPB2 instrument.

\begin{figure}[tbp]
 \postscript{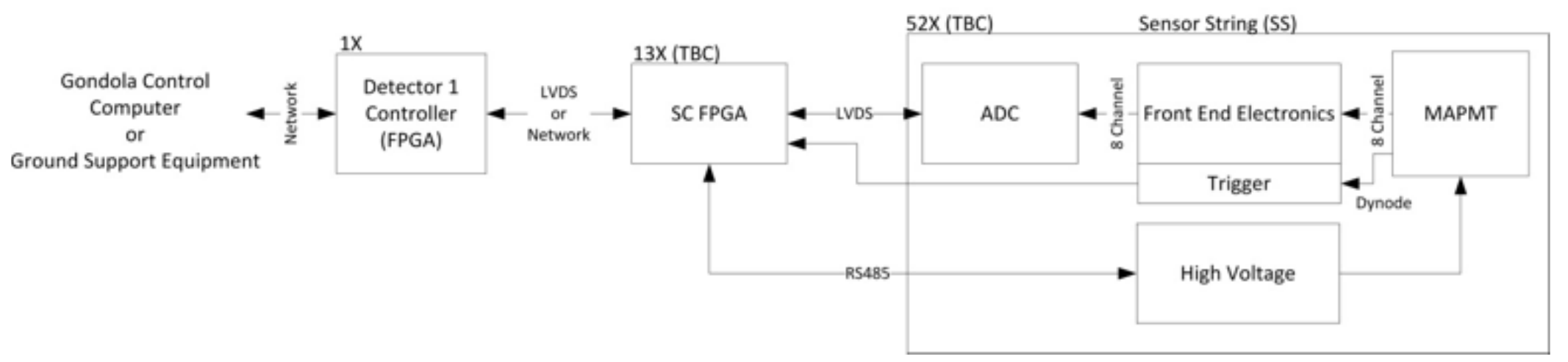}{0.9}
 \caption{Basic element of focal plane detector for imaging the
   Cherenkov signals. \label{fig:10}}
\end{figure}

\begin{figure}[tbp]
 \postscript{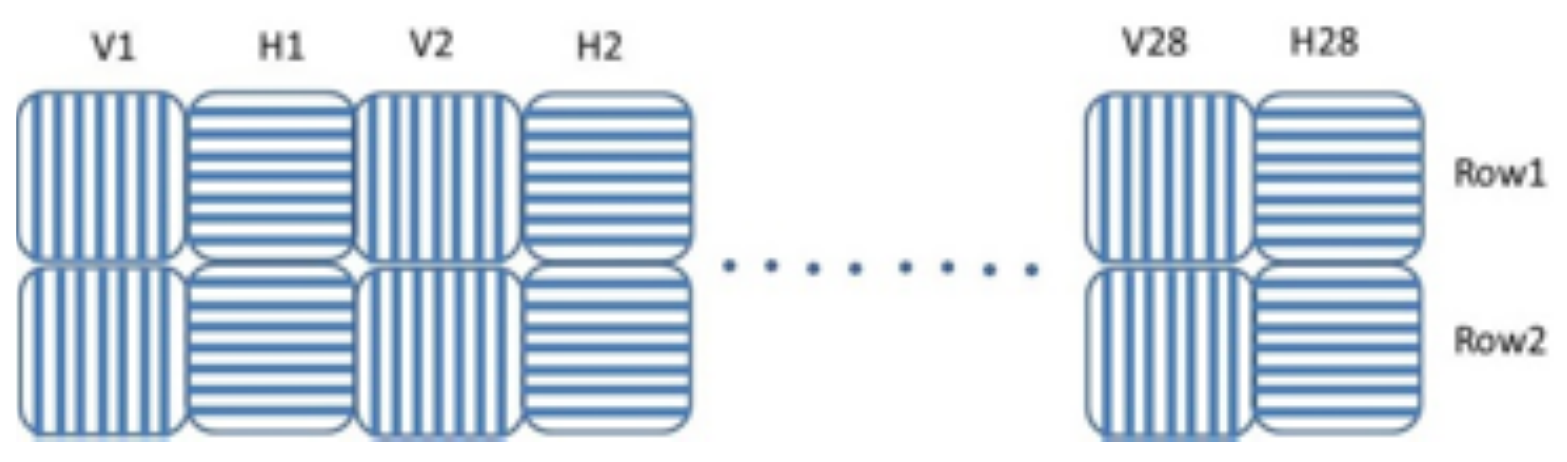}{0.9}
 \caption{MAPMT arrangement concept to measure profiles of the image
   in the vertical and horizontal directions. \label{fig:11}}
\end{figure}

The focal plane detector is built up from sensor strings based on 
MAPMTs followed by custom front-end electronics to condition the
signals and interface to a commercially available off-the-shelf (COTS)
fast analog-to-digital converter (ADC); e.g. AD9637.  These components
are controlled through firmware in a field-programmable gate array
(FPGA) that interfaces between the sensor string and the data
system. In Fig.~\ref{fig:10} we show a block diagram of the concept. It
 will be replicated, as shown, to
fill the full area of the focal surface. We will consider a 
further development of the readout system to adapt the technique
conventionally used in Cherenkov telescopes (such as the
DigiCam~\cite{Schioppa:2015yla}), which would make possible reading
out at the pixel, rather than the strip level. The upper half of the focal
plane uses MAPMTs with the anodes aligned along the horizon and the
MAPMTs in the lower half of the focal plane are aligned perpendicular
to the horizon. The two images produce by the binocular mirror will fall on these
two halves of the focal plane to enable location of the image with
single pixel resolution, see Fig.~\ref{fig:11}.

An optical filter will be used to limit the bandwidth of the arriving
photons. When convolved with the photo-cathode response, the
combination will provide the strongest detectable signal from muons 
relative to background. We have chosen the Schott VG9 filter which peaks at 525~nm, with 70\%
transmission and excludes wavelengths below 400~nm. This together with
Hamamatsu's extend green photocathode improves the response in the
intended band and 
limits the long wavelength response below 675~nm.

These filters are glued to Hamamatsu R11265-64 photomultipliers (PMTs), which  have been
designed for close-pack applications like EUSO. These 64-anode MAPMTs
provide a 78\% fill factor on the focal surface. We bond the 8 pixels in each row
together making a linear device with 8 anodes per PMT. The  dimensions
of each row are 
$3 \times 23~{\rm mm}^2$. Each MAPMT is mounted on a PCboard that
includes an high voltage (HV) supply
(C10940-03-R2) and front-end electronics (FEE). The PMTs are mounted
in a curved fixture to match the optic design and provide attachment
points for mounting at the focal surface.

Each channel's signal is conditioned by the FEE and relayed to the ADC
that continuously samples the PMT output. Digitized values are
processed in a pipeline with a depth of 16 samples. The ADC has
selectable speeds from 10 to 80~Msps. Our studies show that 10~Msps
will provide the required timing information  and fit within
the limit of the pipeline depth. The FPGA will monitor each MAPMT and
 trigger on the signal. This trigger will initiate a download of the
 data from the pipeline. The data will then be transmited to the
 onboard computer. Following the readout,  the sensor string will be
 re-enable to
continue operations. Periodically the FPGA also stores data from each of
the sensors strings to record the unbiased background signals from the telescope.

Image processing is required to recover  the event
from the two recorded linear profiles. This processing is done on the
ground in data analysis, which uses the linear data acquired together
with the timing information, measured background signals, and
simulations of shower signals (triggered by photon, protons, and iron
primaries). The  data are sufficient for event
identification for events with large impact parameters,  as illustrated in the right column of Fig.~\ref{fig:7}. The
``halo'' signal for events with small impact parameters are readily
measured using the linear anodes orientated normal to the horizon, as shown in
the left column of Fig.~\ref{fig:7}. The data from both images will be
analyzed using imaging reconstruction techniques to improve the
sensitivity to the primary composition.

The time profile of the Cherenkov signal contains information on the
development that supplements the analysis of the linear data. The
arrival time of Cherenkov photons depends on where along the shower
track they
were generated, with the earliest arriving photons originating from
the point on the shower nearest to the detector. The arrival time of
the signals indicated from where along the shower they originated. This
 provides leverage for identification of the muon content and thus the
 identity of the primary particle. The sharpness in time of the strong (electron) shower signal
is influenced by the interaction processes. As discussed above, the
absence of knowledge of the shower impact parameter makes it necessary
to use ensemble analysis techniques to extract information of the
primary particles.

\subsubsection{Upward directed Cherenkov signals}

The focal surface detector for the upward directed Cherenkov signals
is based on the sensor string described in the preceding
section. Since no image information is required, the dynodes are used
to form a trigger between paired MAPMTS to reduce false triggers from
charged particle interactions in a single PMT. The filter is selected
 to accommodate more signal bandwidth (300~nm to
550~nm from ground base Cherenkov telescope range). The ADC for each
triggered PMT records the intensity of the signals for analysis of the
Cherenkov pulse.

\subsubsection{Fluorescence sensor}
\label{fluorescence}

The fluorescence detector of EUSO-SPB2 will built on the experience of
EUSO-SPB. The 2017 flight will carry a full photon detection module (PDM)
with 2304 pixels and an optical system for a square FoV $12^\circ
\times 12^\circ$. The EUSO-SPB PDM is self-triggering. It
captures video clips of the slow (10-100~microseconds) fluorescence
signal from UHECR events. The lens system is not expected to survive
termination and recovery. It will be replaced by mirror with higher
optical through put and tighter focusing that will sharpen the
contrast of tracks signals to background by a factor of 2. This in
turn increase the sensitivity of the instrument to horizontal
extensive air showers viewed with the detector pointed near the
horizon.  To match the new FoV now looking at the horizon requires
repackaging of the PDM. The PDM is built from 9 elementary cells (EC),
each of which  is a self-contained detector comprised of 4 MAPMTs, HV, FEE and
trigger electronics that can be connected in parallel. In order to
cover a larger observation area we will change the EC layout for the PDM
from $3 \times 3$ to $1 \times 9$ configuration. Additional options will be considered based, in part,
on the performance of the PDM during  the 2017 EUSO-SPB flight.

\subsubsection{Infrared camera} 

EUSO-SPB2 will have a set of infrared cameras to monitor the cloud
coverage in the field of view of the Cherenkov and fluorescence
detectors. The design will be an update of the University of Chicago
Infrared Camera (UCIRC) built for monitoring cloud coverage during the
EUSO-SPB flight. UCIRC uses two infrared cameras with different
wavelength filters (10 micron and 12 micron) to capture images of cloud
cover in EUSO-SPB FoV. The two infrared images taken every minute
determine the  temperature (and therefore the height) of clouds between EUSO-SPB and
the ground. The design, construction, and testing of UCIRC2 for
EUSO-SPB2 will be done at Chicago with similar image reconstruction
and pixel-by-pixel temperature calibration procedures done for UCIRC.

\subsubsection{Testing  plans}
 
Testing is done at the component and subsystem level to confirm performance
and workmanship quality. It is continued as we integrate to higher levels. System level
testing of the telescopes will include ground-based and high altitude
tests to evaluate overall performance of the instrument. The EUSO-SPB
instrument was co-located at the TA in Utah and tested using
fluorescence signals from extensive air showers and a calibrated laser to permit studies
of the trigger levels, trajectory reconstruction and signal
resolution. The EUSO-SPB2 telescopes will be tested using a similar
approach for Cherenkov signals in addition to fluorescence. The CONUS
balloon flight will provide essential data for further tuning of the
instrument parameters before the science flight.

\subsubsection{Advanced sensor development}

An advanced sensor will be developed to take full advantage of
the information contained within the images of the Cherenkov signal in
discriminating primary particle identification. The imaging system
currently (SPACIROC3) being developed by the EUSO Collaboration is
intended for fluorescence measurements. It  uses a cadence time of
2.5~$\mu$s to measure the slow fluorescence signal. An
application-specific integrated circuit (ASIC) compatible
with shorter signals is needed for the Cherenkov detectors. One option is
to test the functionality and performance of the SPACIROC3
 with faster signals and possibly at a higher clock speed. We will
evaluate this and other available ASICs to develop a full imaging
focal surface that ideally would include timing resolution. A
proto-type of the concept will be developed and tested in the
laboratory. If successful, it will be flown on the test flight as
part of the focal surface. Further plans for development will be
formed after completing this stage.

\subsubsection{Mission operations}

The EUSO-SPB2 flight includes both local and remote operations. The
field team will carry out final integration and check-out of the
instrument at the launch site and commence operations once launch
occurs. Remote communication between the science team and the CSBF
command center will be though a server located at CSBF and used for
transferring data. Progress of the flight will monitored at the
science team home stations with commanding originating from the
Science Operation Center (SOC) at Colorado School of Mines. The science team will provide
input for the observation plan based on data from operational weather
satellites and forecasting. The SOC will also serve as the data
archive during the mission. The processed data, along with the
publications generated by this investigation, will be archived
according to the Data Management Plan.

\section{SIMULATION AND ANALYSIS SOFTWARE}

For the EUSO-SPB2 project, we are employing two independent simulation and
reconstruction packages, one called \Offline\cite{Argiro:2007qg,Abreu:2011fb}
and the other know as {\sc ESAF} (EUSO Simulation and Analysis
Framework)~\cite{Berat:2009va}. Both of these packages support development of
event simulation and reconstruction. Most importantly, the detailed simulation
capacity of these codes will be employed early on in the EUSO-SPB2 project to refine
the final instrument design.

\Offline was originally developed for the Pierre Auger
Observatory~\cite{ThePierreAuger:2015rma}, but has since been adapted
to the needs of EUSO and associated pathfinders. {\sc ESAF} was
specifically designed for EUSO and its pathfinders. These codes are
used to simulate the cosmic ray shower development in the atmosphere,
the production fluorescence and Cherenkov photons, and their
propagation up to the detector. In both frameworks the various EUSO
pathfinders are implemented and are used to simulate the detector
response. The two frameworks also contain algorithms for
reconstruction of data gathered by the pathfinders. We feel it is
quite advantageous to have two simulation and reconstruction packages
available, as it affords an opportunity for detailed cross--checks and
ensures EUSO performance estimates are reliable.

The \Offline software, has been used for simulation and reconstruction for the
Auger Observatory since the first physics results were published in 2004. At the
time of writing the software comprises some 360~000 lines of code and 35~000
lines of configuration information, representing a roughly 100 person-year
investment according to the Constructive Cost Model~\cite{cocomo}.

\Offline includes the latest fluorescence and Cherenkov
light-yield models~\cite{Bohacova:2008vg} atmospheric models and interfaces to
many air shower simulation packages, including {\sc AIRES}~\cite{Sciutto:1999jh},
{\sc CORSIKA}~\cite{Heck:1998vt}, {\sc SENECA}~\cite{Drescher:2002cr} and {\sc
CONEX}~\cite{Bergmann:2006yz}. As Auger analyses using fluorescence
measurements are quite mature at this point, the simulation algorithms have been
well vetted with real data. For realistic Monte Carlo simulation (and real data
analysis), the \Offline code provides simple access to a collection of databases
in which atmospheric monitoring data from a variety of sources can be
stored. Raytracing in the optical systems is performed using
Geant4~\cite{g4}. Electronics and noise simulations can be performed using
parametric models or from field measurements of the instrument in
question. Several algorithms prepared by different teams have been developed to
perform shower reconstruction.

The \Offline framework provides many utilities and conveniences, to be
discussed later, which have been exercised for over a decade by a
large collaboration conducting data analysis. Furthermore, parts of
the \Offline framework have been adopted by other collaborations,
including CODALEMA~\cite{Ardouin:2006gj}, TUNKA~\cite{Antokhonov:2011zz},
HAWC~\cite{Abeysekara:2013tka}, LOFAR~\cite{lofar} and
NA61/SHINE~\cite{Wyszynski:2012fa,Sipos:2012hs}, allowing for mutually beneficial
collaboration among scientists working on different experiments. The
\Offline framework is freely available upon request and is distributed
under an open source BSD license~\cite{bsd}.

The {\sc ESAF} package was specifically designed for the performance assessment of
space based cosmic ray observatories. It was developed in the framework of
the EUSO mission~\cite{Berat:2009va} during the ESA phase A study.  This
software has been written mainly in C++ and makes use of the {\sc ROOT}
package~\cite{root}.  The software was developed following an object oriented
approach and is structured in a modular way.

The compilation of the {\sc ESAF} software produces two distinct executable files
called respectively Simu and Reco. The first one performs the simulation of the
entire physical process from shower to telemetry. In this context, several air
shower generators like {\sc SLAST}~\cite{NaumovSlast}, {\sc CONEX}~\cite{Bergmann:2006yz},
{\sc CORSIKA}~\cite{Heck:1998vt} and others are available for use. An atmospheric
model according to the 1976 Standard US Atmosphere~\cite{USStandard1976} is
implemented as well as different parameterizations for Fluorescence and
Cherenkov yield. Both the NKSA~\cite{Nagano:2004am} and the KLNOTU~\cite{Kakimoto:1995pr}
fluorescence yield models have been implemented in the software. Standard
Cherenkov theory is used in the {\sc ESAF} computations of Cherenkov light emission.
The Rayleigh scattering and ozone absorption processes are simulated in {\sc ESAF} by
means of the {\sc LOWTRAN 7}~\cite{lowtran7} atmosphere software.  Several versions of
the optics Monte Carlo simulator, developed at RIKEN (RIKEN ray trace code)
have been interfaced with {\sc ESAF}. The optics simulators for all the pathfinders
like EUSO-TA, EUSO-Balloon, EUSO-SPB, and Mini-EUSO have been implemented and
tested in {\sc ESAF}. All the space-based detectors like TUS, K-EUSO and JEM-EUSO
(in several configurations) are available.  In addition, a GEANT 4 optics
interface and a parametrical optics simulator are implemented.  Both the PMT and
the EC electric signal treatment is performed in a
parametrical way. The last part of the simulation chain consists of the trigger
sequence. A multiple stage trigger scheme is implemented in {\sc ESAF}
in order to maximize the ratio of real events to background.  Once the trigger
sequence has been applied the Simu executable  produces an
output ROOT file.

The Reco executable activates the reconstruction chain. If the event has
been selected by the trigger algorithms, the first task is the recognition of
pixel-GTUs with signal
within the detector response table. Several algorithms have been implemented for
this purpose.\footnote{A gate time unit (GTU) $= 2.5~\mu{\rm s}$.}
Once a clear shower-like pattern has been identified several time and space
fits are performed for the arrival direction recognition.  Eventually, the
profile reconstruction is performed in order to fit the $X_{\rm max}$ and energy
of the shower.

\subsection{\Offline design}

\begin{figure}[ht]
\centering
\includegraphics[width=0.5\textwidth]{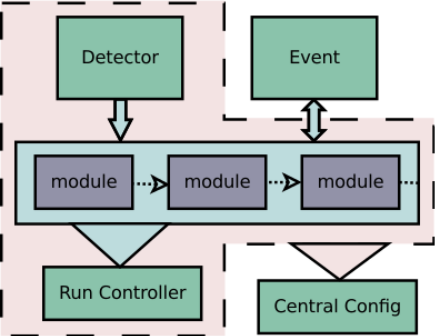}
\caption{General organization of the \Offline framework. See the text for detailed
explanation.}
\label{f:modules}
\end{figure}

The \Offline framework comprises a collection of physics algorithms contained in
{\em modules}; a {\em RunController} which commands the modules to execute in a
particular sequence; a read/write {\em Event Data Model} from which modules read
information and to which they write their results; a {\em Detector Description}
which provides an interface to conditions data, such as detector calibration,
performance and atmospheric conditions; and a {\em CentralConfig} which directs
the modules and framework components to their configuration data and which
tracks provenance. The general scheme is illustrated in Fig.~\ref{f:modules},
and discussed in more detail below.

Simulation and reconstruction tasks are factorized into sequences of processing
steps which can be simply pipeli\-ned. Physicists prepare processing algorithms
in {\em modules}, which they register with the \Offline framework via a one-line
macro. This modular design facilitates comparison of algorithms and building a
variety of applications by combining modules in various sequences. One can, for
instance, very easily swap out a module for reading in simulated showers with a
module to simulate laser shots in the instrument FoV. Control of module
sequences is implemented with a {\em Run Controller}, which directs module
execution according to a set of user provided instructions. We devised
an XML-based language as one option for specifying sequencing instructions; this
approach has proved sufficiently flexible for the majority of our applications,
and it is simple to use, though a Python-based module control is also possible.

The \Offline framework includes parallel hierarchies for accessing data: the
{\em detector description} for retrieving conditions data, including detector
geometry, calibration constants and atmospheric conditions; a plug-in mechanism
in the atmosphere description allowing various techniques for computing
fluorescence and Cherenkov yields, both from parametric models and from
measurements stored in databases; and an {\em event data model} for reading and
writing information that changes for each event.

The detector description provides an interface from which module authors can
retrieve the conditions data. Data requests are relayed to a back-end comprising
a registry of so-called {\em managers}, each of which is capable of extracting a
particular sort of information from various data sources. The manager mechanism
is configurable and relieves authors of the physics code from having to deal
with the details of selecting and decoding the correct data source.  Managers
are arranged in a ``chain of responsibility'' such that if an upstream manager
cannot answer a request, it is passed along to downstream managers for another
try.

The event data model contains raw, calibrated, reconstructed and Monte
Carlo information, and serves as the backbone for communication between modules.
The event is instrumented with a protocol allowing modules to discover its
constituents at any point in processing, and thereby determine whether the input
data required to carry out the desired processing are available.
\Offline is also equipped to read formats employed by the most popular air shower
simulation packages~\cite{Sciutto:1999jh,Heck:1998vt}.

The \Offline framework includes a system to organize and track data used to
configure the software for different applications as well as parameters used in
the physics modules.  The \emph{Central Config} configuration tool points
modules and framework components to the location of their configuration data,
and connects to Xerces-based~\cite{xerces} XML parsers to assist in reading
information from these locations.  We have wrapped Xerces with our own interface
which provides ease of use at the cost of somewhat reduced flexibility, and
which also adds functionality such as automatic units conversion and casting to
various types, including commonly used container types.om typographical errors.

The {\em Central Config} keeps track of all configuration data accessed during a
run and stores them in an XML log file, which can be subsequently used to
reproduce a run with an identical configuration. This allows collaborators to
exchange configuration data for comparing results. The logging mechanism is
also used to record the versions of modules and external libraries which are
used for each run.

Syntax and content checking of the configuration files is implemented using W3C
XML Schema validation~\cite{xml-schema}.  Schema validation is used not only for internal
configuration prepared by framework developers, but also to check the contents
of physics module configuration files.  This approach reduces the amount of code
users and developers must prepare, and supports very robust checking.

As in many large software projects, each low level component of the \Offline framework is
verified with a small test program, known as a {\em unit test}.  We have adopted
the CppUnit testing framework~\cite{cppunit} to help with implementing these
tests. In addition to unit tests, a set of higher level acceptance tests has
been developed which is used to verify that complete physics applications
continue to function as expected, within some tolerance, during ongoing
development.  We employ a BuildBot system~\cite{buildbot} to automatically
compile the \Offline software, run the unit and acceptance tests, and
email developers in case of problems. The BuildBot runs each time the software
repository is modified.

The \Offline build system is based on the CMake cross-platform
build tool~\cite{cmake}, which has proven adequate to manage this project. In order to ease
installation of \Offline and its external dependencies, we have adopted
a tool known as APE (Auger Package and Environment)~\cite{ape}. APE is a
dependency resolution engine co-developed by the Auger and HAWC
collaborations. It automatically downloads a vetted combination of external
packages required by \Offline, builds them in whatever native build system
applies for each package, and sets up the user's environment accordingly. APE is
freely available, and used by the NA61/SHINE Collaboration, as well as Auger and HAWC.

\subsection{ ESAF design}
The {\sc ESAF} simulation code is structured in several independent modules the higher of which is
the so called \textit{SimuApplication}. An instance of this class is created in the
\textit{simu\_main.cc} file where the method \textit{SimuApplication::DoAll()} is
called. This method performs the iterative call of the \textit{SimuApplication::DoEvent
  ()} method which takes care of the entire physical process on
a single event basis. Such a method will create an instance of the
\textit{LightToEuso} class which executes the entire process from primary
particle to photons on pupil. Several choices are available on which simulator is to be
used but the default option is the so called \textit{StandardLightToEuso}
class. By calling the \textit{StandardLightToEuso::Get()},
the virtual \textit{Get()} methods of the shower generator, of the light  production
and transport will be called. Each one of the mentioned \textit{Get()} methods will deliver
 output objects describing the shower profile, photons in atmosphere and
photons on pupil. The choice of the object oriented
approach shows its power here where the call of several polymorphic
\textit{Get()} methods allows great flexibility.

Always inside the \textit{SimuApplication::DoEvent()} method the
virtual \textit{Detector::Get()} method will be
called. This method takes care of the entire detector simulation. Several
choices are open at this stage between various detector configurations. The
most important of them can be considered to be the
\textit{EusoDetector}, (activating the RIKEN ray trace code),  the
\textit{G4Detector}  (activating the Geant 4 optics) and other testing or
debugging detector simulators. Calling one of the above described methods will
activate both optics and electronics simulators.
As final output of the entire procedure a \textit{Telemetry} object is
produced.

\begin{figure}[ht]
\centering
\includegraphics[width=0.6\textwidth]{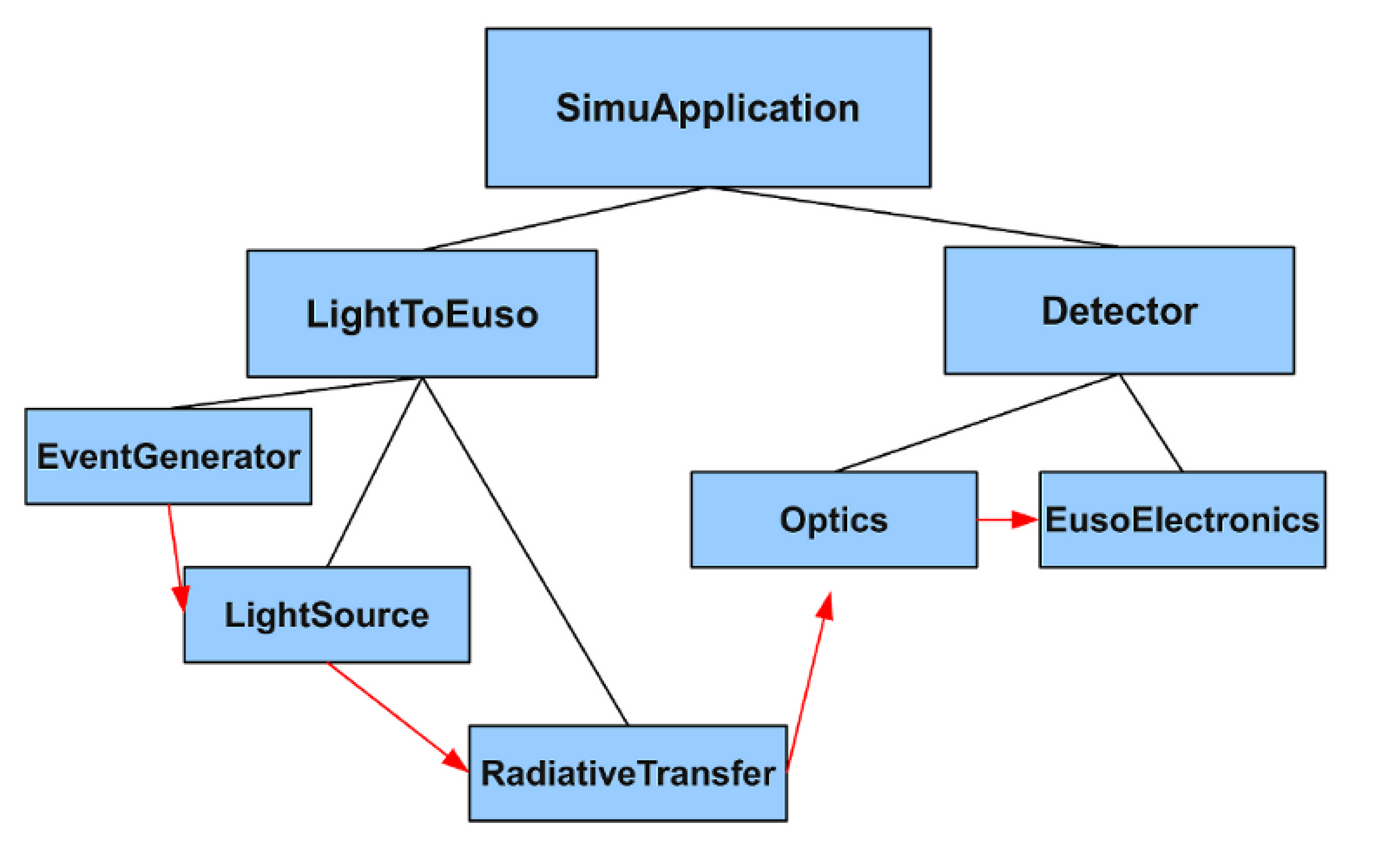}
\caption{a schematic view of the {\sc ESAF} Simu application
structure. The main application is the so
called \textit{SimuApplication}. The \textit{LightToEuso} application
takes care of all the physical process from shower to
detector. The \textit{EusoDetector} application performs the simulation
of optics and electronics.}
\label{fig:esafStruct}
\end{figure}

\begin{figure}[ht]
\centering
\includegraphics[width=0.7\textwidth]{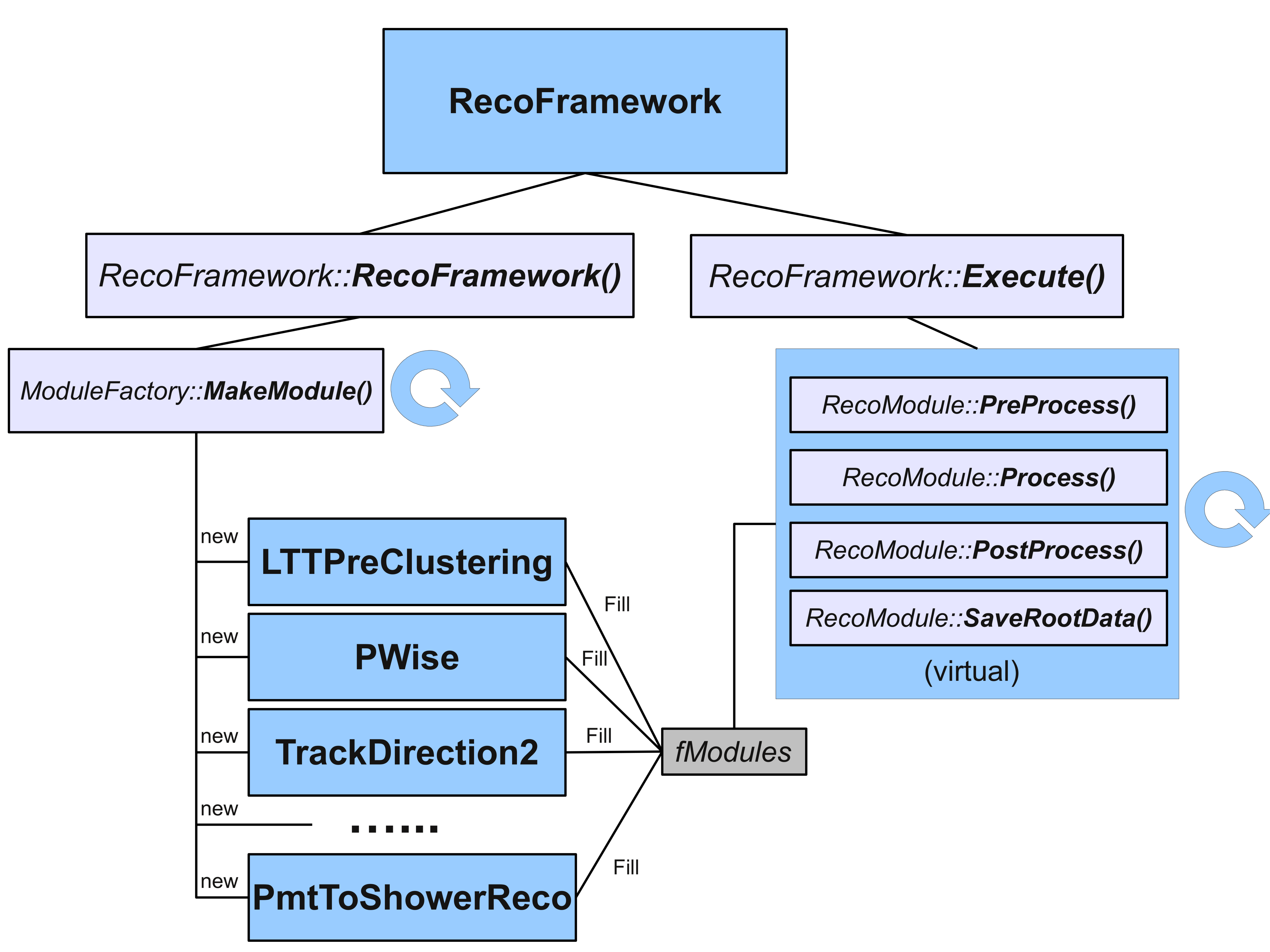}
\caption{A sketch of the reconstruction framework.
The main application \textit{RecoFramework} calls iteratively
the \textit{MakeModule} method which allocates all the required
modules. A \textit{vector} of pointers to the allocated objects is
saved under the name \textit{fModules}. In the \textit{Execute} method
the operations of all the modules are performed. All the modules are
inheriting from the \textit{RecoModule} class.  The virtual
methods \textit{PreProcess}, \textit{Process}, \textit{PostProcess}
and \textit{SaveRootData} are called for all the allocated
modules. Note that blue boxes represent classes, blue-gray boxes
methods, the gray box is a C++ \textit{vector} and the circular arrow
indicates iterative repetition of some method or sequence of
methods.}
\label{fig:recoStruct}
\end{figure}

The reconstruction procedure is activated in the \textit{reco\_main.cc}
file. Here an instance of the \textit{RecoFramework} class is created and the
method \textit{RecoFramework::Execute()} is called. While in the constructor
function
\textit{RecoFramework::RecoFramework()} the module chain is built, the \textit{RecoFramework::Execute()} method
performs the entire sequence of calls to reconstruct the event.
In fact, the module sequence is first initialized through an iterative call of the
\textit{ModuleFactory::MakeModule()} method which allocates all the
\textit{RecoModule} objects requested by parameter files. A
\textit{vector} named
\textit{fModules} with all the pointers to the created \textit{RecoModule}
objects is created.
In the \textit{RecoFramework::Execute()} method all the modules (which inherit from
\textit{RecoModule}) are initialized, called and cleared. Eventually all the
output data are saved in the ROOT file.
For performing all the mentioned operations, the polymorphic methods \textit{RecoModule::PreProcess()},
\textit{RecoModule::Process()}, \textit{RecoModule::PostProcess()} and \textit{RecoModule::SaveRootData()} are
declared in each module.
Each module has a specific function which can be either pattern
recognition, direction fitting, profile reconstruction or $X_{\rm max}$ and energy
reconstruction.
Several modules have been implemented in the course of the years but the most
actual and currently updated are the \textit{LTTPreClustering} and \textit{PWISE} for
the pattern recognition, the \textit{TrackDirection2} for the direction reconstruction
and the \textit{PmtToShowerReco} for the energy reconstruction.
A schematic view of the above mentioned structure is shown in
Figs.~\ref{fig:esafStruct} and \ref{fig:recoStruct}.

\subsection{Use of \Offline and {\sc ESAF} for EUSO pathfinders}

Both the \Offline and the {\sc ESAF} packages were designed to allow a great deal of
flexibility and to easily change the detector configuration.  This flexibility
made it relatively straightforward to use the codes for the various EUSO
pathfinders.  Both packages have been used for simulation and reconstruction of
data for the EUSO-Balloon detector~\cite{Ballmoos:2015spu}, the
EUSO-TA~\cite{Piotrowski:2014tsa} instrument as well as for simulation of the
pending mini-EUSO and SPB missions~\cite{Ricci:2015srs,Wiencke:2015oko}.

A EUSO pathfinder was flown aboard a balloon on August  2015 from the
Timmins Stratospheric Balloon Facility in Ontario. During this flight, the
instrument recorded data for about 5 hours. A laser and flasher were carried
aboard a helicopter which flew beneath the balloon payload to test the
instrument.

\begin{figure}[tbp]
\begin{minipage}[t]{0.49\textwidth}
    \postscript{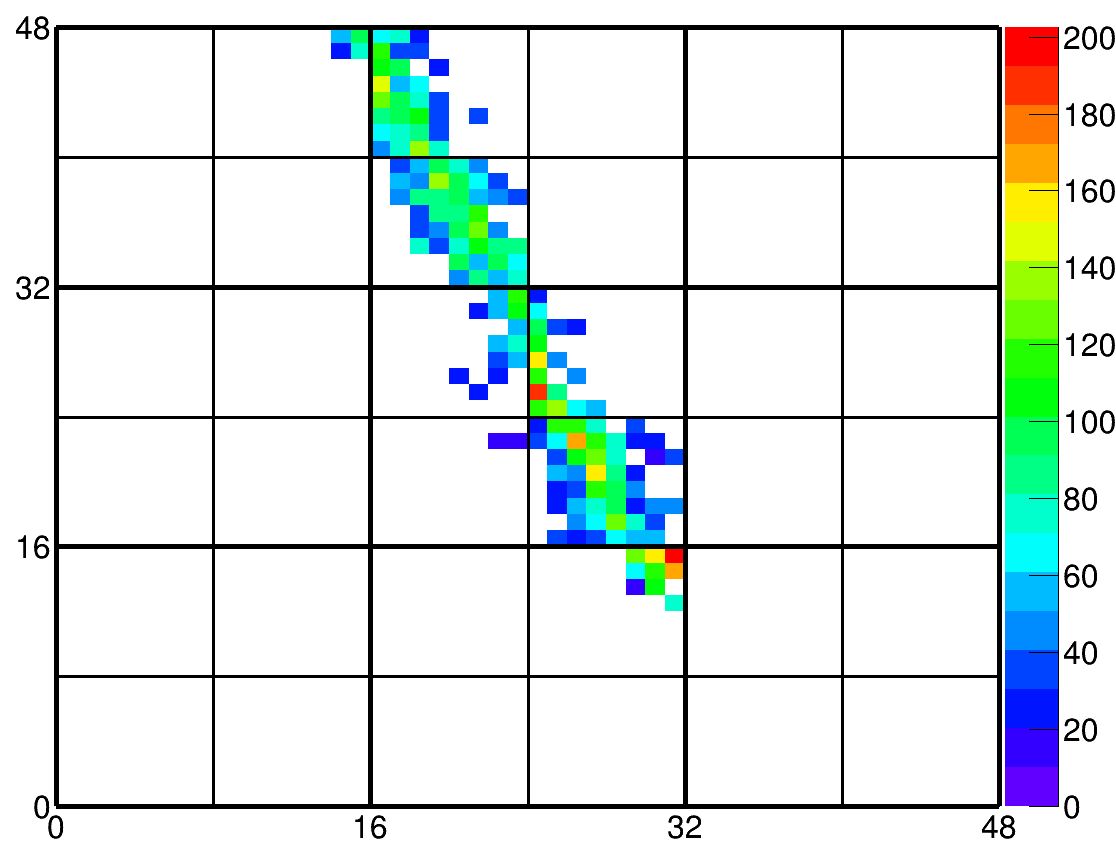}{0.99}
\end{minipage}
\begin{minipage}[t]{0.49\textwidth}
    \postscript{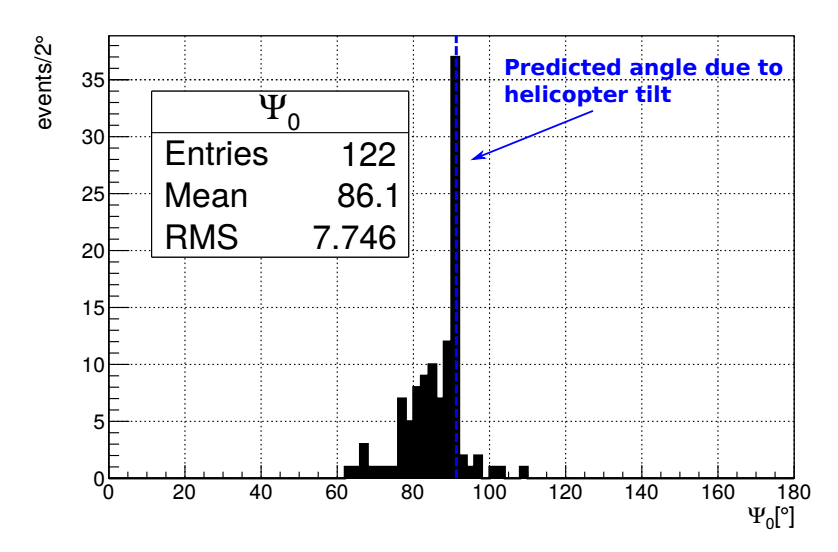}{0.99}
\end{minipage}
\caption{{\it Left.} Focal surface image of a flasher and laser shot taken during the 1'st stratospheric EUSO
  balloon flight as reconstructed using the \Offline software. The horizontal and
  vertical axes label the pixel number, and the color indicates the number of
  flash ADC counts. The image is integrated over all of the time
  gates (2.5 microseconds each) for which flasher or laser data appeared to be
  present. {\it Right.} Zenith angle of reconstructed laser tracks fired across the EUSO-Balloon field
of view.
\label{f:balloon}}
\end{figure}

The \Offline software was used to simulate the instrument and to
reconstruct data taken during the flight.  Figure~\ref{f:balloon} contains
an image of a flasher and laser shot fired across the field of view of the
payload. Figure~\ref{f:balloon} also shows the zenith angle distribution of
reconstructed laser shots gathered during the campaign. As expected, the
distribution peaks near $90^\circ$ as the laser was shot horizontally across the
downward pointing telescope.

Similar studies were performed using the {\sc ESAF} package, including
simulation of the EUSO-Balloon response and detailed study of trigger
performance. For instance, description of the simulation and reconstruction procedure used to
predict the reconstruction performance is detailed in~\cite{cambursanoThesis}.

The EUSO-TA~\cite{Piotrowski:2014tsa} pathfinder measurements were also simulated and
reconstructed in a joint effort using both the \Offline and {\sc ESAF} packages~\cite{Bisconti:2017bla}.
In this experiment, the Black Rock Mesa TA (BRM-TA) telescopes were used to trigger a EUSO prototype
  instrument when an air shower was detected.  Data recorded by the BRM-TA instrument
  were then used to reconstruct the shower distance, angle and energy. The simulation packages were then used to generate
 showers with the appropriate parameters to reproduce the EUSO-TA detected signal.
The mapping of the real detector was introduced in both \Offline and {\sc ESAF} in order to take into account for the different
efficiency of the pixels and for the dead PMTs. Figure~\ref{fig:5} shows a comparison of a measurement
  and simulation of an air-shower recorded by EUSO-TA on 13 May 2015~\cite{Bisconti:2017bla}.
Several test sources like stars, flashers and laser shots were implemented in the packages
in order to validate the detector response and check reconstruction.

\begin{figure}[ht]
\centering
\includegraphics[width=0.9\textwidth]{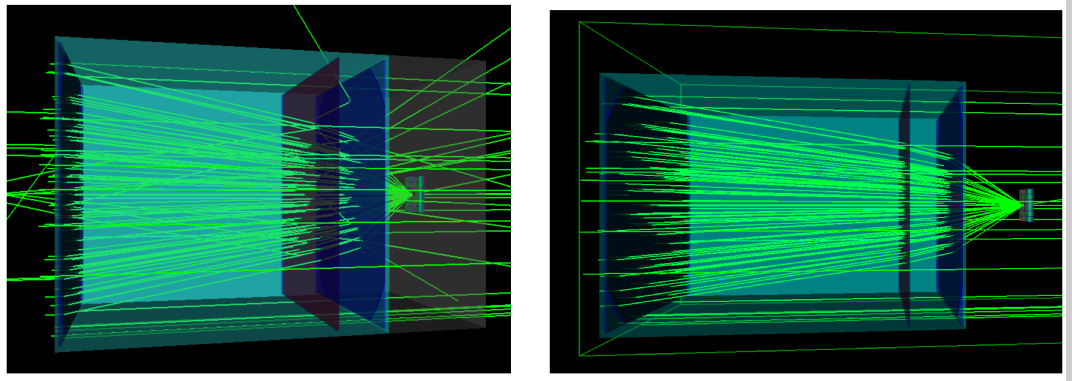}
\caption{EUSO-SPB simulated in Geant 4. {\it Left.} Parallel rays focusing on the Photo Detector Module
(PDM). In this case reflections are included in the simulation. {\it Right.} Same situation
with all reflections switched off. Note that the center (diffractive) lens may not actually be flown on the
EUSO-SPB flight due to excessive photon absorption.}
\label{f:euso-spb}
\end{figure}

A simulation for the upcoming EUSO-SPB flight has also been prepared, as well as
the codes to read the data.  A campaign of laser shots were reconstructed
in \Offline.  Figure~\ref{f:euso-spb} depicts a Geant4~\cite{g4} simulation of
the nominal EUSO-SPB design employing 3 lenses, and illustrates the ray tracing
of a few photons entering the telescope. In these simulations, we re-implemented
the native Geant4 modules which handle Fresnel reflections, total internal
reflection, and reflections from ground or painted surfaces.  This
re-implementation was conducted in order to search for possible artifacts on the
focal surface resulting from reflections in the optical system, and serves as a
very convenient debugging tool. Figure~\ref{f:euso-spb} depicts images with
reflections turned on and reflections turned off.

\begin{figure}[ht]
\centering
\includegraphics[width=0.5\textwidth]{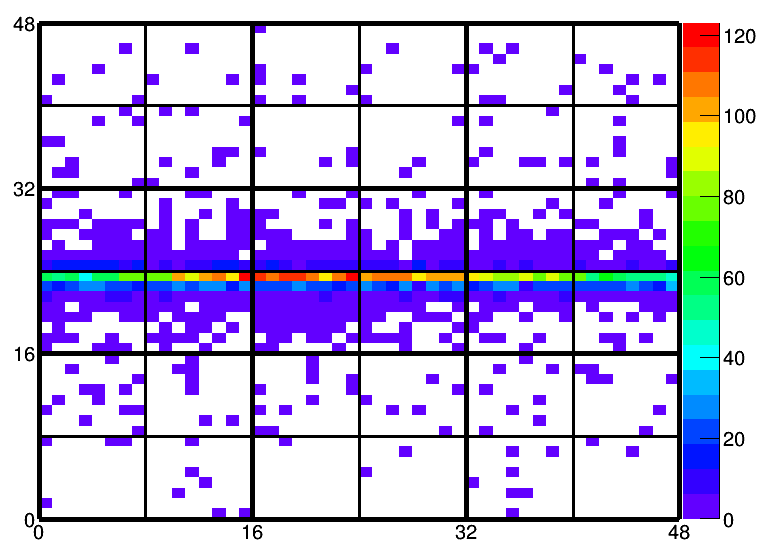}
\caption{Simulated $10^{11}~{\rm GeV}$ shower at a zenith angle of $50^\circ$ as
seen on the EUSO-SPB focal surface. The signal is integrated over 127~GTUs. No noise
is included in this particular simulation.}
\label{f:euso-spb-shower}
\end{figure}

In Fig.~\ref{f:euso-spb-shower} we show a simulation of a
$10^{11}~{\rm GeV}$ air shower image on the EUSO-SPB focal surface,
integrated over 127~GTUs.

A very comprehensive study on the expected trigger rate was performed with {\sc ESAF}.
The efficiency of the trigger has been studied in different cloud conditions,
background, altitude and detector configurations~\cite{cambursanoThesis}.  A
study of the cloud coverage and moon phase expected at the latitude and in the
season where the balloon will fly has been performed~\cite{venezianiThesis}.  We
estimated the number of detectable particles to be between roughly 5 and 11
events depending on conditions the feasible SPB flight durations.  As in
the EUSO-Balloon case, we tested the energy reconstruction performance and estimated the
fraction of reconstructable events. 

Simulations for Mini-EUSO have also been implemented in both \Offline
and {\sc ESAF}.  Using {\sc ESAF}, we found that the energy threshold
is over $10^{12}$~{\rm GeV}~\cite{bertolaThesis}.  Furthermore we tested the
response of the detector to meteors, Transient Luminous Events (TLEs),
lightnings, space debris and planes.  The trigger scheme has been
tested on slow-moving events, relevant, for instance, for meteor
detection. The Mini-EUSO trigger scheme is being tested on a wide
range of events of different times scale and sizes.

\subsection{EUSO-SPB2}

Developing tools for simulation and reconstruction of a variety of
instruments within the same overarching software framework is
challenging.  Fortunately, a large amount of the code developed so far
can be straightforwardly recycled for use by EUSO-SPB2.  It will (of
course) be necessary to update the modules to model the new optical
systems and electronics of SPB2, but the existing code provides an
excellent opportunity to leverage a great deal of previous development
effort.  The adaption of both {\sc ESAF} and \Offline to the needs of
EUSO-SPB2 is underway.  Both of these packages will be employed for
simulations in early phases of the EUSO-SPB2 project in order to
refine design of the final instrument.  Indeed, preliminary tests are
now being carried out to assess the capability to detect the direct
Cherenkov light and the fluorescence light from distant events.
Again, the two packages will be used to cross-check one another.
We have begun developing a pipeline to simulate the Cherenkov
signal resulting from $\nu_\tau$'s skimming the Earth and producing
$\tau$ leptons which generate upgoing showers~\cite{neutrino2014}.

\section{EXPECTED PERFORMANCE}
\label{sec:performance}
\subsection{Duty cycle}

\begin{figure}[ht]
\postscript{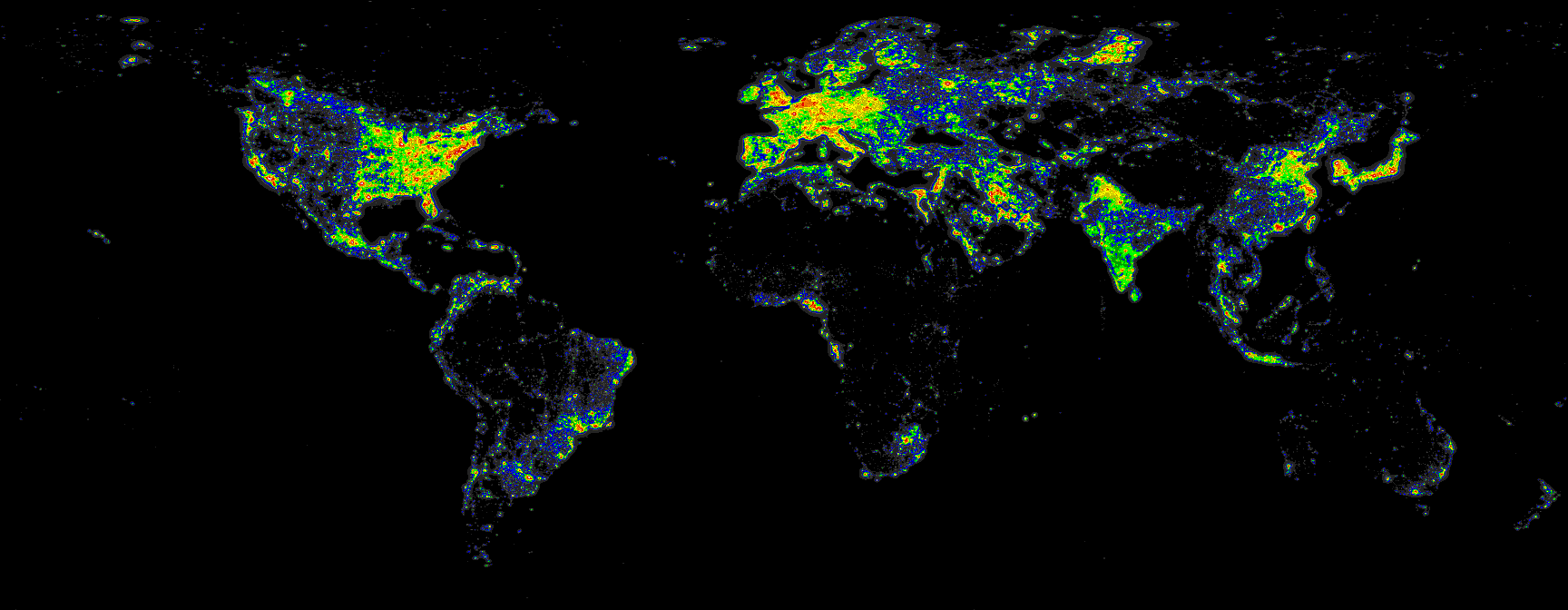}{0.99}
\caption{Light pollution heat map of the world~\cite{Lorenz}.}
\label{dark-night}
\end{figure}

EUSO-SPB2 observations can only be done on clear moonless nights. The
balloon will fly in the Southern hemisphere and thus subject to less
light pollution than if it were exposed to the Northern hemisphere,
see Fig.~\ref{dark-night}. A thorough study suggests that a 50 day
flight launched at Wanaka latitude of $45^\circ~{\rm S}$ and about the
expected time of EUSO-SPB (March/April) would see between 190 and
260~hr of dark time, depending on when the launch happens relative to
the moon phase~\cite{Wiencke:2015oko}.  For a 100~day flight, the
fluctuations would smooth out a bit, and hence we estimate 500~hr
would be a realistic number of dark hours, with no moon and between
the end and start of astronomical twilight at 33~km.  Relative to a
detector on ground, there is a loss of about 50~minutes per day
because the horizon is further away and this loss factor is
included. Then, 500~hr/2400~hr yields a 20.8\% duty cycle.  Since the
SPBs from Wanaka go east, there is a jet lag effect that reduces the
500~hr by a few percent. We account for this correction by taking a
duty cycle $\delta \approx 0.2$.  This estimate does not take into account
possible reduction of the duty cycle due to obscuration by clouds, and
further assumes an operationally perfect detector. In our calculations
we acount for the effect of cloud obscuration by introducing an
scaling factor $\varkappa$, which combines the trigger effects and
reconstruction efficiency in the presence of clounds, as well as the
fraction of time during which data are taken in clear and cloudy
conditions; namely, if the entire data sample were to be taken in
clear sky conditions we have $\varkappa =1$. Preliminary ESAF studies
for EUSO-SPB suggest that $\varkappa \approx
0.75$~\cite{cambursanoThesis,venezianiThesis}.  Since roughly 2/3 of
the time the FoV will be obscured by clouds, the previous $\varkappa$
estimate indicates that half of the events obscured by clouds can be
considered of sufficient quality to have the same reconstruction
and trigger efficiency as clear sky conditions. This agrees with
previous studies for the JEM-EUSO mission~\cite{Cano,Adams:2014tnr,SaezCano:2012km,Saez-Cano:2014zka}. To estimate the event
rate for EUSO-SPB2 we conveniently  define an effective duty cycle
$\delta_{\rm eff} = \delta \, \varkappa \approx 0.15$.

\subsection{Event rates} \label{sec:rates}

By comparing the red and blue lines in the left panel of Fig.~\ref{fig:2}  one can
discern that the number of events detected via Cherenkov
radiation, in the range $10^7 < E/{\rm GeV} < 10^8$, would be around
$1,000$ for a 100 day mission. There may be small variations in this ratio arriving from the
effective duty cycle achieved by the instrument. 

The final configuration for the EUSO-SPB2 fluorescence detector will 
depend on the performance of EUSO-SPB. Therefore, we cannot yet derive with
certainty the expected rate of UHECR events.
Obviously, we expect EUSO-SPB2 to perform better than
EUSO-SPB. In the following we provide a rough estimate of the number of
events using the expected EUSO-SPB performance as a guideline.

\subsubsection{Estimate of UHECR events using {\sc ESAF}}

\begin{figure}[tbp]
\postscript{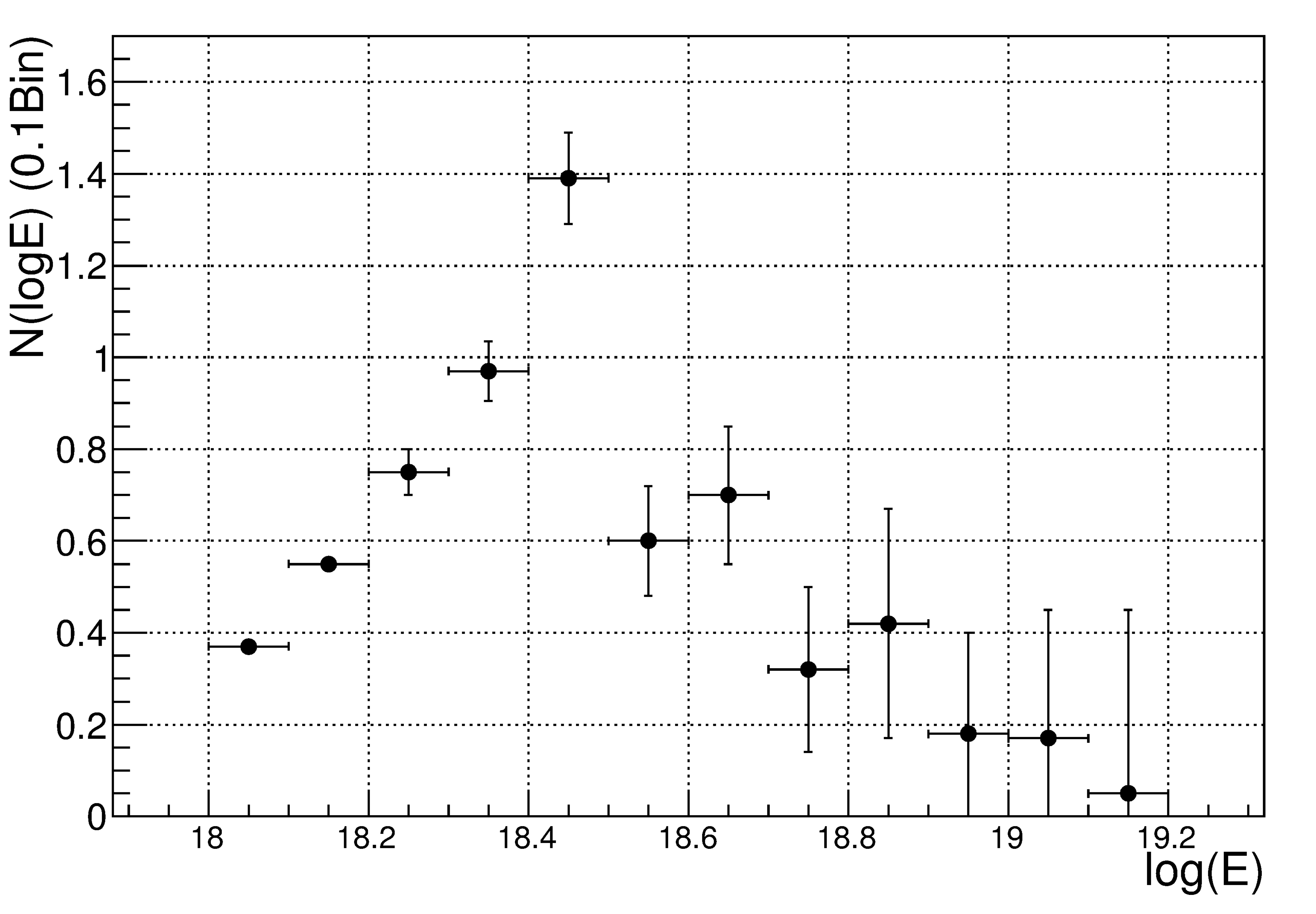}{0.55} \caption{Spectrum of triggered
  events for 118~hr flight as would be detected by EUSO-SPB in clear
  sky conditions. \label{fig:esaf-cloudy}}
\end{figure}

We estimated the event rate of EUSO-SPB with {\sc ESAF} taking into
account clear and cloudy sky conditions, and the effect of Cherenkov reflection from
high level clouds~\cite{venezianiThesis, bertolaThesis}.  As an
illustration, in Fig.~\ref{fig:esaf-cloudy} we show the expected number of
triggered events as a function of $\log(E)$, for clear
sky conditions. Taking into account the reconstruction efficiency
$\epsilon \sim 0.6$ and correcting by the $\varkappa$ factor due to
obscuration by clouds, we find that for a flight with 138~hr of dark time, we expect $6.3
\pm 0.9$ events, whereas for a flight of 211~hr of dark time, we expect $10.6 \pm
2.3$ events.

\subsubsection{Estimate of UHECR event rates using CONEX + \Offline}

\begin{figure}[tbp]
 \postscript{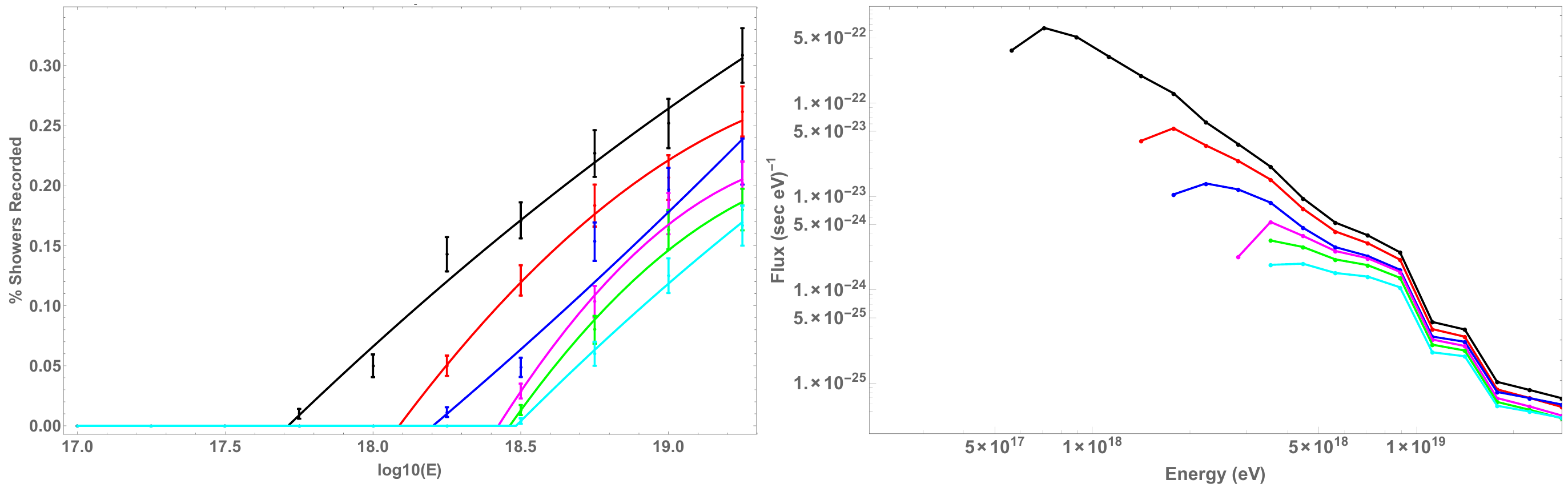}{0.9}
 \caption{{\it Left}. The fraction of observable extensive air showers relative to the
   number simulated events for different photon thresholds: 100+ photons/GTU,
   {\color{red} 200+ photons/GTU}, {\color{blue} 300+ photons/GTU},
   {\color{magenta} 400+ photons/GTU}, {\color{green} 500+
     photons/GTU}, and {\color{cyan} 600+ photons/GTU}. {\it
     Right}. The flux of extensive air showers observable by EUSO-SPB
   for the various photon thresholds~\cite{Cummings}. \label{fig:spb}}
\end{figure}

The expected number of detectable UHECR events has also been estimated by
Monte Carlo techniques using {\sc CONEX}~\cite{Bergmann:2006yz} to
simulate shower development and \Offline to simulate the detector
response. The results from these simulations for various pseudo-triggers are shown in
Fig.~\ref{fig:spb}.

\begin{figure}[tbp]
 \postscript{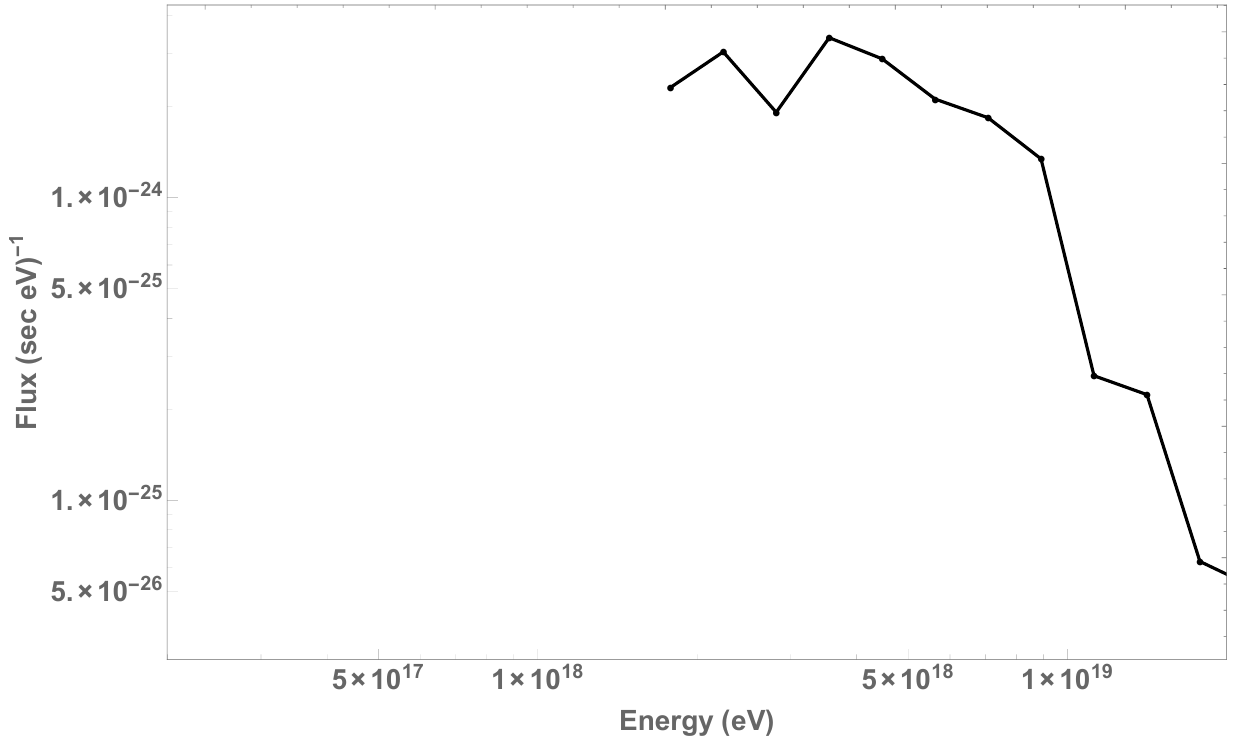}{0.9}
 \caption{Summation of flux curves shown in Fig.~\ref{fig:spb} with
   corresponding trigger efficiency~\cite{Cummings}. \label{total-flux}}
\end{figure}

A laser campaign was performed to assess the trigger
threshold for EUSO-SPB. The results of this campaign indicate that a 100\%
trigger efficiency requires a pseudo-trigger with 500+ photons at the
EUSO-SPB aperture per GTU, for 5 GTUs. However, using the flux curve
for the 500+ photons/GTU threshold alone does not fully capture how
many events EUSO-SPB might record. This is because 400+ photons/GTU
corresponds to an 85\% triggering efficiency, whereas 300+ photons/GTU
corresponds to 22\% triggering efficiency.  The 200+/GTU and 100+/GTU
pseudo-triggers are satisfied only by laser energies that did not
trigger during the field tests. Therefore, the total flux curve would
be the sum of the 500+ photons/GTU curve, the portion of the
400+ curve not captured by the 500+ curve scaled by 0.85, and the
portion of the 300+ curve not captured by the 400+ curve scaled by
0.22. A plot of the estimated total flux for EUSO-SPB is exhibited in Fig.~\ref{total-flux}.

The flux curves shown in Figs.~\ref{fig:spb} and \ref{total-flux} can be
integrated to estimate the expected trigger rate. The results of these
integrations are given in Table.~\ref{event-rates} and can be summarized as
follows: we expect 4.5 triggered events per 55 hours of operation, with an
energy threshold at about $10^{9.7}~{\rm GeV}$ and a photon threshold of 500 to
600 photons/GTU. This corresponds to a trigger event rate $n_{\rm tr} =
0.081~{\rm hr}^{-1}$.

As discussed above independent estimates of the trigger rates have
been performed including the effect of clouds of varying altitude and
thickness (in {\sc ESAF}) and assuming a clear sky (in both {\sc ESAF}
and \Offline). Using the ESAF reconstruction efficiency we can now
translate the CONEX + \Offline trigger rate into a rate of
reconstructed events, $n_{\rm re} = n_{\rm tr} \, \epsilon \sim 0.0486~{\rm
  hr}^{-1}$. Moreover, using the ESAF $\varkappa$ scaling factor we can
take account of cloud obscuration, and for a 138 dark hour flight we
expect a total of about 5 events to be recosntructed. Note that this agrees with the
estimate based soley on {\sc ESAF} simulation at the $1 \sigma$ level.

\begin{table}
\caption{Expected UHECR event rate for EUSO-SPB using various
   pseudo-triggers. The row highlighted in green corresponds to the nominal event
   rate estimated in~\cite{Fenn}. The rows highlighted in
   blue show the expected event rates for two possible pseudo-triggers
   with 100\% efficiency. The last row,  highlighted in red,
   indicates the best estimate, including the different trigger efficiencies~\cite{Cummings}. \label{event-rates}}
\begin{tabular}{|c|c|c|c|}
\hline
\hline
~~~~trigger threshold (photons/GTU)~~~~ & ~~~~event/hr~~~~ & ~~~~event/night~~~~ & ~~~~event/week~~~~ \\
\hline 
100 & ~~~~$2.300 \pm 0.005$~~~~ & ~~~~$18.1 \pm 0.04$~~~~ & ~~~~$127.0 \pm 0.3$~~~~ \\
200 & $0.430\pm 0.020$ & $3.40 \pm 0.10$ & $24.0 \pm 1.0$ \\
{\color{green} 300} & {\color{green} $0.180 \pm 0.010$} &
{\color{green} $1.40 \pm 0.08$} & {\color{green} $10.0 \pm 0.6$} \\
400 & $0.089 \pm 0.008$ & $0.71 \pm 0.07$ & $5.0 \pm 0.5$ \\
{\color{blue} 500} & {\color{blue} $0.065 \pm 0.007$} & {\color{blue}
  $0.52 \pm 0.05$} & {\color{blue} $3.7 \pm 0.4$} \\
{\color{blue} 600} & {\color{blue} $0.047 \pm 0.005$} & {\color{blue}
  $0.37 \pm 0.04$} & {\color{blue} $2.6 \pm 0.3$} \\
{\color{red} complete} & {\color{red} $0.081 \pm 0.010$} & {\color{red}
$0.64 \pm 0.08$} & {\color{red} $4.5 \pm 0.6$} \\
\hline
\hline
\end{tabular}
\end{table}

For EUSO-SPB2, it is likely the photon threshold will be improved by about a
factor of 2 compared to EUSO-SPB. The argument for this is as follows.  The
mirrors should deliver about twice as much light to the focal surface as the
EUSO-SPB lens system, and the spot size should cover about 2 pixels rather than
4-6 pixels. EUSO-SPB has roughly a threshold of 500 photons/GTU and 40 photons
from backgound, which yields 50 counts/GTU from signal and 4 counts/GTU from
background, assuming 0.1 overall efficiency.  Doubling the throughput of the
optical system would yield 100 counts/GTU from signal but still 4 counts from
background (2 pixels multiplied by 2 counts/pixel).   This 
would imply an improvement on the trigger rate by factor of 2. 
As the EUSO-SPB2 FoV could be $3.2^\circ \times 28.8^\circ$ (as opposed to the $12^\circ \times 12^\circ$~FoV
for EUSO-SPB), the trigger rate will also increase by a factor of 2.5 for showers of interest. With
this in mind, the number of reconstructed events in a $T \sim 100$~day mission,
which will be useful for physics analysis, is found to be
\begin{equation}
N_{\rm re} = 5 \, n_{\rm tr}  \ \epsilon \, \delta_{\rm eff}  \  T  \sim 87 \, .
\end{equation}
All in all, we expect a sample size sufficient to perform interesting analysis.

\section{DEVELOPMENT SCHEDULE AND DATA MANAGEMENT}

\begin{figure}[tbp]
 \postscript{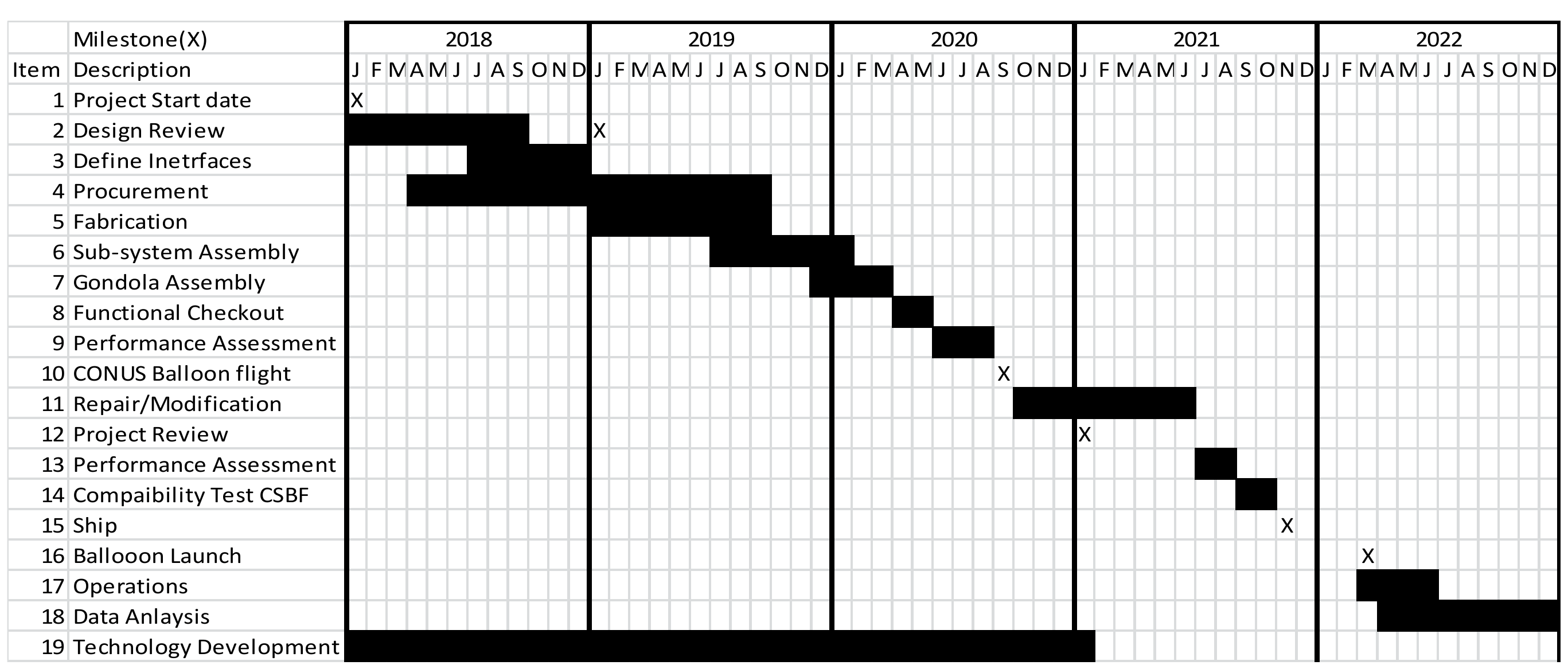}{0.9}
 \caption{Development schedule for EUSO-SPB2. \label{fig:gantt}}
\end{figure}

A Gantt chart showing the development schedule is shown in Fig.~\ref{fig:gantt}.

Low-level processing: The \Offline and {\sc ESAF} software packages will be used to
reconstruct events recorded by EUSO-SPB2, classify and reject the
majority of events that are of no scientific value. At this stage,
preliminary calibrations will have been applied, but not corrections
for atmospheric conditions. This allows the flexibility to make a
revision of the Level 2 files later, without repeating Level
1 processing.

Higher level processing: Physically meaningful data such as arrival
directions, energies, and background estimates will be extracted from
the data sample using \Offline and {\sc ESAF}, and refined later
using atmospheric and calibration databases validated or developed
from observational data. In addition to these Level 2 data, we will
produce a Level 3 product containing an exposure and efficiency
map. This will allow scientists outside the core team to investigate
the data.

Software and Data Archiving and Release: The EUSO US Data Center located at
the SOC will have a Redundant Array of Independent Disks (RAID) for
storage of Level 1-3 software. After approximately a 1 year period for
calibration of data analysis, Level 2 data files and Level 3 products
will be delivered to the scientific community along with software and
documentation required to perform analysis.


\begin{thebibliography}{99}


\bibitem{Neronov:2016iax}
  A.~Neronov, D.~V.~Semikoz, I.~Vovk and R.~Mirzoyan,
  {\color{rossoCP3}   Cosmic-ray composition measurements and cosmic ray background-free $\gamma$-ray observations with Cherenkov telescopes},
  Phys.\ Rev.\ D {\bf 94},  123018 (2016)
  doi:10.1103/PhysRevD.94.123018
  [arXiv:1610.01794 [astro-ph.IM]].



\bibitem{Neronov:2016zou}
  A.~Neronov, D.~V.~Semikoz, L.~A.~Anchordoqui, J.~Adams and A.~V.~Olinto,
  {\color{rossoCP3} Sensitivity of a proposed space-based Cherenkov astrophysical-neutrino telescope},
  Phys.\ Rev.\ D {\bf 95},  023004 (2017)
  doi:10.1103/PhysRevD.95.023004
  [arXiv:1606.03629 [astro-ph.IM]].


\bibitem{Aab:2016hkv}
  A.~Aab {\it et al.} [Pierre Auger Collaboration],
   {\color{rossoCP3}  Testing hadronic interactions at ultrahigh energies with air showers measured by the Pierre Auger Observatory},
  Phys.\ Rev.\ Lett.\  {\bf 117}, 192001 (2016)
  doi:10.1103/PhysRevLett.117.192001
  [arXiv:1610.08509 [hep-ex]].


\bibitem{Stecker:2004wt} 
  F.~W.~Stecker, J.~F.~Krizmanic, L.~M.~Barbier, E.~Loh, J.~W.~Mitchell, P.~Sokolsky and R.~E.~Streitmatter,
  {\color{rossoCP3} Observing the ultrahigh-energy universe with OWL eyes},
  Nucl.\ Phys.\ Proc.\ Suppl.\  {\bf 136C}, 433 (2004)
  doi:10.1016/j.nuclphysbps.2004.10.027
  [astro-ph/0408162].



\bibitem{Anchordoqui:2013eqa}
  L.~A.~Anchordoqui, G. R. Farrar, J. F. Krizmanic, J. Matthews,
  J. W. Mitchell, D. Nitz, A. V. Olinto, T. C. Paul, P. Sokolsky,
  G. B. Thomson, and T. J. Weiler,
   {\color{rossoCP3}  Roadmap for Ultra-High Energy Cosmic Ray Physics and Astronomy (whitepaper for Snowmass 2013)},
  arXiv:1307.5312 [astro-ph.HE].


\bibitem{nasa1} 2011 NASA Strategic Plan. 
Available at:\\ {\tt https://www.nasa.gov/pdf/516579main$_-$NASA2011StrategicPlan.pdf}

\bibitem{nasa2} 2010 NASA Science Plan. (for the Science Mission
  Directory).
Available at:\\  {\tt
  https://smd-prod.s3.amazonaws.com/science-green/s3fs-public/mnt/medialibrary\\}{\tt /2010/08/30/2010SciencePlan$_-$TAGGED.pdf}

\bibitem{cqc}  {\color{rossoCP3} Connecting Quarks with the Cosmos: Eleven Science Questions for the New Century} Washington, DC. The National Academies Press, 2003. 1.


\bibitem{roadmap} {\tt
    https://smd-prod.s3.amazonaws.com/science-pink/s3fs-public/atoms/files/secure}\\
  {\tt
    -Astrophysics$_-$Roadmap$_-$2013.pdf}

\bibitem{Beatty:2013lza}
  J.~J.~Beatty {\it et al.},
  {\color{rossoCP3} Working Group Report: Cosmic probes of fundamental physics},
  arXiv:1310.5662 [hep-ph].




\bibitem{Antoni:2005wq} 
  T.~Antoni {\it et al.} [KASCADE Collaboration],
  {\color{rossoCP3} KASCADE measurements of energy spectra for elemental groups of cosmic rays: Results and open problems},
  Astropart.\ Phys.\  {\bf 24}, 1 (2005)
  doi:10.1016/j.astropartphys.2005.04.001
  [astro-ph/0505413].





\bibitem{Abraham:2010mj} 
  J.~Abraham {\it et al.} [Pierre Auger Collaboration],
  {\color{rossoCP3} Measurement of the energy spectrum of cosmic rays above $10^{18}$~eV using the Pierre Auger Observatory},
  Phys.\ Lett.\ B {\bf 685}, 239 (2010)
  doi:10.1016/j.physletb.2010.02.013
  [arXiv:1002.1975 [astro-ph.HE]].


\bibitem{Bird:1993yi} 
  D.~J.~Bird {\it et al.} [HiRes Collaboration],
  {\color{rossoCP3} Evidence for correlated changes in the spectrum and composition of cosmic rays at extremely high-energies},
  Phys.\ Rev.\ Lett.\  {\bf 71}, 3401 (1993)
  doi:10.1103/PhysRevLett.71.3401.

\bibitem{Abbasi:2007sv} 
  R.~U.~Abbasi {\it et al.} [HiRes Collaboration],
  {\color{rossoCP3} First observation of the Greisen-Zatsepin-Kuzmin suppression},
  Phys.\ Rev.\ Lett.\  {\bf 100}, 101101 (2008)
  doi:10.1103/PhysRevLett.100.101101
  [astro-ph/0703099].


\bibitem{Abraham:2008ru} 
  J.~Abraham {\it et al.} [Pierre Auger Collaboration],
  {\color{rossoCP3} Observation of the suppression of the flux of cosmic rays above $4\times 10^{19}$~eV},
  Phys.\ Rev.\ Lett.\  {\bf 101}, 061101 (2008)
  doi:10.1103/PhysRevLett.101.061101
  [arXiv:0806.4302 [astro-ph]].

\bibitem{Apel:2012tda} 
  W.~D.~Apel {\it et al.},
  {\color{rossoCP3} The spectrum of high-energy cosmic rays measured with KASCADE-Grande},
  Astropart.\ Phys.\  {\bf 36}, 183 (2012).
  doi:10.1016/j.astropartphys.2012.05.023


\bibitem{Aartsen:2013wda} 
  M.~G.~Aartsen {\it et al.} [IceCube Collaboration],
 {\color{rossoCP3} Measurement of the cosmic ray energy spectrum with IceTop-73},
  Phys.\ Rev.\ D {\bf 88}, no. 4, 042004 (2013)
  doi:10.1103/PhysRevD.88.042004
  [arXiv:1307.3795 [astro-ph.HE]].


\bibitem{Knurenko:2013dia} 
  S.~P.~Knurenko, Z.~E.~Petrov, R.~Sidorov, I.~Y.~Sleptsov, S.~K.~Starostin and G.~G.~Struchkov,
  {\color{rossoCP3} Cosmic ray spectrum in the energy range $10^{15}-10^{18}$~eV and the second knee according to the small Cherenkov setup at the Yakutsk EAS array},
  arXiv:1310.1978 [astro-ph.HE].



\bibitem{Prosin:2014dxa} 
  V.~V.~Prosin {\it et al.},
 {\color{rossoCP3} Tunka-133: Results of 3 year operation},
  Nucl.\ Instrum.\ Meth.\ A {\bf 756}, 94 (2014).
  doi:10.1016/j.nima.2013.09.018


\bibitem{AbuZayyad:2000ay} 
  T.~Abu-Zayyad {\it et al.} [HiRes-MIA Collaboration],
  {\color{rossoCP3} Measurement of the cosmic ray energy spectrum and composition from $10^{17}$-eV to $10^{18.3}$-eV using a hybrid fluorescence technique},
  Astrophys.\ J.\  {\bf 557}, 686 (2001)
  doi:10.1086/322240
  [astro-ph/0010652].



\bibitem{Bergman:2007kn} 
  D.~R.~Bergman and J.~W.~Belz,
  {\color{rossoCP3} Cosmic rays: The second knee and beyond},
  J.\ Phys.\ G {\bf 34}, R359 (2007)
  doi:10.1088/0954-3899/34/10/R01
  [arXiv:0704.3721 [astro-ph]].

\bibitem{Hoerandel:2002yg} 
  J.~R.~Hoerandel,
 {\color{rossoCP3} On the knee in the energy spectrum of cosmic rays},
  Astropart.\ Phys.\  {\bf 19}, 193 (2003)
  doi:10.1016/S0927-6505(02)00198-6
  [astro-ph/0210453].

\bibitem{Olive:2016xmw} 
  C.~Patrignani {\it et al.} [Particle Data Group],
     {\color{rossoCP3}  Review of Particle Physics},
  Chin.\ Phys.\ C {\bf 40}, no. 10, 100001 (2016).
  doi:10.1088/1674-1137/40/10/100001



\bibitem{Allard:2005ha} D.~Allard, E.~Parizot, E.~Khan, S.~Goriely and
  A.~V.~Olinto, 
{\color{rossoCP3} UHE nuclei propagation and the
    interpretation of the ankle in the cosmic-ray spectrum}, Astron.\
  Astrophys.\ {\bf 443}, L29 (2005) doi:10.1051/0004-6361:200500199
  [astro-ph/0505566].



\bibitem{Allard:2005cx} 
  D.~Allard, E.~Parizot and A.~V.~Olinto,
  {\color{rossoCP3} On the transition from galactic to extragalactic cosmic-rays: spectral and composition features from two opposite scenarios},
  Astropart.\ Phys.\  {\bf 27}, 61 (2007)
  doi:10.1016/j.astropartphys.2006.09.006
  [astro-ph/0512345].



\bibitem{Greisen:1966jv} 
  K.~Greisen,
 {\color{rossoCP3} End to the cosmic ray spectrum?},
  Phys.\ Rev.\ Lett.\  {\bf 16}, 748 (1966)
  doi:10.1103/PhysRevLett.16.748.

\bibitem{Zatsepin:1966jv} 
  G.~T.~Zatsepin and V.~A.~Kuzmin,
  {\color{rossoCP3} Upper limit of the spectrum of cosmic rays},
  JETP Lett.\  {\bf 4}, 78 (1966)
  [Pisma Zh.\ Eksp.\ Teor.\ Fiz.\  {\bf 4}, 114 (1966)].




\bibitem{Hillas:1967} M. Hillas,
  {\color{rossoCP3} The energy spectrum of cosmic rays in an evolving universe},
  Phys. Lett. A {\bf 24}, 677 (1967)
doi:10.1016/0375-9601(67)91023-7.



\bibitem{Berezinsky:2002nc} 
  V.~Berezinsky, A.~Z.~Gazizov and S.~I.~Grigorieva,
  {\color{rossoCP3} On astrophysical solution to ultrahigh-energy cosmic rays},
  Phys.\ Rev.\ D {\bf 74}, 043005 (2006)
  doi:10.1103/PhysRevD.74.043005
  [hep-ph/0204357].










\bibitem{Kampert:2012mx} 
  K.~H.~Kampert and M.~Unger,
  {\color{rossoCP3} Measurements of the cosmic ray composition with air shower experiments},
  Astropart.\ Phys.\  {\bf 35}, 660 (2012)
  doi:10.1016/j.astropartphys.2012.02.004
  [arXiv:1201.0018 [astro-ph.HE]].


\bibitem{Aab:2014kda} 
  A.~Aab {\it et al.} [Pierre Auger Collaboration],
  {\color{rossoCP3} Depth of maximum of air-shower profiles at the Pierre Auger Observatory I: Measurements at energies above $10^{17.8}$ ~eV},
  Phys.\ Rev.\ D {\bf 90}, no. 12, 122005 (2014)
  doi:10.1103/PhysRevD.90.122005
  [arXiv:1409.4809 [astro-ph.HE]].

\bibitem{Aab:2014aea} 
  A.~Aab {\it et al.} [Pierre Auger Collaboration],
  {\color{rossoCP3} Depth of maximum of air-shower profiles at the Pierre Auger Observatory II: Composition implications},
  Phys.\ Rev.\ D {\bf 90}, no. 12, 122006 (2014)
  doi:10.1103/PhysRevD.90.122006
  [arXiv:1409.5083 [astro-ph.HE]].





\bibitem{Aab:2016htd} 
  A.~Aab {\it et al.} [Pierre Auger Collaboration],
   {\color{rossoCP3} Evidence for a mixed mass composition at the ‘ankle’ in the cosmic-ray spectrum},
  Phys.\ Lett.\ B {\bf 762}, 288 (2016)
  doi:10.1016/j.physletb.2016.09.039
  [arXiv:1609.08567 [astro-ph.HE]].



\bibitem{Aab:2016zth} 
  A.~Aab {\it et al.} [Pierre Auger Collaboration],
  {\color{rossoCP3}  Combined fit of spectrum and composition data as measured by the Pierre Auger Observatory},
  [arXiv:1612.07155 [astro-ph.HE]].




\bibitem{Auger:2012an} 
  P.~Abreu {\it et al.} [Pierre Auger Collaboration],
  {\color{rossoCP3} Large scale distribution of arrival directions of cosmic rays detected above $10^{18}$ eV at the Pierre Auger Observatory},
  Astrophys.\ J.\ Suppl.\  {\bf 203}, 34 (2012)
  doi:10.1088/0067-0049/203/2/34
  [arXiv:1210.3736 [astro-ph.HE]].


\bibitem{ThePierreAuger:2014nja} 
  A.~Aab {\it et al.} [Pierre Auger Collaboration],
  {\color{rossoCP3} Large scale distribution of ultrahigh energy cosmic rays detected at the Pierre Auger Observatory with zenith angles up to $80^\circ$},
  Astrophys.\ J.\  {\bf 802}, no. 2, 111 (2015)
  doi:10.1088/0004-637X/802/2/111
  [arXiv:1411.6953 [astro-ph.HE]].






\bibitem{Abbasi:2014sfa} 
  R.~U.~Abbasi {\it et al.},
  {\color{rossoCP3} Study of ultrahigh energy cosmic ray composition using Telescope Array’s Middle Drum detector and surface array in hybrid mode},
  Astropart.\ Phys.\  {\bf 64}, 49 (2014)
  doi:10.1016/j.astropartphys.2014.11.004
  [arXiv:1408.1726 [astro-ph.HE]].





\bibitem{Abbasi:2015xga} 
  R.~Abbasi {\it et al.} [Pierre Auger and Telescope Array Collaborations],
  {\color{rossoCP3} Report of the working group on the composition of ultrahigh energy cosmic rays},
  arXiv:1503.07540 [astro-ph.HE].

\bibitem{Abbasi:2014lda} 
  R.~U.~Abbasi {\it et al.} [Telescope Array Collaboration],
 {\color{rossoCP3} Indications of intermediate-scale anisotropy of cosmic rays with energy greater than 57 EeV in the Northern sky measured with the surface detector of the Telescope Array experiment},
  Astrophys.\ J.\  {\bf 790}, L21 (2014)
  doi:10.1088/2041-8205/790/2/L21
  [arXiv:1404.5890 [astro-ph.HE]].

\bibitem{Apel:2013ura} 
  W.~D.~Apel {\it et al.},
  {\color{rossoCP3} Ankle-like feature in the energy spectrum of light elements of cosmic rays observed with KASCADE-Grande},
  Phys.\ Rev.\ D {\bf 87}, 081101 (2013)
  doi:10.1103/PhysRevD.87.081101
  [arXiv:1304.7114 [astro-ph.HE]].



\bibitem{Aloisio:2013hya}
  R.~Aloisio, V.~Berezinsky and P.~Blasi,
  {\color{rossoCP3} Ultrahigh energy cosmic rays: implications of Auger data for source spectra and chemical composition},
  JCAP {\bf 1410}, no. 10, 020 (2014)
  doi:10.1088/1475-7516/2014/10/020
  [arXiv:1312.7459 [astro-ph.HE]].


\bibitem{Unger:2015laa} 
  M.~Unger, G.~R.~Farrar and L.~A.~Anchordoqui,
 {\color{rossoCP3} Origin of the ankle in the ultrahigh energy cosmic ray spectrum, and of the extragalactic protons below it},
  Phys.\ Rev.\ D {\bf 92}, no. 12, 123001 (2015)
  doi:10.1103/PhysRevD.92.123001
  [arXiv:1505.02153 [astro-ph.HE]].


\bibitem{Globus:2015xga} 
  N.~Globus, D.~Allard and E.~Parizot,
  {\color{rossoCP3} A complete model of the cosmic ray spectrum and composition across the Galactic to extragalactic transition},
  Phys.\ Rev.\ D {\bf 92}, no. 2, 021302 (2015)
  doi:10.1103/PhysRevD.92.021302
  [arXiv:1505.01377 [astro-ph.HE]].





\bibitem{Torres:2004hk}
  D.~F.~Torres and L.~A.~Anchordoqui,
  {\color{rossoCP3} Astrophysical origins of ultrahigh energy cosmic rays},
  Rept.\ Prog.\ Phys.\  {\bf 67}, 1663 (2004)
  doi:10.1088/0034-4885/67/9/R03
  [astro-ph/0402371].





\bibitem{Kotera:2011cp} 
  K.~Kotera and A.~V.~Olinto,
  {\color{rossoCP3} The astrophysics of ultrahigh energy cosmic rays},
  Ann.\ Rev.\ Astron.\ Astrophys.\  {\bf 49}, 119 (2011)
  doi:10.1146/annurev-astro-081710-102620
  [arXiv:1101.4256 [astro-ph.HE]].


\bibitem{Biermann:1987ep} 
  P.~L.~Biermann and P.~A.~Strittmatter,
    {\color{rossoCP3} Synchrotron emission from shock waves in active galactic nuclei},
  Astrophys.\ J.\  {\bf 322}, 643 (1987).
  doi:10.1086/165759


\bibitem{Rachen:1992pg} 
  J.~P.~Rachen and P.~L.~Biermann,
    {\color{rossoCP3} Extragalactic ultrahigh-energy cosmic rays I: Contribution from hot spots in FR-II radio galaxies},
  Astron.\ Astrophys.\  {\bf 272}, 161 (1993)
  [astro-ph/9301010].


\bibitem{Neronov:2007mh} 
  A.~Y.~Neronov, D.~V.~Semikoz and I.~I.~Tkachev,
    {\color{rossoCP3} Ultra-high energy cosmic ray production in the polar cap regions of black hole magnetospheres},
  New J.\ Phys.\  {\bf 11}, 065015 (2009)
  doi:10.1088/1367-2630/11/6/065015
  [arXiv:0712.1737 [astro-ph]].


\bibitem{Moncada:2017hvq} 
  R.~J.~Moncada, R.~A.~Colon, J.~J.~Guerra, M.~J.~O'Dowd and L.~A.~Anchordoqui,
   {\color{rossoCP3}  Ultrahigh energy cosmic ray nuclei from remnants of dead quasars},
  JHEAp (to be published) arXiv:1702.00053 [astro-ph.HE].


\bibitem{Cavallo:1978}
G. Cavallo,
  {\color{rossoCP3} On the sources of ultra-high energy cosmic rays},
Astron.\ Astrophys.\ {\bf 65}, 415 (1978).

\bibitem{Romero:1995tn} 
  G.~E.~Romero, J.~A.~Combi, L.~A.~Anchordoqui and S.~E.~Perez Bergliaffa,
    {\color{rossoCP3} A possible source of extragalactic cosmic rays with arrival energies beyond the GZK cutoff},
  Astropart.\ Phys.\  {\bf 5}, 279 (1996)
  doi:10.1016/0927-6505(96)00029-1
  [gr-qc/9511031].


\bibitem{PierreAuger:2014yba} 
  A.~Aab {\it et al.} [Pierre Auger Collaboration],
    {\color{rossoCP3} Searches for anisotropies in the arrival directions of the highest energy cosmic rays detected by the Pierre Auger Observatory},
  Astrophys.\ J.\  {\bf 804}, no. 1, 15 (2015)
  doi:10.1088/0004-637X/804/1/15
  [arXiv:1411.6111 [astro-ph.HE]].



\bibitem{Anchordoqui:1999cu} 
  L.~A.~Anchordoqui, G.~E.~Romero and J.~A.~Combi,
    {\color{rossoCP3} Heavy nuclei at the end of the cosmic ray spectrum?},
  Phys.\ Rev.\ D {\bf 60}, 103001 (1999)
  doi:10.1103/PhysRevD.60.103001
  [astro-ph/9903145].



\bibitem{Blasi:2000xm} 
  P.~Blasi, R.~I.~Epstein and A.~V.~Olinto,
   {\color{rossoCP3} Ultrahigh-energy cosmic rays from young neutron star winds},
  Astrophys.\ J.\  {\bf 533}, L123 (2000)
  doi:10.1086/312626
  [astro-ph/9912240].


\bibitem{Fang:2012rx} 
  K.~Fang, K.~Kotera and A.~V.~Olinto,
   {\color{rossoCP3} Newly-born pulsars as sources of ultrahigh energy cosmic rays},
  Astrophys.\ J.\  {\bf 750}, 118 (2012)
  doi:10.1088/0004-637X/750/2/118
  [arXiv:1201.5197 [astro-ph.HE]].


\bibitem{Fang:2013cba} 
  K.~Fang, K.~Kotera and A.~V.~Olinto,
   {\color{rossoCP3} Ultrahigh energy cosmic ray nuclei from extragalactic pulsars and the effect of their Galactic counterparts},
  JCAP {\bf 1303}, 010 (2013)
  doi:10.1088/1475-7516/2013/03/010
  [arXiv:1302.4482 [astro-ph.HE]].





\bibitem{Kotera:2015pya} 
  K.~Kotera, E.~Amato and P.~Blasi,
   {\color{rossoCP3} The fate of ultrahigh energy nuclei in the immediate environment of young fast-rotating pulsars},
  JCAP {\bf 1508}, no. 08, 026 (2015)
  doi:10.1088/1475-7516/2015/08/026
  [arXiv:1503.07907 [astro-ph.HE]].


\bibitem{Ackermann:2012vca} 
  M.~Ackermann {\it et al.} [Fermi-LAT Collaboration],
   {\color{rossoCP3} GeV observations of star-forming galaxies with \textit{Fermi}-LAT},
  Astrophys.\ J.\  {\bf 755}, 164 (2012)
  doi:10.1088/0004-637X/755/2/164
  [arXiv:1206.1346 [astro-ph.HE]].



\bibitem{Anchordoqui:2002dj} 
  L.~A.~Anchordoqui, H.~Goldberg and D.~F.~Torres,
    {\color{rossoCP3} Anisotropy at the end of the cosmic ray spectrum?},
  Phys.\ Rev.\ D {\bf 67}, 123006 (2003)
  doi:10.1103/PhysRevD.67.123006
  [astro-ph/0209546].



\bibitem{Anchordoqui:2014yva} 
  L.~A.~Anchordoqui, T.~C.~Paul, L.~H.~M.~da Silva, D.~F.~Torres and B.~J.~Vlcek,
    {\color{rossoCP3} What IceCube data tell us about neutrino emission from star-forming galaxies (so far)},
  Phys.\ Rev.\ D {\bf 89}, no. 12, 127304 (2014)
  doi:10.1103/PhysRevD.89.127304
  [arXiv:1405.7648 [astro-ph.HE]].


\bibitem{He:2014mqa} 
  H.~N.~He, A.~Kusenko, S.~Nagataki, B.~B.~Zhang, R.~Z.~Yang and Y.~Z.~Fan,
    {\color{rossoCP3} Monte Carlo Bayesian search for the plausible source of the
  Telescope Array hot spot},
  Phys.\ Rev.\ D {\bf 93}, 043011 (2016)
  doi:10.1103/PhysRevD.93.043011
  [arXiv:1411.5273 [astro-ph.HE]].


\bibitem{Pfeffer:2015idq} 
D.~N.~Pfeffer, E.~D.~Kovetz and  M.~Kamionkowski, 
{\color{rossoCP3} Ultra-high-energy-cosmic-ray hot
    spots from tidal disruption events}, 
Mon.\ Not.\ Roy.\ Astron.\ Soc.\ {\bf 466}, 2922 (2017)
 doi:10.1093/mnras/stw3337
  [arXiv:1512.04959 [astro-ph.HE]].




\bibitem{Nemmen:2010bp} 
  R.~S.~Nemmen, C.~Bonatto and T.~Storchi-Bergmann,
    {\color{rossoCP3}  A correlation between the highest energy cosmic rays and nearby active galactic nuclei detected by Fermi},
  Astrophys.\ J.\  {\bf 722}, 281 (2010)
  doi:10.1088/0004-637X/722/1/281
  [arXiv:1007.5317 [astro-ph.HE]].





\bibitem{Piran:1999kx} 
  T.~Piran,
    {\color{rossoCP3}  Gamma-ray bursts and the fireball model},
  Phys.\ Rept.\  {\bf 314}, 575 (1999)
  doi:10.1016/S0370-1573(98)00127-6
  [astro-ph/9810256].

\bibitem{Waxman:1995vg} 
  E.~Waxman,
   {\color{rossoCP3}  Cosmological gamma-ray bursts and the highest energy cosmic rays},
  Phys.\ Rev.\ Lett.\  {\bf 75}, 386 (1995)
  doi:10.1103/PhysRevLett.75.386
  [astro-ph/9505082].

\bibitem{Vietri:1995hs} 
  M.~Vietri,
    {\color{rossoCP3} On the acceleration of ultrahigh-energy cosmic rays in gamma-ray bursts},
  Astrophys.\ J.\  {\bf 453}, 883 (1995)
  doi:10.1086/176448
  [astro-ph/9506081].


\bibitem{Globus:2014fka} 
  N.~Globus, D.~Allard, R.~Mochkovitch and E.~Parizot,
  {\color{rossoCP3} UHECR acceleration at GRB internal shocks},
  Mon.\ Not.\ Roy.\ Astron.\ Soc.\  {\bf 451}, no. 1, 751 (2015)
  doi:10.1093/mnras/stv893
  [arXiv:1409.1271 [astro-ph.HE]].

\bibitem{Globus:2017ehu} 
  N.~Globus, D.~Allard, E.~Parizot and T.~Piran,
   {\color{rossoCP3} Probing the extragalactic cosmic rays origin with gamma-ray and neutrino backgrounds},
  arXiv:1703.04158 [astro-ph.HE].



\bibitem{Clay:2003pv} 
  R.~W.~Clay [for the Pierre Auger Collaboration],
    {\color{rossoCP3} The anisotropy search program for the Pierre Auger Observatory},
astro-ph/0308494.











\bibitem{Anchordoqui:1998nq} 
  L.~A.~Anchordoqui, M.~T.~Dova, L.~N.~Epele and S.~J.~Sciutto,
 {\color{rossoCP3} Hadronic interactions models beyond collider energies},
  Phys.\ Rev.\ D {\bf 59}, 094003 (1999)
  doi:10.1103/PhysRevD.59.094003
  [hep-ph/9810384].

\bibitem{Gondolo:1995fq} 
  P.~Gondolo, G.~Ingelman and M.~Thunman,
   {\color{rossoCP3} Charm production and high-energy atmospheric muon and neutrino fluxes},
  Astropart.\ Phys.\  {\bf 5}, 309 (1996)
  doi:10.1016/0927-6505(96)00033-3
  [hep-ph/9505417].




\bibitem{Anchordoqui:2004xb} 
  L.~Anchordoqui, M.~T.~Dova, A.~G.~Mariazzi, T.~McCauley, T.~C.~Paul, S.~Reucroft and J.~Swain,
 {\color{rossoCP3}  High energy physics in the atmosphere: Phenomenology of cosmic ray air showers},
  Annals Phys.\  {\bf 314}, 145 (2004)
  doi:10.1016/j.aop.2004.07.003
  [hep-ph/0407020].



\bibitem{Ulrich:2010rg} 
  R.~Ulrich, R.~Engel and M.~Unger,
  Phys.\ Rev.\ D {\bf 83}, 054026 (2011)
  doi:10.1103/PhysRevD.83.054026
  [arXiv:1010.4310 [hep-ph]].




\bibitem{Aartsen:2013bka}
  M.~G.~Aartsen {\it et al.} [IceCube Collaboration],
   {\color{rossoCP3} First observation of PeV-energy neutrinos with IceCube},
  Phys.\ Rev.\ Lett.\  {\bf 111}, 021103 (2013)
  doi:10.1103/PhysRevLett.111.021103
  [arXiv:1304.5356 [astro-ph.HE]].



\bibitem{Aartsen:2013jdh}
  M.~G.~Aartsen {\it et al.} [IceCube Collaboration],
   {\color{rossoCP3} Evidence for high-energy extraterrestrial neutrinos at the IceCube detector},
  Science {\bf 342}, 1242856 (2013)
  doi:10.1126/science.1242856
  [arXiv:1311.5238 [astro-ph.HE]].






\bibitem{Aartsen:2014muf}
  M.~G.~Aartsen {\it et al.} [IceCube Collaboration],
     {\color{rossoCP3} Atmospheric and astrophysical neutrinos above 1 TeV interacting in IceCube},
  Phys.\ Rev.\ D {\bf 91},  022001 (2015)
  doi:10.1103/PhysRevD.91.022001
  [arXiv:1410.1749 [astro-ph.HE]].


\bibitem{Aartsen:2014gkd}
  M.~G.~Aartsen {\it et al.} [IceCube Collaboration],
   {\color{rossoCP3} Observation of high-energy astrophysical neutrinos in three years of IceCube data},
  Phys.\ Rev.\ Lett.\  {\bf 113}, 101101 (2014)
  doi:10.1103/PhysRevLett.113.101101
  [arXiv:1405.5303 [astro-ph.HE]].

\bibitem{Aartsen:2015zva}
  M.~G.~Aartsen {\it et al.} [IceCube Collaboration],
  {\color{rossoCP3}  The IceCube neutrino observatory - Contributions
    to ICRC 2015 Part II: Atmospheric and astrophysical diffuse neutrino searches of all flavors},
  arXiv:1510.05223 [astro-ph.HE].



\bibitem{Gaisser:1994yf} 
  T.~K.~Gaisser, F.~Halzen and T.~Stanev,
    {\color{rossoCP3} Particle astrophysics with high-energy neutrinos},
  Phys.\ Rept.\  {\bf 258}, 173 (1995)
  Erratum: [Phys.\ Rept.\  {\bf 271}, 355 (1996)]
  doi:10.1016/0370-1573(95)00003-Y
  [hep-ph/9410384].



\bibitem{Learned:2000sw} 
  J.~G.~Learned and K.~Mannheim,
    {\color{rossoCP3} High-energy neutrino astrophysics},
  Ann.\ Rev.\ Nucl.\ Part.\ Sci.\  {\bf 50}, 679 (2000).
  doi:10.1146/annurev.nucl.50.1.679

\bibitem{Halzen:2002pg} 
  F.~Halzen and D.~Hooper,
    {\color{rossoCP3} High-energy neutrino astronomy: The Cosmic ray connection},
  Rept.\ Prog.\ Phys.\  {\bf 65}, 1025 (2002)
  doi:10.1088/0034-4885/65/7/201
  [astro-ph/0204527].


\bibitem{Anchordoqui:2009nf} 
  L.~A.~Anchordoqui and T.~Montaruli,
    {\color{rossoCP3} In search for extraterrestrial high energy neutrinos},
  Ann.\ Rev.\ Nucl.\ Part.\ Sci.\  {\bf 60}, 129 (2010)
  doi:10.1146/annurev.nucl.012809.104551
  [arXiv:0912.1035 [astro-ph.HE]].



\bibitem{Anchordoqui:2013dnh} 
  L.~A.~Anchordoqui {\it et al.},
    {\color{rossoCP3} Cosmic neutrino pevatrons: A brand new pathway to astronomy, astrophysics, and particle physics},
  JHEAp {\bf 1-2}, 1 (2014)
  doi:10.1016/j.jheap.2014.01.001
  [arXiv:1312.6587 [astro-ph.HE]].



\bibitem{Aartsen:2015dml} 
  M.~G.~Aartsen {\it et al.} [IceCube and Pierre Auger and Telescope Array Collaborations],
    {\color{rossoCP3} Search for correlations between the arrival directions of IceCube neutrino events and ultrahigh-energy cosmic rays detected by the Pierre Auger Observatory and the Telescope Array},
  JCAP {\bf 1601}, no. 01, 037 (2016)
  doi:10.1088/1475-7516/2016/01/037
  [arXiv:1511.09408 [astro-ph.HE]].

\bibitem{Fang:2014uja} 
  K.~Fang, T.~Fujii, T.~Linden and A.~V.~Olinto,
    {\color{rossoCP3} Is the ultra-high energy cosmic-ray excess observed by the Telescope Array correlated with IceCube neutrinos?},
  Astrophys.\ J.\  {\bf 794}, no. 2, 126 (2014)
  doi:10.1088/0004-637X/794/2/126
  [arXiv:1404.6237 [astro-ph.HE]].



\bibitem{GonzalezGarcia:2007ib} 
  M.~C.~Gonzalez-Garcia and M.~Maltoni,
   {\color{rossoCP3} Phenomenology with massive neutrinos},
  Phys.\ Rept.\  {\bf 460}, 1 (2008)
  doi:10.1016/j.physrep.2007.12.004
  [arXiv:0704.1800 [hep-ph]].


\bibitem{Aartsen:2015ivb}
  M.~G.~Aartsen {\it et al.} [IceCube Collaboration],
 {\color{rossoCP3} Flavor ratio of astrophysical neutrinos above 35~TeV in IceCube},
  Phys.\ Rev.\ Lett.\  {\bf 114},  171102 (2015)
  doi:10.1103/PhysRevLett.114.171102
  [arXiv:1502.03376 [astro-ph.HE]].




\bibitem{Palomares-Ruiz:2015mka}
  S.~Palomares-Ruiz, A.~C.~Vincent and O.~Mena,
   {\color{rossoCP3} Spectral analysis of the high-energy IceCube neutrinos},
  Phys.\ Rev.\ D {\bf 91},  103008 (2015)
  doi:10.1103/PhysRevD.91.103008
  [arXiv:1502.02649 [astro-ph.HE]].

\bibitem{Learned:1994wg} 
  J.~G.~Learned and S.~Pakvasa,
   {\color{rossoCP3}  Detecting tau-neutrino oscillations at PeV energies},
  Astropart.\ Phys.\  {\bf 3}, 267 (1995)
  doi:10.1016/0927-6505(94)00043-3
  [hep-ph/9405296, hep-ph/9408296].


\bibitem{Xu:2017yxo}
  D.~Xu [for the IceCube Collaboration],
  {\color{rossoCP3}    Search for astrophysical tau neutrinos with IceCube},
  arXiv:1702.05238 [astro-ph.HE].



\bibitem{Fargion:2000iz} 
  D.~Fargion,
    {\color{rossoCP3} Discovering ultra high energy neutrinos by horizontal and upward tau air-showers: Evidences in terrestrial gamma flashes?},
  Astrophys.\ J.\  {\bf 570}, 909 (2002)
  doi:10.1086/339772
  [astro-ph/0002453].


\bibitem{Feng:2001ue} 
  J.~L.~Feng, P.~Fisher, F.~Wilczek and T.~M.~Yu,
    {\color{rossoCP3} Observability of earth skimming ultrahigh-energy neutrinos},
  Phys.\ Rev.\ Lett.\  {\bf 88}, 161102 (2002)
  doi:10.1103/PhysRevLett.88.161102
  [hep-ph/0105067].


\bibitem{Bertou:2001vm} 
  X.~Bertou, P.~Billoir, O.~Deligny, C.~Lachaud and A.~Letessier-Selvon,
  {\color{rossoCP3}   Tau neutrinos in the Auger Observatory: A new window to UHECR sources},
  Astropart.\ Phys.\  {\bf 17}, 183 (2002)
  doi:10.1016/S0927-6505(01)00147-5
  [astro-ph/0104452].




\bibitem{Ackermann:2014usa} 
  M.~Ackermann {\it et al.} [Fermi-LAT Collaboration],
   {\color{rossoCP3} The spectrum of isotropic diffuse gamma-ray emission between 100~MeV and 820~GeV},
  Astrophys.\ J.\  {\bf 799}, 86 (2015)
  doi:10.1088/0004-637X/799/1/86
  [arXiv:1410.3696 [astro-ph.HE]].


\bibitem{TheFermi-LAT:2015ykq} 
  M.~Ackermann {\it et al.} [Fermi-LAT Collaboration],
   {\color{rossoCP3} Resolving the extragalactic $\gamma$-ray background above 50~GeV with the Fermi Large Area Telescope},
  Phys.\ Rev.\ Lett.\  {\bf 116}, no. 15, 151105 (2016)
  doi:10.1103/PhysRevLett.116.151105
  [arXiv:1511.00693 [astro-ph.CO]].





\bibitem{Kang:2015gpa} 
  D.~Kang {\it et al.},
     {\color{rossoCP3}  A limit on the diffuse gamma-rays measured with KASCADE-Grande},
  J.\ Phys.\ Conf.\ Ser.\  {\bf 632}, no. 1, 012013 (2015).
  doi:10.1088/1742-6596/632/1/012013



\bibitem{Aab:2015bza} 
  A.~Aab {\it et al.} [Pierre Auger Collaboration],
     {\color{rossoCP3}  The Pierre Auger Observatory: Contributions to the 34th International Cosmic Ray Conference (ICRC 2015)},
  arXiv:1509.03732 [astro-ph.HE].


\bibitem{Aab:2016agp} 
  A.~Aab {\it et al.} [Pierre Auger Collaboration],
     {\color{rossoCP3}  Search for photons with energies above 10$^{18}$~eV using the hybrid detector of the Pierre Auger Observatory},
  [arXiv:1612.01517 [astro-ph.HE]].






\bibitem{AlvarezMuniz:2012dd}
  J.~Alvarez-Muniz, L.~Cazon, R.~Conceicao, J.~D.~de Deus, C.~Pajares and M.~Pimenta,
    {\color{rossoCP3}  Muon production and string percolation effects in cosmic rays at the highest energies},
  arXiv:1209.6474 [hep-ph].




\bibitem{Farrar:2013sfa}
  G.~R.~Farrar and J.~D.~Allen,
     {\color{rossoCP3}  A new physical phenomenon in ultra-high energy collisions},
  EPJ Web Conf.\  {\bf 53}, 07007 (2013)
  doi:10.1051/epjconf/20135307007
  [arXiv:1307.2322 [hep-ph]].


\bibitem{Anchordoqui:2016oxy} 
  L.~A.~Anchordoqui, H.~Goldberg and T.~J.~Weiler,
   {\color{rossoCP3} Strange fireball as an explanation of the muon excess in Auger data},
  Phys.\ Rev.\ D {\bf 95}, no. 6, 063005 (2017)
  doi:10.1103/PhysRevD.95.063005
  [arXiv:1612.07328 [hep-ph]].





\bibitem{Tomar:2017mgc}
  G.~Tomar,
   {\color{rossoCP3}  Lorentz invariance violation as an explanation of muon excess in Auger data},
  arXiv:1701.05890 [hep-ph].

\bibitem{Fomin:2016kul}
  Y.~A.~Fomin, N.~N.~Kalmykov, I.~S.~Karpikov, G.~V.~Kulikov, M.~Y.~Kuznetsov, G.~I.~Rubtsov, V.~P.~Sulakov and S.~V.~Troitsky,
  {\color{rossoCP3}   No muon excess in extensive air showers at 100-500 PeV primary energy: EAS-MSU results},
  arXiv:1609.05764 [astro-ph.HE].


\bibitem{AbuZayyad:1999xa}
  T.~Abu-Zayyad {\it et al.} [HiRes and MIA Collaborations],
    {\color{rossoCP3}  Evidence for changing of cosmic ray composition between $10^{17}$-eV and $10^{18}$-eV from multicomponent measurements},
  Phys.\ Rev.\ Lett.\  {\bf 84}, 4276 (2000)
  doi:10.1103/PhysRevLett.84.4276
  [astro-ph/9911144].


\bibitem{Haungs} A. Haungs [for the KASCADE-Grande Collaboration],
 {\color{rossoCP3} KASCADE-Grande: Composition and post-LHC models},
talk given at the 19th International Symposium on Very High Energy
Cosmic Ray Interactions, Moskow, Russia, August 22-27, 2016.





\bibitem{Beresinsky:1969qj} 
  V.~S.~Berezinsky and G.~T.~Zatsepin,
  {\color{rossoCP3} Cosmic rays at ultrahigh-energies (neutrino?)},
  Phys.\ Lett.\  {\bf 28B}, 423 (1969).
  doi:10.1016/0370-2693(69)90341-4


\bibitem{Stecker:1978ah} 
  F.~W.~Stecker,
  {\color{rossoCP3} Diffuse fluxes of cosmic high-energy neutrinos},
  Astrophys.\ J.\  {\bf 228}, 919 (1979).
  doi:10.1086/156919



\bibitem{Kotera:2010yn} 
  K.~Kotera, D.~Allard and A.~V.~Olinto,
  {\color{rossoCP3} Cosmogenic neutrinos: parameter space and detectabilty from PeV to ZeV},
  JCAP {\bf 1010}, 013 (2010)
  doi:10.1088/1475-7516/2010/10/013
  [arXiv:1009.1382 [astro-ph.HE]].


\bibitem{Ahlers:2010fw} 
  M.~Ahlers, L.~A.~Anchordoqui, M.~C.~Gonzalez-Garcia, F.~Halzen and S.~Sarkar,
  {\color{rossoCP3} GZK neutrinos after the \textit{Fermi}-LAT diffuse photon flux measurement},
  Astropart.\ Phys.\  {\bf 34}, 106 (2010)
  doi:10.1016/j.astropartphys.2010.06.003
  [arXiv:1005.2620 [astro-ph.HE]].


\bibitem{Aloisio:2015ega} 
  R.~Aloisio, D.~Boncioli, A.~di Matteo, A.~F.~Grillo, S.~Petrera and F.~Salamida,
  {\color{rossoCP3} Cosmogenic neutrinos and ultrahigh-energy cosmic ray models},
  JCAP {\bf 1510}, no. 10, 006 (2015)
  doi:10.1088/1475-7516/2015/10/006
  [arXiv:1505.04020 [astro-ph.HE]].




\bibitem{Aab:2015kma} 
  A.~Aab {\it et al.} [Pierre Auger Collaboration],
  {\color{rossoCP3} Improved limit to the diffuse flux of ultrahigh energy neutrinos from the Pierre Auger Observatory},
  Phys.\ Rev.\ D {\bf 91}, no. 9, 092008 (2015)
  doi:10.1103/PhysRevD.91.092008
  [arXiv:1504.05397 [astro-ph.HE]].


\bibitem{Gorham:2010kv} 
  P.~W.~Gorham {\it et al.} [ANITA Collaboration],
  {\color{rossoCP3} Observational constraints on the ultrahigh-energy cosmic neutrino flux from the second flight of the ANITA experiment},
  Phys.\ Rev.\ D {\bf 82}, 022004 (2010)
  Erratum: [Phys.\ Rev.\ D {\bf 85}, 049901 (2012)]
  doi:10.1103/PhysRevD.82.022004, 10.1103/PhysRevD.85.049901
  [arXiv:1011.5004 [astro-ph.HE]].

\bibitem{Decerprit:2011qe} 
  G.~Decerprit and D.~Allard,
  {\color{rossoCP3} Constraints on the origin of ultrahigh energy cosmic rays from cosmogenic neutrinos and photons},
  Astron.\ Astrophys.\  {\bf 535}, A66 (2011)
  doi:10.1051/0004-6361/201117673
  [arXiv:1107.3722 [astro-ph.HE]].


\bibitem{Kuzmin:1998uv} 
  V.~Kuzmin and I.~Tkachev,
  {\color{rossoCP3} Ultrahigh-energy cosmic rays, superheavy long living particles, and matter creation after inflation},
  JETP Lett.\  {\bf 68}, 271 (1998)
  [Pisma Zh.\ Eksp.\ Teor.\ Fiz.\  {\bf 68}, 255 (1998)]
  doi:10.1134/1.567858
  [hep-ph/9802304].



\bibitem{Dubovsky:1998pu} 
  S.~L.~Dubovsky and P.~G.~Tinyakov,
  {\color{rossoCP3} Galactic anisotropy as signature of CDM related ultrahigh-energy cosmic rays},
  JETP Lett.\  {\bf 68}, 107 (1998)
  doi:10.1134/1.567830
  [hep-ph/9802382].





\bibitem{Berezinsky:1998ed} 
  V.~Berezinsky and M.~Kachelriess,
  {\color{rossoCP3} Limiting SUSY QCD spectrum and its application for decays of superheavy particles},
  Phys.\ Lett.\ B {\bf 434}, 61 (1998)
  doi:10.1016/S0370-2693(98)00728-X
  [hep-ph/9803500].


\bibitem{Birkel:1998nx} 
  M.~Birkel and S.~Sarkar,
  {\color{rossoCP3} Extremely high-energy cosmic rays from relic particle decays},
  Astropart.\ Phys.\  {\bf 9}, 297 (1998)
  doi:10.1016/S0927-6505(98)00028-0
  [hep-ph/9804285].


\bibitem{Blasi:2001hr} 
  P.~Blasi, R.~Dick and E.~W.~Kolb,
  {\color{rossoCP3} Ultrahigh-energy cosmic rays from annihilation of superheavy dark matter},
  Astropart.\ Phys.\  {\bf 18}, 57 (2002)
  doi:10.1016/S0927-6505(02)00113-5
  [astro-ph/0105232].


\bibitem{Sarkar:2001se} 
  S.~Sarkar and R.~Toldra,
  {\color{rossoCP3} The high-energy cosmic ray spectrum from relic particle decay},
  Nucl.\ Phys.\ B {\bf 621}, 495 (2002)
  doi:10.1016/S0550-3213(01)00565-X
  [hep-ph/0108098].

\bibitem{Aloisio:2015lva} 
  R.~Aloisio, S.~Matarrese and A.~V.~Olinto,
  {\color{rossoCP3} Super heavy dark matter in light of BICEP2, Planck and ultrahigh energy cosmic ray observations},
  JCAP {\bf 1508}, no. 08, 024 (2015)
  doi:10.1088/1475-7516/2015/08/024
  [arXiv:1504.01319 [astro-ph.HE]].


\bibitem{Kierans:2017bmv} 
  C.~A.~Kierans {\it et al.},
   {\color{rossoCP3} The 2016 Super Pressure Balloon flight of the Compton Spectrometer and Imager},
  arXiv:1701.05558 [astro-ph.IM].

\bibitem{Adams:2012hr} 
  J.~H.~Adams, Jr. {\it et al.},
    {\color{rossoCP3} Summary Report of JEM-EUSO Workshop at KICP in Chicago},
  arXiv:1203.3451 [astro-ph.IM].

\bibitem{Bergmann:2006yz} 
  T.~Bergmann, R.~Engel, D.~Heck, N.~N.~Kalmykov, S.~Ostapchenko, T.~Pierog, T.~Thouw and K.~Werner,
   {\color{rossoCP3}  One-dimensional hybrid approach to extensive air shower simulation},
  Astropart.\ Phys.\  {\bf 26}, 420 (2007)
  doi:10.1016/j.astropartphys.2006.08.005
  [astro-ph/0606564].







\bibitem{Zemax} {\tt http://www.zemax.com}


\bibitem{Schioppa:2015yla} 
  E.~J.~Schioppa {\it et al.},
   {\color{rossoCP3} The SST-1M camera for the Cherenkov Telescope Array},
  PoS ICRC {\bf 2015}, 930 (2016)
  [arXiv:1508.06453 [astro-ph.IM]].







\bibitem{Argiro:2007qg} 
S.~Argiro, S.~L.~C.~Barroso, J.~Gonzalez, L.~Nellen, T.~C.~Paul, T.~A.~Porter, L.~Prado, Jr. and M.~Roth, R. Ulrich, and D. Veberi\u{c},  
 {\color{rossoCP3} The Offline software framework of the Pierre Auger Observatory},
  Nucl.\ Instrum.\ Meth.\ A {\bf 580}, 1485 (2007)
  doi:10.1016/j.nima.2007.07.010
  [arXiv:0707.1652 [astro-ph]].


\bibitem{Abreu:2011fb} 
  P.~Abreu {\it et al.} [Pierre Auger Collaboration],
  {\color{rossoCP3} Advanced functionality for radio analysis in the Offline software framework of the Pierre Auger Observatory},
  Nucl.\ Instrum.\ Meth.\ A {\bf 635}, 92 (2011)
  doi:10.1016/j.nima.2011.01.049
  [arXiv:1101.4473 [astro-ph.IM]].


\bibitem{Berat:2009va} 
  C.~Berat {\it et al.},
   {\color{rossoCP3} ESAF: Full simulation of space-based extensive air showers detectors},
  Astropart.\ Phys.\  {\bf 33}, 221 (2010)
  doi:10.1016/j.astropartphys.2010.02.005
  [arXiv:0907.5275 [astro-ph.IM]].



\bibitem{ThePierreAuger:2015rma}
  A.~Aab {\it et al.} [Pierre Auger Collaboration],
   {\color{rossoCP3} The Pierre Auger cosmic ray observatory},
  Nucl.\ Instrum.\ Meth.\ A {\bf 798}, 172 (2015)
  doi:10.1016/j.nima.2015.06.058
  [arXiv:1502.01323 [astro-ph.IM]].


\bibitem{cocomo}
  B. Boehm,
  {\color{rossoCP3} Software engineering economics},
  Englewood Cliffs, NJ:Prentice-Hall, 1981. ISBN 0-13-822122-7.



\bibitem{Bohacova:2008vg} 
  M.~Bohacova {\it et al.} [AIRFLY Collaboration],
   {\color{rossoCP3} A novel method for the absolute fluorescence yield measurement by AIRFLY},
  Nucl.\ Instrum.\ Meth.\ A {\bf 597}, 55 (2008)
  doi:10.1016/j.nima.2008.08.049
  [arXiv:0812.3649 [astro-ph]].


\bibitem{Sciutto:1999jh} 
  S.~J.~Sciutto,
 {\color{rossoCP3}  AIRES: A System for air shower simulations},
  astro-ph/9911331.



\bibitem{Heck:1998vt} 
  D.~Heck, G.~Schatz, T.~Thouw, J.~Knapp and J.~N.~Capdevielle,
  {\color{rossoCP3} CORSIKA: A Monte Carlo code to simulate extensive air showers},
  FZKA-6019.


\bibitem{Drescher:2002cr} 
  H.~J.~Drescher and G.~R.~Farrar,
  {\color{rossoCP3}  Air shower simulations in a hybrid approach using cascade equations},
  Phys.\ Rev.\ D {\bf 67}, 116001 (2003)
  doi:10.1103/PhysRevD.67.116001
  [astro-ph/0212018].

\bibitem{g4}
{\tt https://geant4.cern.ch}



\bibitem{Ardouin:2006gj} 
  D.~Ardouin {\it et al.},
    {\color{rossoCP3} CODALEMA: A cosmic ray air shower radio detection experiment},
  Int.\ J.\ Mod.\ Phys.\ A {\bf 21S1}, 192 (2006).
  doi:10.1142/S0217751X0603360X


\bibitem{Antokhonov:2011zz} 
  B.~A.~Antokhonov {\it et al.},
    {\color{rossoCP3} A new 1-km$^2$ EAS Cherenkov array in the Tunka valley},
  Nucl.\ Instrum.\ Meth.\ A {\bf 639}, 42 (2011).
  doi:10.1016/j.nima.2010.09.142


\bibitem{Abeysekara:2013tka} 
  A.~U.~Abeysekara {\it et al.} [HAWC Collaboration],
    {\color{rossoCP3} The HAWC gamma-ray observatory: Design, calibration, and operation},
  arXiv:1310.0074 [astro-ph.IM].

\bibitem{lofar}
{\tt http://www.lofar.org/}



\bibitem{Wyszynski:2012fa} 
  O.~Wyszynski, A.~Laszlo, A.~Marcinek, T.~Paul, R.~Sipos, M.~Szuba, M.~Unger and D.~Veberi\u{c},
    {\color{rossoCP3} Legacy code: Lessons from NA61/SHINE offline software upgrade adventure},
  J.\ Phys.\ Conf.\ Ser.\  {\bf 396}, 052076 (2012).
  doi:10.1088/1742-6596/396/5/052076


\bibitem{Sipos:2012hs} 
  R.~Sipos, A.~Laszlo, A.~Marcinek, T.~Paul, M.~Szuba, M.~Unger, D.~Veberi\'u{c} and O.~Wyszynski,
    {\color{rossoCP3} The offline software framework of the NA61/SHINE experiment},
  J.\ Phys.\ Conf.\ Ser.\  {\bf 396}, 022045 (2012).
  doi:10.1088/1742-6596/396/2/022045

\bibitem{bsd}
  {\tt http://www.opensource.org/}

\bibitem{root}
{\tt http://root.cern.ch/}



\bibitem{NaumovSlast}
D. Naumov,
{\color{rossoCP3} SLAST, Shower Light Attenuated to the Space Telescope}, EUSO
mission internal document (2003).

\bibitem{USStandard1976}
National Oceanic Atmospheric Administration et al.,
{\color{rossoCP3} U.S standard atmosphere, 1976}, public document (1976).


\bibitem{Nagano:2004am} 
  M.~Nagano, K.~Kobayakawa, N.~Sakaki and K.~Ando,
  {\color{rossoCP3} New measurement on photon yields from air and the application to the energy estimation of primary cosmic rays},
  Astropart.\ Phys.\  {\bf 22}, 235 (2004)
  doi:10.1016/j.astropartphys.2004.08.002
  [astro-ph/0406474].


\bibitem{Kakimoto:1995pr} 
  F.~Kakimoto, E.~C.~Loh, M.~Nagano, H.~Okuno, M.~Teshima and S.~Ueno,
   {\color{rossoCP3} A measurement of the air fluorescence yield},
  Nucl.\ Instrum.\ Meth.\ A {\bf 372}, 527 (1996).
  doi:10.1016/0168-9002(95)01423-3

\bibitem{lowtran7}
Air Force Geophysics Laboratory,
{\color{rossoCP3}Users guide to LOWTRAN 7},
public document (1988).





\bibitem{xerces}
{\tt http://xerces.apache.org/}

\bibitem{xml-schema}
  {\tt http://www.w3.org/standards/xml/schema}


\bibitem{cppunit}
  {\tt http://sourceforge.net/projects/cppunit/}

\bibitem{buildbot}
  {\tt http://buildbot.net/}

\bibitem{cmake}
  {\tt http://www.cmake.org}

\bibitem{ape}
  {\tt https://svn.auger.unam.mx/trac/projects/ape/}


\bibitem{Ballmoos:2015spu} 
  P.~V.~Ballmoos,
   {\color{rossoCP3}  The EUSO-BALLOON mission},
  PoS ICRC {\bf 2015}, 322 (2016).


\bibitem{Piotrowski:2014tsa} 
  L.~W.~Piotrowski {\it et al.},
   {\color{rossoCP3}  On-line and off-line data analysis for the EUSO-TA experiment},
  Nucl.\ Instrum.\ Meth.\ A {\bf 773}, 164 (2015).
  doi:10.1016/j.nima.2014.08.045


\bibitem{Bisconti:2017bla} 
  F.~Bisconti [JEM-EUSO Collaboration],
 {\color{rossoCP3}  EUSO-TA fluorescence detector},
  arXiv:1701.07091 [astro-ph.IM].


\bibitem{Ricci:2015srs} 
  M.~Ricci,
   {\color{rossoCP3}  Mini-EUSO: a pathfinder for JEM-EUSO to measure Earth’s UV background from the ISS},
  PoS ICRC {\bf 2015}, 599 (2016).

\bibitem{Wiencke:2015oko} 
  L.~Wiencke,
  {\color{rossoCP3}  EUSO-Balloon mission to record extensive air showers from near space},
  PoS ICRC {\bf 2015}, 631 (2016).

\bibitem{cambursanoThesis} S. Cambursano
{\color{rossoCP3} A simulation study of the EUSO-SPB trigger and reconstruction performances},
MSc Dissertation, Torino University (2016).





\bibitem{venezianiThesis} A. Veneziani,
{\color{rossoCP3} Study on the expected sky conditions during Super Pressure Balloon flights
and observation of cosmic rays},
Bachelor Thesis, Torino University (2016)

\bibitem{bertolaThesis} E. Bertola,
{\color{rossoCP3} Study on the respoonse of the Mini--EUSO first trigger level}
Bachelor Thesis, Torino University (2016).

\bibitem{neutrino2014} L. A. Anchordoqui, K. Islo, A. V. Olinto,
  T. C. Paul, and B. Vlcek, 
{\color{rossoCP3}  Cosmic neutrino detection from the International
  Space Station}, XXVI International Conference on Neutrino Physics
and Astrophysics (Neutrino 2014), Boston, June 2-7, 2014. {\tt http://neutrino2014.bu.edu/files/2013/08/Neutrino-2014-Poster-contributions- book.pdf}




\bibitem{Lorenz}
D. Lorenz, {\color{rossoCP3} Dark Site Finder}, {\tt http://darksitefinder.com/map/}.



\bibitem{Cano}
  G.~S\'aez-Cano {\it et al.} [JEM-EUSO Collaboration],
    {\color{rossoCP3} Observation of ultra-high energy cosmic rays in cloudy conditions
by the JEM-EUSO Space Observatory},
  doi:10.7529/ICRC2011/V03/1034


\bibitem{Adams:2014tnr}
  A.~Guzman {\it et al.} [JEM-EUSO Collaboration],
    {\color{rossoCP3} The JEM-EUSO observation in cloudy conditions},
  Exper.\ Astron.\  {\bf 40}, no. 1, 135 (2015).
  doi:10.1007/s10686-014-9377-2


\bibitem{SaezCano:2012km}
  G.~S\'aez-Cano {\it et al.} [JEM-EUSO Collaboration],
    {\color{rossoCP3} Observation of ultra-high energy cosmic rays in cloudy conditions
by the space-based JEM-EUSO Observatory},
  J.\ Phys.\ Conf.\ Ser.\  {\bf 375}, 052010 (2012).
  doi:10.1088/1742-6596/375/1/052010


\bibitem{Saez-Cano:2014zka}
  G.~S\'aez-Cano {\it et al.} [JEM-EUSO Collaboration],
  {\color{rossoCP3}  Observation of extensive air showers in cloudy conditions by the
JEM-EUSO Space Mission},
  Adv.\ Space Res.\  {\bf 53}, 1536 (2014).
  doi:10.1016/j.asr.2013.07.015

\bibitem{Cummings} 
A. Cummings,
{\color{rossoCP3} Field testing for extreme universe space observatory
  aboard a super pressure balloon (EUSO-SPB): Logistics and first results},
MSc Dissertation, Colorado School of Mines (2017). 

\bibitem{Fenn}
J. S. Fenn,  
{\color{rossoCP3}  Estimating the cosmic ray extensive air shower
  detection rate for the EUSO Super Pressure Balloon mission}, 
MSc Dissertation, Colorado School of Mines (2015). {\tt http://hdl.handle.net/11124/17150}


\end{thebibliography}
\end{document}